\documentclass[runningheads]{llncs}
\usepackage{newtxtext,newtxmath}
\usepackage{listings}
\usepackage{algorithm}
\usepackage{algpseudocode}
\usepackage{graphicx}
\usepackage{xspace}
\usepackage{enumitem}
\usepackage{multirow}
\usepackage{url} 
\usepackage{eucal}
\usepackage{subfig}
\usepackage{rotating}
\usepackage{colortbl} 
\usepackage{amsmath}
\usepackage{tablefootnote}
\usepackage{threeparttable}
\usepackage{soul}
\usepackage[font={footnotesize}]{caption}
\usepackage{makecell}

\newcommand{\redtext}[1]{}

\newcommand{\GB}{Gummy Browsers\xspace} 
\newcommand{\RLA}{RLA\xspace}
\newcommand{\HLA}{HLA\xspace}
\newcommand{\AOSO}{Acquire-Once-Spoof-Once\xspace}
\newcommand{\AOSF}{Acquire-Once-Spoof-Frequently\xspace}
\newcommand{\AFSF}{Acquire-Frequently-Spoof-Frequently\xspace}

\begin{document}
\title{Gummy Browsers: Targeted Browser Spoofing against State-of-the-Art Fingerprinting Techniques}

\author{Zengrui Liu\inst{1} \and
Prakash Shrestha\inst{2} \and Nitesh Saxena\inst{3}}
%

% First names are abbreviated in the running head.
% If there are more than two authors, 'et al.' is used.
%
\institute{Texas A\&M University, College Station TX 77843 \\\email{lzr@tamu.edu}\and University of Florida, Gainesville, FL 32611 \\\email{prakash.shrestha@ufl.edu}
\and
Texas A\&M University, College Station TX 77843 \\\email{nsaxena@tamu.edu}}

%\author{}

%\institute{}

\maketitle
\begin{abstract}
Here goes the abstract of the paper.

\end{abstract}
\section{Introduction}
%\secspace
\label{sec:introduction}

%\vspace{-2mm}
Many websites and web services leverage browser fingerprinting techniques to track their
users for various purposes, including targeted advertisements
\cite{hoofnagle2012behavioral} based on browsing history and habits, user authentication
\cite{identification2015,ref54,ref55}, and fraud detection
\cite{florian_2019,ipqualityscore}. Browser fingerprinting aims to
uniquely identify web browsers. 
%\textcolor{red}{without using stateful identifiers (e.g.,cookies)}. 
Specifically, browser fingerprinting uses a stateless identifier for
web browsers composed of a set of browser and system attributes, including browser
vendor and version, plugins and extensions, canvas rendering, available fonts,
performance characteristics, platform, clock skews and screen resolutions.
These attributes are collected through JavaScript APIs and HTTP
headers. 

Based on different combinations of browser and system attributes, and
their uniqueness to the browser, researchers 
and practitioners 
have proposed a
myriad of {browser fingerprinting} techniques \cite{mdnwebdocs,boda2011user,olejnik2015leaking,mowery2012pixel,mowery2011fingerprinting,das2016tracking,fifield2015fingerprinting,saleh_saleh_1960,ref46,ref47,alvestrand2002content,roesner2012detecting,tutorialrepublic,mdnwebdocs2,norte2016advanced,mesbah2012automated} .  However,  the uniqueness of the
fingerprint alone is not sufficient for prolonged user tracking because the browser
fingerprint changes over time, potentially when the browsers are updated or configured differently \cite{vastel2018fp}. For a successful long-term user tracking,
changes to the fingerprints need to be tracked to link the current fingerprint with
previously recorded fingerprints \cite{vastel2018fp,eckersley2010unique},
using what is referred to as a \textit{tracking technique}.

% Briefly describe Panopticlick and FPStalker linking algorithms.
The fingerprint linking algorithm \textit{Panopticlick}, proposed by Eckersley, 
%(from PETS 2010), 
 \cite{eckersley2010unique}, and
\textit{FP-Stalker} developed by Vastel et al. \cite{vastel2018fp}, 
%(from IEEE S\&P 2018)
are representative instantiations of such tracking techniques.  Panopticlick
showed that its visitors can be uniquely identified from a fingerprint composed
of only eight browser and system attributes. It follows a very simple heuristic
based on the comparison of the string representation of browser characteristics.
FP-Stalker consists of two variants of fingerprint linking algorithms -- a
\textit{rule-based variant} and a \textit{hybrid variant}, which leverage ruleset and machine learning
algorithms. These algorithms aim  to link browser fingerprint evolutions for
tracking the user.  The experiment conducted in the FP-Stalker paper \cite{vastel2018fp}  showed that its
linking algorithm, especially the hybrid variant, can track a given browser
instance for a long period of time, significantly better than Panopticlick. 
% for more than 51 days on average.  Details on each of these algorithms are
% presented in Section \ref{sec:background}.

In this paper, we 
%set out to 
closely investigate the potential privacy leakage
and security vulnerability associated with state-of-the-art browser fingerprint
linking algorithms, Panopticlick and FP-Stalker to be specific, motivated by
their very appealing applications and practicality features.  Unfortunately, we
identify a significant threat vector against such linking algorithms.
Specifically, we find that an attacker can capture and spoof the browser
characteristics of a victim's browser, and hence can ``present'' its own
browser as the victim's browser when connecting to a website.  The browser
attributes can be easily captured (one-time or frequently based on the
application) by luring the victim into visiting a benign-looking website
controlled by the attacker (or a malicious website). 
%\textcolor{blue}{This looks like easier for attackers to get the victim's browser cookies by using cross-script attacks. We may not discuss further comparisons between browser fingerprinting and cookies as this is not the focus of our work. }
Then, all (or most of)
these attributes can be spoofed (once, or continually based on the intended level
of adversarial impact on the victim), for example, by injecting a web script,
modifying the existing web script, or utilizing the browser's built-in settings
and debugging tools.  By spoofing the victim's browser characteristics, which
are used to construct its fingerprint, the attacker's browser would be
recognized as the victim's browser when visiting a targeted website. 
%and is
%highly likely to be deemed as the victim browser by the remote webserver
%hosting the targeted website.

Exploiting this general threat, we introduce \textit{\GB}, an attack system that can
fully compromise the security and privacy of the schemes that leverage browser
fingerprinting techniques. For instance, if the browser fingerprinting is employed for personalized and targeted ads, 
%In \GB, the attacker exposes its browser as the victim's browser.  Given this,
%the Since browser fingerprinting are often used by most of the websites for
%the purpose of personalized ads, 
the web server, hosting a benign website, would push the same or similar ads to the attacker's browser like the ones that would have been pushed to the victim's browser because the web server considers the attacker's browser as the victim's browser.  Based on the personalized ads (e.g., related to pregnancy products, medications and brands),  the attacker can infer various sensitive information about the victim (e.g, gender, age group, health condition, interests, salary level, etc.), even build a personal behavioral profile of the victim. Leakage of such personal and private information can raise a  frightful privacy threat to the user.  
The study of Castelluccia et al. \cite{castelluccia2012betrayed} has demonstrated that the knowledge of the ads the user is provided in targeted advertising can indeed leak significant sensitive information about the user. Similarly, if browser fingerprinting is used for security purposes, such as user authentication and fraud detection (e.g., clickbot detection),   
% Further, in a broader context, since browser fingerprinting techniques are
% often used for fraud detection and user authentication purposes, 
our fingerprint spoofing attacker can circumvent the security functionality of such defensive schemes. The authentication system may be based on some other factors beyond browser fingerprinting. In this paper, we only show how to defeat the fingerprinting factor.

\GB can remain hidden and invisible to the targeted user and the
targeted website.  Since the capturing and spoofing of the browser attributes
is done fully transparently and remotely, \GB can be launched easily and
effectively without being noticed by the user or the website.  In this light,
given the fact that browser fingerprinting techniques are getting deployed
widely in the real world, \GB   can have a devastating and lasting impact on
the online privacy and security of the users. Capturing the victim's fingerprinting information just once allows the attacker to spoof the victim for a long period of time. The process can be repeated for further impact.
Given the fundamental nature of the attack, it would be very difficult to
defeat. 

Our experiments consider that the website only uses
browser fingerprinting for tracking, and does not employ cookies
(or cookies are blocked by the user). Therefore
our attacks and implications of our attacks are only limited
to fingerprint spoofing.

%\vspace{-1mm}
%\smallskip
\noindent \textbf{Our Contributions:}
We believe that our work makes the following contributions: 

\begin{enumerate}[leftmargin = *]
%\vspace{-4mm}

	\item  \textit{\textbf{A Novel Threat of Spoofing Browser Fingerprints:}} We
		introduce a novel and serious threat raised due to the use of
		browser fingerprinting techniques to track the user, referred
		to as \textit{\GB}. Specifically, this attacker with the ability
		to capture and spoof the browser fingerprint can learn various
		personal and sensitive information about the user 
		%(e.g.,
		%gender, age group, profession, health condition, interests,
		%etc.) 
		based on personalized ads 
		%(e.g., pregnancy products,
		%medications, etc.)  
		and compromise the security of
		browser-fingerprinting based defensive applications, such as
		user authentication and fraud detection. The ease with which this threat 
		can be perpetrated is a strength of our work since it can be deployed in real world by even 
		naive attackers. 
	%	presented based on the user's browsing history and browser
		%	fingerprint. 
	
%\vspace{-3mm}
	\item \textit{\textbf{Design and Implementation of  \GB:}} We provide the design
		and implementation of \GB that enable an attacker to glean
		sensitive information about the user and compromise the browser
		fingerprinting based defensive schemes. \GB leverages a
		benign-looking fake website to capture the victim's browser
		characteristics (could also be a malicious, attacker-controlled website). 
		\GB then utilizes spoofing methods, such as
		\textit{script injection}, \textit{script modification}, or
		\textit{browser's built-in setting and debugging tool} to
		orchestrate its browser to appear as the victim's browser.
%\vspace{-3mm}	
	\item \textit{\textbf{Evaluation against Notable Fingerprinting
		Techniques:}} We employ state-of-the-art browser fingerprinting
		algorithms, specifically \textit{Panopticlick}
		\cite{eckersley2010unique} and \textit{FP-Stalker}
		\cite{vastel2018fp}, and evaluate the performance of \GB
		against them. Based on a dataset of 200+ users, our results
		show that the attacker can successfully spoof the fingerprint
		of the browser instance to match with that of the targeted
		victim's browser instance for a long period of time
		without any significant impact on the tracking of the victim. %We also demonstrate via visual analysis that \GB can be
%		used to effectively mimic many different types of browsers
%		(including mobile browsers and the Tor browser). 
		
%We further successfully tested our attack against the FingerprintJS demo site, which means our attack can bypass any web service that deploys FingerpringJS in real-world.
	
\end{enumerate}

%\textcolor{blue}{Our experiments considered that the website only uses browser
%fingerprinting for user tracking, and did not use cookies (or that the cookies were blocked by the user), and therefore our
%attacks and implications of our attacks are only limited to 
%fingerprint only.}

%\vspace{-5mm}
%\vspace{-2mm}
\section{Background and Related Work}
\label{sec:background}
%\secspace

%\vspace{-1mm}
\subsection{Browser Fingerprinting}
\label{sec:browser_fingerprinting}
%% PS 2/22: Why do you need browser fingerprinting? Its applications? Just list them. Detail description will come in later sections. Also, try to revise the definition, perhaps add citations.
% ZL 3/9: Updated
%\vspace{-2mm}
Different combinations of the browser and system attributes can be used to generate a unique identifier for a given browser, referred to as
the \textit{browser fingerprint}. 
\redtext{Websites and web services often leverage
various browser fingerprinting techniques to track its users for
various purposes, such as to push targeted advertisements, user
authentication, and fraud detection. 
}Based on different combinations
of attributes, various browser fingerprinting techniques have
been proposed
\cite{mdnwebdocs,boda2011user,olejnik2015leaking,mowery2012pixel,mowery2011fingerprinting,das2016tracking,norte2016advanced,mesbah2012automated}. 
These attributes can be grouped into three different categories~\cite{alaca2016device} as presented in Table \ref{tbl:features_categories}. 
%Below we briefly describe the three main categories these features. 
%Since all these features are independent and different from each other, they are given different weights to generate a unique device fingerprint.

%PS 2/22: Can you please check following paper for the details on each of the fingerprinting features and try to summarize them in your own word? We do not need detail discussion like in that paper, but one para for each category like we have now is sufficient. Also, check how they have presented/written. As a sample example, I have revised "Browser-Provided Information" para. Please add citation to this para wherever necessary and follow similar structure for other category of fingerprinting. You can limit your description to the features that we are using in our paper. 

% ZL 3/9: Updated

% Paper: "Device Fingerprinting for Augmenting Web Authentication: Classification and Analysis of Methods"
%\vspace{-2mm}
%\smallskip
	\noindent\textbf{{(C1) Browser-Provided Information:}}
	JavaScript API can be used to extract a wide range of system information, referred to as browser-provided information, that can be employed to fingerprint a device. A set of such features are listed in the first row of Table~\ref{tbl:features_categories}.
	The feature set in this category includes software and hardware details (e.g., browser/OS vendor and version, system language \cite{alvestrand2002content}, platform \cite{ref47}, user-agent string \cite{boda2011user}, resolution, etc.), device timezone and clock drift \cite{boda2011user} from Coordinated Universal Time (UTC), battery information \cite{olejnik2015leaking} (e.g., battery charge level, discharge rate), and password autofill \cite{ref46} (e.g., the password is user-typed or auto-filled by a browser or password manager).
	The information corresponding to
	\textit{WebGL} \cite{mdnwebdocs}, a JavaScript API for rendering graphics within web browsers, and \textit{WebRTC} \cite{beltran2014user}, a set of W3C standards that supports browser-to-browser applications, e.g., voice and video chat, can also be used to fingerprint a browser. 
	WebGL information includes the WebGL vendor and version, maximum texture size, supported WebGL extensions, renderer strings, etc. WebRTC information includes connected media devices (e.g., webcam and microphones) information.
%	Resolution gives the current screen resolution values. 
	The support for \textit{local storage}, which enables the browser to store data without any expiration \cite{mdnwebdocs3}, 
	and the status of \textit{do not track}, which blocks (or allows) 
	the website from tracking \cite{mdnwebdocs2} 	
	are also often used in browser fingerprinting.
	\noindent{\textbf{(C2) Inference based on Device Behavior:}}
		The device information can also be extracted by executing a specially crafted JavaScript code on the browser and observing the resulting effect.
		This category of the fingerprinting features is based on the fact that  
		the execution of JavaScript code creates different effects based on the software and hardware configuration of the device, and hence can be used to infer various characteristics of the device.
		For instance, HTML5 canvas renders the text and graphics differently based on OS, available fonts, and the video driver \cite{mowery2012pixel}.
		The elapsed time to execute the JavaScript code can be used to infer the performance characteristics of the device \cite{mowery2011fingerprinting}. 
%		\bluetext{Deleted this sentence ``Hardware sensors can be fingerprinted based on the variation in the calibration error of motion-position sensors (e.g., accelerometer) or frequency-response of audio sensors (e.g., speaker, microphone)''}
		Various aspects of a pointing device can be inferred by monitoring the scroll events generated by the mouse wheel or touchpad \cite{das2016tracking}.
		The browser vendor and version can be inferred by testing CSS features \cite{mesbah2012automated}.
		The presence (or absence) of different fonts can be inferred by rendering a text with a predefined list of fonts \cite{fifield2015fingerprinting}.

	\noindent  {\textbf{(C3) Browser Extensions and Plugins:}}
	The aforementioned approaches
%	, i.e., leveraging JavaScript APIs and 
%	inference based on the execution of a specially crafted JavaScript code,
	can be used to extract information about the browser extensions and plugins
	to build a browser fingerprint.
	Various browser plugins, e.g., Java, Flash and Silverlight, can be queried through JavaScript APIs to reveal system information \cite{faizkhademi2015fpguard}. For instance, Flash can provide the OS kernel version. Both Java and Flash can provide an enumerated list of system fonts. Installed NoScript (that disables JavaScript) and its blacklisted website can be detected by loading a large set of websites. Similarly, AdBlocker can be detected by monitoring if fake ads are loaded on the websites \cite{iqbal2017ad} or not. Other extensions can also be detected by other methods.

\begin{table*}[t]
\centering
\vspace{-3mm}
	\caption{Three different categories of browser fingerprinting features \cite{alaca2016device}, and a summary of how they can be spoofed via our attack.}\label{tbl:features_categories}
\begin{threeparttable}
	\centering
	\scriptsize
%	\caption{Three different categories of browser fingerprinting features \cite{alaca2016device}, and a summary of how they can be spoofed via our attack.}
\begin{tabular}{|c|l|c|l|c| }
\hline
\textbf{Category} & \textbf{Feature Name} & \textbf{Spoofable} & \textbf{Spoofing Approach} & \textbf{Detectable by Targeted Websites}
\\

%& Detection Approach & Solution\\

\hline
\hline
C1  & 1. User-agent + - * \cite{boda2011user} &  & a, b, c & \\

%&\\

 & 2. WebGL information - * \cite{mdnwebdocs} &  & b & \\
 %(if incorrect data format/data response time) & The site may detect the webgl value format: panopticlick use webgl hash, and webgl API can return GPU brand and type. Also the script may contain function settimeout, so the response time would be the detection approach & When we re-write the function, first inspect their original codes, find out all “set time” related functions, and add the same response time in our re-write function. Make sure we copy the correct format feature values.\\
 & 3. System time + - *\cite{boda2011user} &  & a, b & \\
 % (if incorrect data format/data response time) & If the date or timezone format is incorrect, the site would detect. If we change the script json and ignore the relative response time, the site may detect. If the spoof time has large different to the current time, the site would detect. & When we re-write the function, first inspect their original codes, find out all “set time” related functions, and add the relative response time (same benign response time+current time-spoofed time) in our re-write function. Make sure we copy the correct format feature values.\\
 & 4. Battery information \cite{olejnik2015leaking} &  & a, b & \\
 % (if incorrect data format/data response time) & If the battery API value format is incorrect, the website would detect. If we change the script json and ignore the relative response time, the site may detect. & When we re-write the function, first inspect their original codes, find out all “set time” related functions, and add the same response time in our re-write function. Make sure we copy the correct format feature values.\\
 & 5. Cookie enabled + - * \cite{saleh_saleh_1960} &  & a, b, c & \\
 % (if incorrect data format/data response time) & The site would detect cookieEnabled API value and if they could create real cookies in the browser. & When we re-write the function, first inspect their original codes, find out all “set time” related functions, and add the same response time in our re-write function. Make sure we copy the correct format feature values.\\
 & 6. WebRTC \cite{beltran2014user} &  & b & \\
 % (if incorrect data format/data response time) & The website would detect the browser version, platform and webRTC connection to detect if this feature is spoofing. Some browser and OS would support webRTC, some are not. If the website detected the webRTC connection is exisiting but the browser and OS didn't support the webRTC, this means the user may use spoofing features. & We can change the detector value of the webRTC connection to keep it, the browser and the OS consistent.\\
 & 7. Password autofill \cite{ref46}& Yes & b & Hard\\
 %&\\
 & 8. Platform - * \cite{ref47} &  & a, b & \\
 % (if incorrect data format/data response time) & The site may calculate the time between script response time and visit request time. If this time is different to the setting time in the script, it means their script is attacked and the data is spoofed. As we change the data in the API or in the browser setting, the site cannot detect this change if we keep the spoof data format correct. & When we re-write the function, first inspect their original codes, find out all “set time” related functions, and add the same response time in our re-write function. Make sure we copy the correct format feature values.\\
 & 9. Language + - * \cite{alvestrand2002content} &  & a, b, c & \\
 % & We only change the language setting in the browser language setting, so the site will not detect the spoof.\\
 & 10. Local storage + - * \cite{roesner2012detecting} &  & b & \\
 % (if incorrect data format/data response time) & The site would insert data in the browser local storage then detect if those values are inserted. If the return value is not the original value, the site would know this value may be spoofed. & When we re-write the function, first inspect their original codes, find out all “set time” related functions, and add the same response time in our re-write function. Make sure we copy the correct format feature values. \\
 & 11. Resolution + - * \cite{tutorialrepublic} &  & a, b & \\
 % (if incorrect data format/data response time) & The site would detect if the resolution values are all integer. And also detect the script response time. & When we re-write the function, first inspect their original codes, find out all “set time” related functions, and add the same response time in our re-write function. Make sure we copy the correct format feature values. \\
 & 12. Do Not Track - * \cite{mdnwebdocs2} &  &  a, b, c & \\
 % (if incorrect data format/data response time) & The navigator.donottrack value would be 0, 1, or “unspecified”. All other values will indicate the feature has been spoofed. Also can detect the response time to decide if the script is modified. & When we re-write the function, first inspect their original codes, find out all “set time” related functions, and add the same response time in our re-write function. Make sure we copy the correct format feature values.\\
 % (if incorrect data format) & The site may calculate the time between script response time and visit request time. If this time is different to the setting time in the script, it means their script is attacked and the data is spoofed. As we change the data in the API or in the browser setting, the site cannot detect this change if we keep the spoof data format correct. & When we re-write the function, first inspect their original codes, find out all “set time” related functions, and add the same response time in our re-write function. Make sure we copy the correct format feature values.\\
 
\hline
C2 & 1. HTML5 canvas fingerprinting - * \cite{mowery2012pixel} &  & b & \\
% (if incorrect data format/data response time) & Detect the canvas value format: if the format is url, may detect if the url can show the correct image. If the format is hash value, then detect the length of the canvas hash. Also if the script set the response time, the site would detect the response time. & When we re-write the function, first inspect their original codes, find out all “set time” related functions, and add the same response time in our re-write function. Make sure we copy the correct format feature values.\\
 & 2. System performance \cite{mowery2011fingerprinting} &  & b & \\
 % (if incorrect data format/data response time) & Based on the code running time (‘end time’-‘start time’), they site may compare this running time with other fingerprinting features, like useragent+platform+webgl. Also the response time & When we re-write the function, first inspect their original codes, find out all “set time” related functions, and add the same response time in our re-write function. Make sure we copy the correct format feature values.\\
% & 3. Hardware sensors \cite{das2016tracking}& In progress &&\\
 %&\\
 & 3. Font detection \cite{fifield2015fingerprinting} & Yes & b & Hard\\
 & 4. Scroll wheel fingerprinting \cite{norte2016advanced}&  & b & \\
 %&\\
 & 5. CSS feature detection \cite{mesbah2012automated}&  & b & \\
 %&\\
 %& 6. Font detection \cite{fifield2015fingerprinting} & yes & script modification & Hard\\
 % (if incorrect data format/data response time) & The fonts list would only contain the real fonts name. And the site would compare the fonts list between JS obtained and flash obtained. Also the response time. & When we re-write the function, first inspect their original codes, find out all “set time” related functions, and add the same response time in our re-write function. Make sure we copy the correct format feature values.\\
\hline
C3 & 1. Browser plugin fingerprinting + - * \cite{faizkhademi2015fpguard}&  & a, b & \\
% (if incorrect data format/data response time) & The plugin list would be different in different type of browsers. So incorrect combination between the UA and the plugins would be recognized as the spoofed features. Also the response time. & When we re-write the function, first inspect their original codes, find out all “set time” related functions, and add the same response time in our re-write function. Make sure we copy the correct format feature values.\\
 & 2. Browser extension fingerprinting \cite{iqbal2017ad}& Yes & b & Hard\\
 %&\\
 &&&&\\
 %&\\
\hline

\end{tabular}
    \begin{tablenotes}
      \scriptsize
      \item I. +: Features used in Panopticlick \cite{eckersley2010unique}. -: Features used in Rule-based Linking Algorithm \cite{vastel2018fp}. *: Features used in Hybrid Linking Algorithm \cite{vastel2018fp}. 
      
	  \item II. C1: Browser-provided information. C2: Inference based on device behavior. C3: Extensions and plugins.
	  \item III. a: Script Injection. b: Script Modification. c: Browser Setting and Debugging Tool.
    \end{tablenotes}
\end{threeparttable}

\end{table*}

%\vspace{-3mm}
\vspace{-2mm}
%\vspace{-1mm}
\subsection{Representative Fingerprinting Techniques}
\label{sec:representatives}
%\vspace{-2mm}
% PS 2/22: Do not say we will do attack against these three schemes yet. Just say there are several fingerprinting approaches, mention few and give very very high level idea (with citation). Then say you will describe only a few representative approaches in detail that you will utilize to demonstrate the feasibility of our attack.
% ZL 3/9: Updated
As mentioned earlier, various browser fingerprinting approaches have been proposed in the literature, each utilizing a different set of device characteristics. 
%for various applications, such as for tracking the user, authenticating the user, and detecting and preventing fraudulent activities.
%There are several fingerprinting approaches: user tracking, user authentication, fraud prevention. 
Panopticlick \cite{eckersley2010unique} and FP-Stalker \cite{vastel2018fp}, specifically its Rule-based Linking Algorithm and Hybrid Linking Algorithm, are representative browser fingerprint linking techniques.
%We present a brief description of three representative browser fingerprinting techniques, namely Panopticlick, FP-Stalker Rule-based Linking Algorithm (short as \RLA), and FP-Stalker Hybrid Linking Algorithm (short as \HLA) ~\cite{vastel2018fp} and their applications.

% ZL 3/9: I think we may say `Fu' and `Fk' here. So that we don't need to define them in the following 3 algorithms.
%In the next sections, the fingerprint `Fu' means the current new fingerprint, and the `Fk' is one of the know fingerprintings which has already been stored in the database.

%	\smallskip
%	\begin{itemize}[leftmargin=*]
%		\item 
		\noindent\textbf{Panopticlick:}
		% PS 2/22: Citation missing. Check the link: https://www.usenix.org/conference/enigma2016/conference-program/presentation/budington 
		% ZL 3/9: Citation added. 
		Panopticlick~\cite{eckersley2010unique} leverages eight different browser and system attributes to track the user through browser fingerprinting. 
		% PS 2/22: How these two groups are different from each other and why they divided?
		% ZL 3/9: In the citationn ref12: https://panopticlick.eff.org/static/browser-uniqueness.pdf, section 5.2, the author divided the features into 2 groups. He didn't say why they divided.
		%	 Based on the \cite{ref12}, the algorithm
		It categorizes these attributes into two groups.
		The first group contains cookies enabled (C1-5), screen resolution (C1-11), time zone (C1-3), and partial supercookie test (e.g., local storage, session storage and IE userData) (C1-10). 
		The second group contains user-agent (C1-1), HTTP ACCEPT headers (C1-9), system fonts (C2-3), and browser plugins information (C3-1). 
		To learn the identity of an unknown fingerprint `$F_u$', 
		Panopticlick compares $F_u$  with each of the pre-stored fingerprints `$F_k$'.
		If $F_u$ has all the eight attributes the same as that of 
		$F_k$, Panopticlick marks them as the same fingerprint, i.e., generated from the same browser instance.
		If any of the attributes in the first group and more than one attribute from the second group differs, Panopticlick marks $F_u$ and $F_k$ as different fingerprints. 
%		The partial supercookie here represents C1-10.
		%	When comparing an unknown fingerprint `$F_u$' with previously stored fingerprint `$F_k$', if any of the features in the first group or more than one feature from the second group differs, the `$F_u$' not be recognized as the user who has `$F_k$'. 
		In the case where there is only one difference in the attribute set from the first group, Panopticlick estimates the similarity score of that attribute between $F_u$ and $F_k$. If the similarity score is higher than a set threshold (say 0.85), $F_u$ is marked the same as $F_k$. In the rest of the scenarios,  $F_u$ is marked differently from $F_k$.
		% If there is 1 difference in changeable group, Panopticlick will run the python lab "difflib.SequenceMatcher().ratio()" to estimate the similarity of the 1 different feature. If the score is larger than 0.85, the 'Fu' will be assigned to 'Fk'. Otherwise, 'Fu' will have no relations to 'Fk'.
		
%		\vspace{.5mm}
%\smallskip
		\noindent\textbf{Rule-based Linking Algorithm (\RLA):}
%		Rule-based algorithm is from the FP-stalker paper[2]. 
		% PS 2/22: Like in Panopticlick, before the core algorithm, should list the features that this algorithm is using or at least refer it to the table of feature list. Does this algorithm divide the features into groups? Seems like it does.
%		Rule-based Linking Algorithm~\cite{ref2} is static device fingerprinting techniques. 
		This approach for browser fingerprinting categorizes the fingerprinting attributes under consideration into three sets.
		The first attribute set consists of 
%				Based on \cite{ref2}, the rule-based algorithm have an unchangeable feature set0: 
				operating system (C1-1), platform (C1-8), browser name (C1-1), local storage (C1-10), do not track (C1-12), cookies enable (C1-5), and canvas (C2-1). 
				The second set consists of user-agent (C1-1), GPU vendor (C1-2), renderer (C1-2), 
				% what doe you mean by accept and headers here? resolution, encoding?
% ZL 3/10: it should be http accept headers.
% ZL 3/10: resolution is screen resolution. they don't define the encoding in FPS-talker paper. The example in their dataset is : "gzip, deflate, sdch".
				browser plugins (C3-1), system language (C1-9) and HTTP accept headers (C1-9). 
				The third feature set consists of the resolution (C1-11), time zone (C1-3) and encoding (HTTP header). 
				Similar to Panopticlick, 
%				to map an unknown fingerprint `$F_u$' to one of the known fingerprint `$F_k$' (if any exists), 
				\RLA compares the aforementioned attributes of  an unknown fingerprint $F_u$ with each of the stored fingerprint $F_k$. 
				If all the attributes of both the fingerprints are the same, \RLA marks them as \textit{exact} fingerprints.
%				If two fingerprints `Fk' and `Fu' has same values in set0, set1 and set2, the `Fk' will be added to a set `exact'. 
%				Otherwise, the algorithm will do the following step:  
				If $F_u$ and $F_k$ have differences in at least one of the attributes in the first set, \RLA marks them as different. if $F_u$ has an older version of the browser, the algorithm will mark them as different. Otherwise, it 
%				If two fingerprints `Fk' and `Fu' have differences in unchangeable feature set, `Fu' will not be assigned to the user of `Fk'. 
 estimates the similarity between the remaining attributes from the second and third sets.
 If the similarity score is greater than the set threshold (say 0.75), the algorithm counts the number of features that are different between $F_u$ and $F_k$. 
 % PS 2/3: What does it mean by less than one -- zero, please confirm this?
% ZL 3/10: The value 0.75 is defined by the author in FPS-talker paper, section 4-B, rule 5. They use the python lab difflib.SequenceMatcher().ratio to calculate the similarity ratio.
 All the $F_k$s that have less than one different attribute from the first set and less than two different attributes from the first and second sets are marked as \textit{candidate} fingerprints. 				
%				If these features are same, Rule-based algorithm will compare the similarity of features in changeable feature set using Python library ``difflib.SequenceMatcher().ratio". If the score $>$0.75, the algorithm will compare the number of different features. The number of changes should be less than one in set1, and less than two in set1+set2. If all these conditions are satisfied, the `Fk' will be stored in a set called `candidates'.
% PS 2/3: Please clarify the following statement. It is not clear at all. What is id here now?
% ZL 3/10: ID is user. I updated the following sentence.
%If all $F_k$s marked as exact fingerprint have the same id, $F_k$ is assigned this id.
If all the $F_k$-s that have been marked as \textit{exact} fingerprints correspond to the same user, $F_k$ is assigned to that particular user.
%After comparing the `Fu' to all `Fk', and if all the fingerprints in `exact have the same id, this id would be the id of `Fu'. 
% ZL 3/10: also updated the following sentence.
Similarly, if all the $F_k$-s that have been marked as  \textit{candidate} fingerprints belong to the same user, $F_u$ is assigned to that particular user. 
In the rest of the cases, $F_u$ is recognized as a new user.
%Or if all the fingerprints in `candidates' have the same id, this id would be the id of `Fu'. Otherwise, the `Fu' would be recognized as a new user.
		% PS 2/22: Do you follow Algorithm 1 of FP-Stalker while writing the algorithm? I see differences between your text and Algorithm 1 of FP-Stalker. Where is the concept of 'candidates' vs. 'exact'. The algorithm does not only compare with Fk, it compares with a bunch of Fks that it has stored. 
		
		%ZL 3/10: The authors in FPS-talker mentioned the `exact' and `candicates' in their paper, section 4B for rulebased and section 4C for hybrid.
		
%		\vspace{1mm}
		 \noindent \textbf{Hybrid Linking Algorithm (\HLA):}
		% PS 2/22: Should mention advantages of Hybrid approach over rule-based approach.
		% Please follow same comments as from Rule-based, i.e., should briefly explain Algorithm 2 of FP-Stalker. 
		This approach enhances \RLA with the machine learning technique. 
		%Hybrid algorithm is another algorithm from the FP-stalker paper \cite{ref2}. 
		%It is called ``Hybrid" because this algorithm contains parts of conditions in Rule-based algorithm, and it uses the machine learning to calculate some features similarity to compare two fingerprints. 
		\HLA divides the browser attributes into two sets. 
		The first set consists of 
		%This algorithm have the unchangeable feature set0: 
		the operating system (C1-1), device platform (C1-8), browser information (C1-1), local storage (C1-10), do not track (C1-12), cookies enable (C1-5), and canvas 
	  (C2-1). 
		The second set %changeable set1 
		contains the following nine features -- number of changes, system languages (C1-9), HTTP based user-agent (C1-1), canvas (C2-1), created time (C1-3), browser plugins (C3-1), fonts (C2-3), renderer (C1-2) and resolution (C1-11).  
		\HLA compares an unknown fingerprint `$F_u$' with each of the known fingerprints `$F_k$' to give an identity to $F_u$.  
		$F_k$ is assigned to the set ``\textit{exact}'' if these two fingerprints have the exact same first attribute set, otherwise, to the set ``$F_{k\_sub}$''.
		% PS 2/3: What is id exactly here? Can you please write this algorithm steps by steps -- line a pseudocode. I will then try to revise. 
		% ZL 3/10: exact contains the known fingerprintings, if those fingerprintings are 100% same as the current unknown fingerprinting. I will write the pseudocode in a seperate file.
		If all the fingerprints in the set \textit{exact} have the same id, this id is assigned to $F_u$, otherwise, a new id is given to $F_u$.
		%If there are fingerprints in `exact' after comparing all `Fk' in `F', the algorithm will assign id in `exact' to `Fu' if all fingerprints in `exact' have the same ID, or assign an new ID to `Fu'. 
		% S 2/3: Fu is already assigned id, why do you need the steps below? NOt clear?
		% ZL 3/10The previous is the first loop. First loop compare each Fk in the dataset to the new Fu. This loop is looking for the exact same Fk to Fu. If exact same Fk exist, and those Fk belongs to same user, then Fu belongs to this user. If those Fk belongs to different users, then Fu belongs to a totally different new user.
		% ZL 3/10: The next is the second loop. This loop is looking for the Fk which have most same features to the Fu, not the exact same.
		If there are no fingerprints in the set \textit{exact}, \HLA compares the first attribute set of $F_u$ with that of each of the $F_k$ in $F_{k\_sub}$. Each attribute comparison results in `1' if the attribute is the same in both $F_k$ and $F_u$, otherwise, `0'. 
		If there are less than five different attributes, 
%		the number of attributes that are different is less than 5, 
		\HLA feeds the results to the machine learning model, Random Forest to be specific, resulting in a similarity score (in the range of 0 and 1).  
		The $F_k$ having a score higher than 0.994  is assigned to the set `\textit{candidates}'. 
		The $F_k$-s in the candidate set are sorted in descending order of the score. If the first score is larger than the second one plus 0.1, the id of $F_u$ ID  becomes the top-one id. If the top-one and top-two ids have the same id, this id is assigned to $F_u$,  otherwise, a new ID is given to $F_u$.

\vspace{-4mm}
%\vspace{-1mm}
\subsection{Applications of Browser Fingerprinting}
\label{sec:application}
%\vspace{-2mm}
%\vspace{-2mm}
%\textcolor{blue}{Browser fingerprinting can be used for targeted advertisements, authenticating the user, and detecting fraudulent activities as described below.}

%Websites and online services are widely using the browser fingerprinting for targeted advertisements, authenticating the user and detecting fraudulent activities as described below.

%\smallskip
%\vspace{-1mm}
\noindent{\textbf{Targeted Advertising:}}
The browser fingerprinting can be employed to provide 
	targeted and personalized ads on the user devices (e.g., general desktop PC, handheld mobile device) \cite{hoofnagle2012behavioral}. 
	When a user visits a website, the web server (or the online service provider) extracts and stores the browser fingerprint along with the user's browsing behavior. 
%	Later on, 
	When the user revisits the same website, 
	the web server \redtext{checks if the same user has visited its website earlier by comparing the current fingerprint 
	with the set of fingerprints that it has stored and 
}looks for his fingerprint in its repository and
pushes the relevant ads
 based on the user's prior browsing behavior. 
%	The browser fingerprinting could be used by the website to deliver targeted ads based on this user's activities from this browser, no matter in computer or mobile phone. 
% PS 2/3: Define each approach clearly?
% ZL 3/10: defined the account-based and cookie based.
  Besides browser fingerprinting, there exist other approaches for targeted advertisements, such as account-based targeted ads~\cite{taylor2011friends} and cookie-based targeted ads~\cite{tucker2012economics}.
%20210617 The account-based approach requires the user to log into his online services/applications
\redtext{Based on the account owner's search interests and visit history, the
service provider pushes the relevant ads to the user. 
}
%20210617while the cookie-based approach requires enabling the cookie on his browser.
\redtext{When the user visits a website, the website embeds an unique identifier to the cookie. When the user revisits the website, it recognizes the user
based on the cookie and pushes the relevant and personalized ads based on the user's prior browsing behavior (often stored in the cookie itself).
}Unlike these approaches, the browser fingerprinting neither requires the user to log into his online account, nor
 requires the user to enable the cookie, rather it works transparently.
% PS 2/3: Need to say how fingerprinting can be used for targeted ads.
% ZL 3/10: Updated in the following paragraph.

%Similar to the cookies-based tracking, the website may identify the user through browser fingerprinting when a user visits the website. 
%Unlike storing cookies in the user's browser, the website would store the user's browser fingerprinting information on their servers. 
%When the user visits the website again, the website would extract user’s current browser fingerprinting information and retrieve this user's previous behavioral information from the server, then push the relative ads to this user.

%the website can get the user's browser fingerprinting features without user's notice.
% Then the user's browser fingerprinting can be used directly by the website to push personalized ads, or be sold to third-party ads company for the targeting ads.
	
%\vspace{-3mm}
%	\vspace{1mm}
	\noindent {\textbf{User Authentication:}}
	% PS 2/22: Is browser-fingerprinting alone is used for authentication or they require username-password? Need details and clear statement. Are you using the correct reference? 
	% ZL 3/10: added more details about the authentication. It need the username and password in some section.
	\redtext{User authentication schemes are also incorporating the browser fingerprinting techniques to improve their overall security and usability.
}Various services, such as Oracle~\cite{identification2015},  Inauth~\cite{ref54} and SecureAuth IdP~\cite{ref55} are leveraging the browser fingerprinting technique to enhance the overall security and usability of their authentication mechanisms~\cite{martherus2007user}. 
 	The browser fingerprinting is usually integrated with 
 	existing authentication schemes, such as two-factor authentication (2FA) schemes \cite{ref55}.
%
% 	It still needs the username password every time, and 2-factor authentication when the user logs in for the first time from a new device/browser .
% 	
%20210617 	In such an authentication scheme,
%20210617 	 when the user attempts to log into a web service for the first time, the user has to go through a complete authentication process, i.e., type in the username, password and prove the possession of the second-factor device, e.g., by providing a one-time PIN to the authenticating terminal. 
On successful login, the web server captures and stores the browser fingerprint of the device that the user has used to login. 
 	 Next time, when the user attempts to login to the same web service using the same device, the current browser fingerprint is matched against the stored fingerprints.
 	 If they match with a high score, the second-factor of 2FA process is dropped (i.e., no need to provide the PIN), merely typing in the password is sufficient to login. 
 	Thus, browser fingerprinting approach for authentication lowers the user-effort during the authentication process, and hence improves the system's usability.
% 	 of the system 
% 	 The next time when the user logs in, during the authentication process, the device fingerprint is compared against the stored list of fingerprints of the user device.
% 	If this fingerprint comparison succeeds, i.e., they match with a high score, the user is authenticated successfully.
%	When the user try to log in, the browser fingerprinting based user authentication would compare current browser fingerprinting to the stored browser fingerprinting history. If these two browser fingerprinting are paired, the user authentication will pass this login behaviour. 
	% PS 2/22: alert to website or web-server/service?
	% ZL 3/10: deleted the alert sentences.
%	An alert, indicating fraudulent login attempt, is sent to the service provider if 
%	the fingerprint comparision fails, i.e. the fingerprints does not matches sufficiently.
%	Otherwise, an alert would be sent to the server or this account owner. The browser fingerprinting based user authentication may directly refuse the access if the current browser fingerprinting information can not be paired to any of this account's browser fingerprinting history.
 
%	\vspace{-3mm}
%	\smallskip
	\noindent {\textbf {Fraud Detection:}}
	\redtext{The browser fingerprinting techniques are also being used to  detect fraudulent activities in the online setting.
}Several security services, e.g., Seon~\cite{florian_2019} and IPQualityScore~\cite{ipqualityscore},
are leveraging browser fingerprinting for the purpose of  fraud detection and prevention in the online setting. 
%	Some third party service companies like Seon \cite{ref3} and IPQualityScore \cite{ref4} are using browser fingerprinting in their fraud detection services. 
	% PS 2/22: Blacklist of what? How they are blacklisted? Need more details on the mechanism. Read paper and article, then write.
	% ZL 3/10: deleted the blacklist method. Updated the supervised and unsupervised method.
	The fraud detection techniques can be categorized into two groups -- \textit{supervised} and \textit{unsupervised} methods \cite{bolton2002statistical}. The supervised method 
	leverages the information from the prior fraud behavior
	(i.e., already marked as fraud)  to build a model to infer if the current behavior is fraud or non-fraud. 
	The unsupervised method does not rely on the prior  fraudulent behavior, rather it sets a baseline for normal behavior. If the current behavior significantly deviates from the baseline behavior, the unsupervised method marked the behavior as fraudulent.
%	Then it will look for the most dissimilar user's behavior to this baseline.
%	Fraud detections \cite{ref34} are using the blacklist to block some known devices, or monitoring some suspicious devices. 
The browser fingerprinting can be used to mark the user as a fraudster or a legitimate user. When any of these methods find the user's behavior fraudulent, the service provider
captures and flags the browser fingerprint as fraudulent.
%, so when the user has the fraudulent behavior detected by the supervised detection method, or the user's behavior is way more different than the normal behavior baseline (detected by the unsupervised method), the website would mark this user as the `attacker' through the browser fingerprinting. 
Since the browser fingerprint changes over time, a risk level can be estimated by comparing the browser fingerprint against the flagged fingerprints. If the risk level is significantly high, the current user is flagged as a fraudster. 

%\textcolor{blue}{However, no services are known to use browser fingerprinting only currently, even the service we mentioned above. Using browser fingerprinting only may become the new trend in the future. The purpose of our research is to expose if only using browser fingerprinting on website is safe.}
 
%	With browser fingerprinting, the website could verify if this user is in the blacklist, or estimate the danger levels (comparing the current browser fingerprinting information with the one in the blacklist). 
%	As most of the browser fingerprinting features values remain stable, browser fingerprinting based fraud detection would have a brighter future.

%\vspace{-2mm}
\section{Attack Model \& Spoofing Methods}
\label{sec:out_attack}
%\secspace
%\begin{figure*}[]

\begin{figure}[t]
	\centering
%	\vspace{-2mm}
%\includegraphics[width=0.53\linewidth]{our_attack.pdf}
	\includegraphics[width=0.65\linewidth]{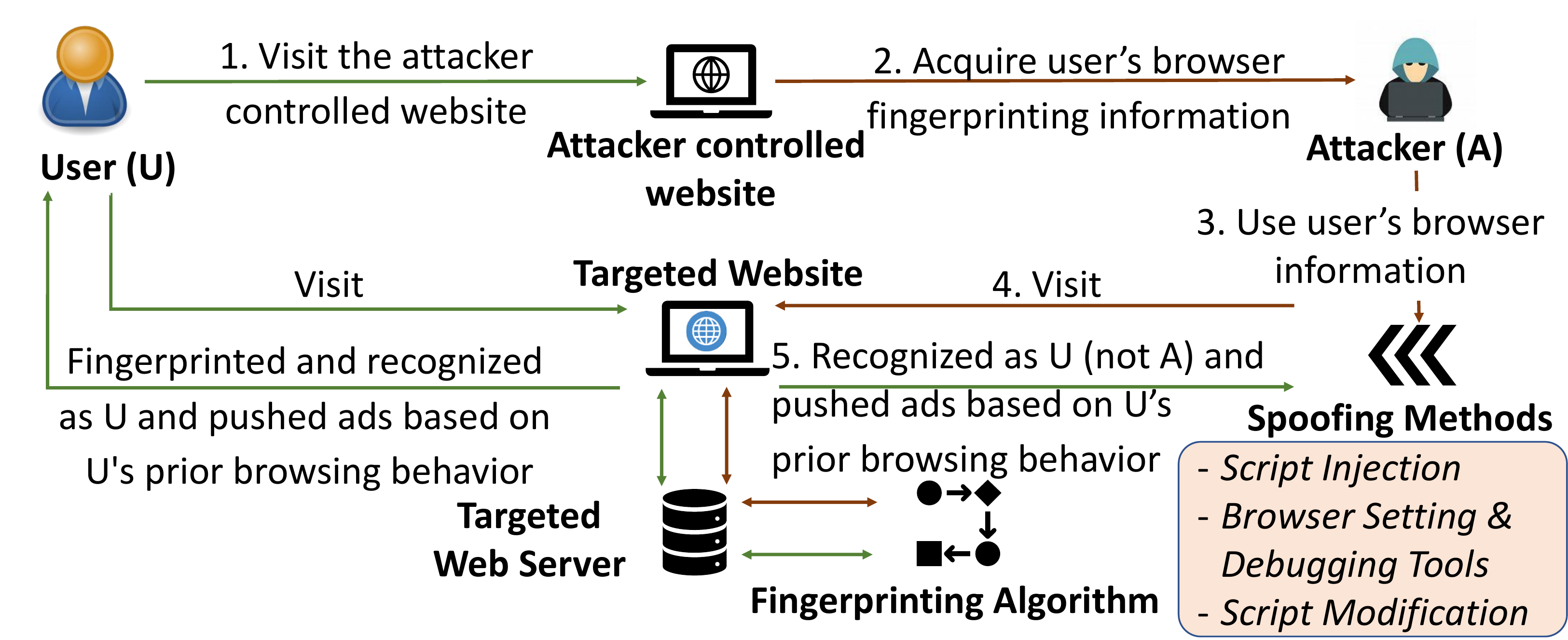}
	\caption{A high-level overview of the \GB attack model.}
	\label{fig:attack_model}
%	\vspace{-3mm}
%\end{figure*}
\end{figure}

% PS 2/23: Explain the attack in brief at a very high level. Then say specificaly our attack consists of two concreted steps as described below.
%\vspace{-1mm}
%\vspace{-2mm}
%\vspace{-1mm}
\subsection{Attack Model}
%\vspace{-2mm}
\GB consider a remote adversary who can spoof the victim's browser to a target remote web service. The main goal of \GB is to fool the web server into believing that a 
legitimate user is accessing its services so that it can 
learn sensitive information about the user (e.g., interests of the user based on the personalized ads), or circumvent various security schemes (e.g., authentication and fraud detection) that rely on the browser fingerprinting.
A high-level overview of the attack is shown in Figure \ref{fig:attack_model}.

We assume that the attacker has obtained the browser fingerprint of the victim. 
%Since the device fingerprint information can be extracted transparently in the background , 
The adversary can easily capture the victim's fingerprinting information by designing a benign-looking website
and luring the victim into visiting his website. 
The adversary can leverage the exact mechanism as that of any fingerprinting website to acquire the browser fingerprint, i.e., via JavaScript APIs.
It is also possible that a compromised web service, running a malicious script, could acquire the victim's browser fingerprint when the victim visits the attacker-owned website. 
%Since the process of acquiring the fingerprinting information is completely transparent, the ongoing attack remains oblivious to the victim.

We also assume that before accessing a target web service, 
the attacker spoofs (or injects) previously acquired victim's browser information into his own fully controlled device to present it as the victim's device. When the attacker visits the target website, the target web server would receive the victim's fingerprint from the attacker's device. Therefore,  for the target web service, it looks like the victim is accessing its services, and can not really recognize the malicious attacker.

%In this section, we introduce the attack model of the browser information acquiring and feature spoofing. The attacker can design a website, and deploy the scripts on it to acquire the visitors' browser information. After letting the user `U' visit the attacker's website, the attacker can obtain `U' 's browser information. Then the attacker put those information in his browser, making his browser same as the `U' 's. After that, the attacker uses this spoofed browser to visit the targeted website. At this moment, the targeted website would recognize the attacker as `U', and the attacker can obtain `U' 's information which are provided by the targeted website browser fingerprinting based functions. The fig 1 shows this attack model.

%\subsubsection{Frequency of Acquiring and Spoofing}

We consider three different modes of executing the attack. 
%When launching \GB against the browser fingerprinting, 
An adversary can retrieve and spoof the victim's browser fingerprint only once, referred to \textit{\AOSO}. \textit{\AOSO} can be used to bypass the security of the user authentication scheme.  Alternatively, to increase the impact of the attack, 
the attacker can spoof the same browser fingerprint instance multiple times over a few days gap, referred to \textit{\AOSF}. Leveraging \textit{\AOSF}, the attacker can track the personalized ads associated with the victim for a long period of time, and can infer various sensitive information about the user, even build a personal profile of the victim.  
Since the browser fingerprint changes over time, to increase the attack success rate, the attacker can also retrieve and spoof the browser fingerprint multiple times, and is referred to  \textit{\AFSF}. With this approach, the attacker could always obtain the latest browser fingerprint of the victim. This can enable the attacker to compromise the security of the fraud detection mechanism.
%Based on how often the attacker acquire and spoof the victim's device fingerprint, we consider three different attack settings.

%\paragraph {\textbf{Acquire Once, Spoof Once}}
%
%The attacker only needs the user to visit his own designed website one time. which means he obtains the user's browser information only one time. And he uses these information to spoof the website one time. 
%
%\paragraph {\textbf{Acquire Once, Spoof Frequent}}
%
%The attacker only needs the user to visit his own designed website one time. which means he obtains the user's browser information only one time. But he will use these information to spoof the website several times, like after 1 day, then after 2 days, etc.
% 
%\paragraph {\textbf{Acquire Frequent, Spoof Frequent}}
% 
%In this setting, the attacker needs the user to visit our his designed website several time. Every time the user visit his website, he will use this time obtained information to spoof our targeted website one time. 

%\secspace
\vspace{-2mm}
%\vspace{-1mm}
\subsection{Spoofing Methods}
\label{sec:spoofing_methods}
%\vspace{-2mm}
The key component of \GB is the ability of the attacker to spoof the victim's browser fingerprint so that the attacker can present its own browser as if it is the victim's browser in front of the web service. Our spoofing methods are only focusing on the features which are listed in Table \ref{tbl:features_categories}, and we did not spoof network level features like IP address.
\redtext{Once the attacker has acquired the victim's browser fingerprinting information, it}
The attacker can leverage the following methods to spoof the fingerprint.

%In this section, we will describe how we use user's browser information to finish the attack.
%
%When we successfully obtain the user's browser fingerprinting features, the next question would be: how could we use those features to in the attack device to spoof the website. In our paper, we will introduce two methods.

%\subsubsection{Settings and API:}
%	The website obtained the device fingerprinting features through the browser. So in this level, the attackers would make changes in the browser setting and Javascript read-only APIs to spoof the website. I will split this section into two parts: 1. Browser property and 2. Browser setting. In ``Browser property'', we used the code to change the values pre-set by the browser. In ``Browser setting'', we changed browser application setting options, and the values in developer tools.
%\secspace
\vspace{-5mm}
%\vspace{-1mm}
\subsubsection{\textbf{Script Injection}}
%\vspace{-2mm}
%\vspace{-1mm}
 In browser fingerprinting,
	when the browser loads a website, the website executes scripts consisting of various JavaScript API calls to extract the browser information. To spoof the browser fingerprint, the values extracted by the JavaScript API calls should be changed before the browser executes the scripts embedded in the website.
	The objects where these extracted values are stored can be overwritten by creating a new object with the same name and constructor as that of the original JavaScript APIs. 
	To implement this method, a browser extension, a specialized and independent software module for customizing a web browser, and/or Selenium~\cite{ref57}, a portable framework for testing web applications, can be utilized. 
	The browser always loads and executes the website scripts in the browser extension prior to loading and executing it to the client machine. Those scripts would not change any scripts contents that are loaded from the visited websites.
	In the case of Selenium, pre-designed scripts are executed, which is followed by launching the browser, loading the website, and executing the embedded scripts. 
	The feature of the browser extension and Selenium to execute 
	the scripts prior to loading the website allows the adversary to overwrite the browser properties extracted through JavaScript API calls.
	%	When the user visits the website, the browser would run the scripts on the website. Those scripts can obtain the browser information through Javascript APIs. So attackers should change those API values before the browser loading website scripts. Attackers can write a new object with same name and construction as the existing APIs, then use the new ones to overwrite the existing ones. Using browser extension and non-proxy software like selenium can be good choices to implement this method. When you use the extension, the browser will always load the scripts in the extension first, then run the scripts obtained from the website. When you use selenium, the selenium will also run the pre-designed scripts, then open the browser and load the website. 
An example is listed in Appendix \ref{sec:efi}.

\vspace{-5mm}
%\vspace{-1mm}
\subsubsection{\textbf{Browser Setting and Debugging Tool}}
%\vspace{-2mm}
%	\vspace{-1mm}
	Many of the browsers offer a mechanism in the form of the \textit{browser setting} and the \textit{debugging tool} that enables its users (the attacker in our case) to change various attributes of the client device and the browser. 
%	Utilizing the setting option provided by the browser application, the user (in our case the attacker) can easily change
%	various properties of the browser that are used in device fingerprinting. 
	For instance, cookies, local storage and ``do not track'' options can be enabled or disabled simply through the browser setting in the Google Chrome browser~\cite{melicher2016not} and the ``about:config'' page in the Firefox browser~\cite{firefox}.
	Further, about:config page in the Firefox browser allows the user to design his own APIs that can overwrite the browser's pre-defined APIs. This approach can completely change the browser's attributes.  
%			In the browser setting, you can make some changes. In Chrome setting, you can really disable the cookies, local storage, or allow the "Do not track" \cite{ref40}. 
%	Firefox provided a page called "about:config".\cite{ref41}.
%	 In this page, you can design your own APIs, or design some APIs to overwrite the browser exisiting ones, and this looks more like "changing Browser property". 
	
	The browser also offers a \textit{debugging tool} intended for web application developers that allows them to debug and improve their web application functionality~\cite{google}. 
	Using the debugging tool, various browser attributes, such as {user-agent}, geolocation, and 
%	network conditions, 
caches disabled
% and orientation setting, 
can be easily changed.
%	For example: In chrome, you can use the debugging tools  to change the user-agent. 
	The changes affect both the JavaScript API (e.g.,  $\mathtt{navigator.userAgent}$) and the corresponding value in the HTTP header (e.g., the value of user-agent field). 
%	You can also change the geolocation, network conditions, caches disabled and orientation setting in this tool. 
	The debugging tool allows the changes on the browser's attributes to any custom value, whether it is a pre-defined valid string, or a random text.
%	You can not only use the existing stored values provided by the browsers, but also type your own designed.(except the caches disabled) 

%	The strength of this method is: the attacker use the tool provided by the browsers. This looks more trustworthy as the browser may not sell you to the website. The weakness is: changing the setting is also change the real function of the browser. This would reduce the attackers the browser using experiences.
	
%\secspace
\vspace{-5mm}
%\vspace{-1mm}
\subsubsection{\textbf{Script Modification}}
%\vspace{-2mm}
%\vspace{-1mm}
	The browser properties can also be changed by modifying the scripts embedded in the website. Once the embedded scripts have 
	extracted the browser information, they can be changed before the website sends it to the web server.
	%	In this level, we are going to change the website scripts before our browser loading them. As the scripts obtain information from our browsers and send information to the server, we can change the values before the scripts sending those information. 
	Utilizing the developer debugging tool (mentioned earlier), a breakpoint can be set at the beginning of each script of the website so that the scripts' execution gets paused at the set breakpoint.
%	In the developer debugging model, we can set the breakpoint at the first line in each scripts. When we open the website, all the scripts will be paused at the breakpoint where we set before. 
By inspecting the embedded scripts, the JavaScript API expression can be replaced with the spoofed values. For instance, 
% We can inspect which Javascript API the script is using. Then we can replace those API expression with spoofed values. 
$\mathtt{platform=navigator.platform}$ can be replaced with $\mathtt{platform=``Win32''}$ that exposes the underlying platform of the device as Win32, instead of the actual platform.
% (e.g., Linux, Mac-OS, etc.).
%that make the website think my operating system is "Windows". But actually my system is "MacOS". 
However, each API expression should be changed very carefully as the use of an incorrect expression (i.e., its value and format) can alert the web service, and the changes can fail. 

A more convenient method to spoof the browser information is to leverage the fact that JavaScript always uses \textit{Ajax} (Asynchronous JavaScript And XML) to transfer the data to the remote server~\cite{garrett2005ajax}. 
Since Ajax employs JSON (JavaScript Object Notation) ~\cite{json}\cite{ref44} format  when transferring data to the web server, the browser information can be 
changed by checking the variable in the JSON object.
%The format of the dataset is "json" \cite{ref43} \cite{ref44}, so you can change the value by checking the index. 
Given that the debugging tool shows current variables and their values at each breakpoint, the values can be changed easily.
Once the changes on the scripts are completed, the breakpoints are removed allowing the execution of the modified scripts. With this approach, the remote web service would receive the spoofed browser attributes.
%After we finish the scripts changes, we can unpause the breakpoint and let the `changed' scripts run. And the targeted website would receive our spoofed feature values. 
As the executed scripts are never sent outside the client machine, the approach remains oblivious to the remote web server.
%As the script would not be sent to the server, changing data on the scripts won't let the website know our behaviours. 
%20210617However, some websites and their scripts may set a response time, which allows the web server to detect  the use of such breakpoints. 
%which means if we pause the scripts running, the website may know it. 
%20210617Fortunately, the modified scripts can be pre-designed and can be used to replace the website script that can defeat the purpose of setting the response time.
%20210617As different websites have unique scripts, different from \textit{Script Injection}, this method should be executed manually.
%We can pre-designed the whole scripts and replace it as we visit the website to reduce this weakness. 

Most websites or web services use JavaScript obfuscation on the scripts, instead of the native ones. The purpose of using obfuscation is to make the scripts difficult to understand. JavaScript Obfuscator Tool~\cite{jot} is an example of such obfuscation methods. JavaScript obfuscation can indeed make script modification harder than native scripts. However, there are JavaScript deobfuscation methods that can help us to get native scripts. A previous study~\cite{lu2012automatic} and deobfuscation service~\cite{de4js} have proved that deobfuscation can work. So obfuscated scripts will not pose a problem in script modification.

We have listed all spoofing approaches for each feature in Table \ref{tbl:features_categories}. More details for spoofing all features are listed in Appendix \ref{sec:example}.
%\secspace

%\vspace{-2mm}

%After proving that we can spoof the browser features, we would test if our spoofing can affect the fingerprinting based tracking algorithms results. We selected three algorithms to do this test, which are listed in the ``background" section. We would analyze the dataset after doing the attacks, to verify if our attacks would affect the tracking of targeted user, and other users, and the tracking algorithms functionalities.

%\vspace{-3mm}

%\vspace{-5mm}
%\vspace{-2mm}
\section{Attack Implementation}
\label{sec:attack_design}
%\secspace

\newcommand{\benigndata}{benign dataset\xspace} 
\newcommand{\attackdata}{attack dataset\xspace} \newcommand{\originaldata}{original dataset\xspace} 

%In this section, we first present the details on our implementation and validation of our spoofing methods. Then, we provide the description of the implementation of our attack against prominent device fingerprinting algorithms (presented in Section \ref{sec:representatives}). 

%\begin{figure}[]
%	\centering
%	\includegraphics[width=.8\linewidth]{windows_firefox.png}
%%	\vspace{2mm}
%	\caption{Victim User: The original features of combination ``Win+Firefox''. Testing website: \textbf{Panopticlick}}
%	\label{fig:win_firefox}
%	\vspace{1.5mm}
%\end{figure}

%we describe how we have the ability to obtain the users' browser fingerprinting features, and how we use those features to spoof the website.
%\vspace{-3mm}
%\vspace{-1mm}
\subsection{Acquiring User Browser Fingerprint} 
\label{sec:a_u_b_f}
%\vspace{-2mm}
To impersonate as the victim in front of the target website, \GB  needs to acquire the device fingerprinting information from the victim's device. \GB  employ the following two methods to capture the victim's browser fingerprint. 

% 4/7: why do we need the following two methods? Can we not simply say that we use the approach similar to standard device fingerprinting approach. Anyway we already provided details on the approaches that you can used to gain the device information in Section 2, no?

% ZL 4/8: we need the following two methods. 
%\vspace{-1mm}
%\vspace{1mm}
\noindent \textbf{With JavaScript:} 
JavaScript provides a variety of APIs that can be utilized to extract the device information. 
The execution of these APIs does not require any permission from the users~\cite{mulazzani2013fast}. 
For instance, the API $\mathtt{navigator.platform}$ retrieves 
the details on the platform (e.g., MacIntel, Win32, Linux, etc.) of the device that the user is using. 
\redtext{The $\mathtt{Date()}$ API  extracts the device's system time and its timezone. }The $\mathtt{cookieEnabled}$ API tells if the browser has disabled cookies or not. These methods are
 exactly the same as deployed by the web service that uses browser fingerprinting. All these APIs are completely transparent to the user.
% PS 4/7: Not clear? can you elaborate this more. What does it mean by trustworthy here?

% ZL 4/8: I comment the following sentence. I tried to point out regular web user wouldn't change those API values, so the attacker can believe that all the values he obtained is the real one, not the spoofed one.
 
%These APIs are \textit{read-only} APIs, i.e., the web users without having a strong JavaScript knowledge cannot easily fake the values returned by these APIs. 
%Hence, these features are trustworthy for the attackers. 
%\vspace{-1mm}
%\vspace{1mm}
\noindent \textbf{Without JavaScript:} 
% ZL 4/8 What I supposed to say before should be "user-agent can be extracted not only Javascript APIs ", not "not Javascript". I added "not Javascript APIs" feature example: Flash.
Some device fingerprinting attributes can also  be extracted through methods other than JavaScript APIs. For instance, user-agent, supported languages and their order can be retrieved from the HTTP header~\cite{kristol2000http}, fonts can be extracted using Flash and CSS. 
% PS 4/7: Any other feature that cannot be extracted using JavaScript
% ZL 4/8: "language" can also be extracted using Javascript and non-Javascript method.
% Why HTTP header is preferred?
% ZL 4/8: I change "preferred" to "needed". Also added the explanation why HTTP header is needed.
Although JavaScript has $\mathtt{navigator.userAgent}$ API, the use of HTTP header is preferred to retrieve  user-agent
because the user can disable the JavaScript, thereby failing the retrieval of user-agent through JavaScript API.  
%If the user disables the JavaScript, the website cannot use the Javascript API to obtain the user-agent. 
Fortunately, in such a situation, the HTTP header can still provide the user-agent attribute of the browser.
%the websites are preferred to use the HTTP header to confirm which browser platform the user is running. 
%PS 4/7: What does it mean by installed languages? and order of languages? Have you defined them earlier?
% ZL 4/8: I change "installed languages" to "browser language list". In the browser setting page, the languages order can be set in the language setting page. So the order of languages is the order of languages...I don't know how to define them...
\redtext{Similar to user-agent, the list of supported languages and their orders can also be extracted using HTTP header.
While the JavaScript has various APIs for extracting many of the fingerprinting attributes, 
it 
}For some of the attributes, such as the list of fonts in the device, JavaScript does not offer any APIs.
% (there is no single API to return a object contain full list of fonts). 
%Some browser features cannot be extracted by the Javascript APIs. 
Flash and CSS are used to list the available fonts in the device. 
%CSS is another method to detect the list of fonts. Javascript does not have API to return a full list of the fonts.
% PS 4/7: Why use additional approach to the JavaScript API? What purpose does it server?

% ZL 4/8: I comment the cookie enable example here. It is similar to "user-agent". 

%The Javascript API $\mathtt{navigator.cookieEnabled}$ is not used alone to detect if the user has disabled the cookie. 
%Oftentimes, a new and random cookie is injected in the browser 
%and by checking the presence/absence of the cookies, it is determined if the cookie has been enabled or disabled

%By employing the aforementioned approaches, one can obtain various browser fingerprinting features. 

% PS 2/23: I do not think we will need the text below. In this para, we can  simply say that we designed a website and implemented various functionality as discussed in earlier section (probably in background section), perhaps list few features. We were able to capture all the highlighted features in Table 1. An attacker can simply design a fake website and fool the user to visit it and collect fingerprinting information. THis is all we need here.

%PS 2/23: This section should be second step of our attack
%i.e., \subsection{Spoofing  User Browser Fingerprint}
% Here, you expalain about your spoofing design details. You explain why you chose Panopticlick (for the evaluating the feasibility of spoofing). Then details on mechanisms you employ to spoof. 

%\secspace
\vspace{-2mm}
%\vspace{-2mm}
\subsection{Visual Attack}
\label{sec:VAAPS}
%\vspace{-2mm}
We utilize the Panopticlick website \cite{panopticlick} and the FingerprintJS demo website \cite{fingerprintjs} to assess the effectiveness of various spoofing methods, 
%(as discussed in Section \ref{sec:spoofing_methods}), 
referred to as the \textit{visual attack}. 
%\textcolor{blue}{Figure \ref{fig:win_firefox} and Figure \ref{fig:fjsvictim} are large and may occupy too much space, so we put those two figures in Appendix \ref{sec:vpfp}.}

%\vspace{-2mm}
%\smallskip
	\noindent\textbf{Attacking Panopticlick Site:} 
% can effectively spoof various device fingerprinting information, 
% we assessed these methods with  Panopticlick. 
%The Panopticlick  website inspects the browser privacy protection and raises awareness about the threat of device fingerprinting. 
Panopticlick provides a dashboard for displaying the browser information, which we leverage to assess our spoofing methods. 
Figure \ref{fig:win_firefox} in Appendix \ref{sec:vpfp} presents a snapshot of 
the Panopticlick dashboard showing fingerprint information 
when a (victim) user uses a Firefox browser on a Windows machine, i.e., ``Win+Firefox''. 
By visually inspecting the information displayed on the dashboard, we validated if the spoofing methods succeed
in injecting spoofed attributes.
We use the browser setting and debugging tool to 
modify the following attributes -- user-agent, HTTP accept header, cookie enabled, and local storage, used in Panopticlick. 
Specifically, we use the
 debugging tool to change the user-agent and the browser's setting option to change the language attribute found in HTTP accept header. 
%PS 4/7: Not clear what you  are trying to convey here?
% ZL 4/8: I want to explain how to change the language feature in http accept header.
\redtext{HTTP accept header contains a list of languages and their  order.  }We change the language category and its order
in the Google Chrome browser to meet target languages combination. 
To modify the \textit{cookie enabled} and \textit{local storage}, we use corresponding options in the privacy setting of the Google Chrome browser.  
%which are the same method category as the previous two in this paragraph. 
To change the remaining attributes used in Panonpticlick,
either the script injection or the script modification approach is used. 
Due to the convenience of using script modification, 
we use this approach for the said purpose. 
%We changed all of the rest features on the ``Panopticlick" site using the method "the website scripts method". 
Specifically, we change the attributes' value in the JSON file of the script such that the Panopticlick would receive the modified values. 
% PS 4/7: Where do you use Script INjection method? Never mentioned here.
% ZL 4/8: Script injection changed the API value. Script modification changed the value that will be sent to the server. You can use either of these two method. Script modification is more convinenet here, so I choose script modification in this attack, not script injection.

\noindent\textbf{Attacking FingerprintJS Site and Real-Life Fingerprint Service:}
We also successfully did the visual attack against FingerprintJS website and the Fingerprintjs pro service. We listed full details in Appendix \ref{sec:vpfp}.

%\vspace{-2mm}
%\smallskip

\begin{table}[t]
\vspace{-5mm}
%	\begin{threeparttable}
		\centering
%		\scriptsize
%		\captionsetup{width=\linewidth}
		\caption{The attacks executed for each user in our evaluation methodology.}
		\label{tbl:nine_attack}
		\begin{tabular}{ | c | c | c | c | c | c | c | c | c | c | }
			\hline
			\textbf{Attack Number} &  1 & 2 & 3 & 4 & 5 & 6 & 7 & 8 & 9\\
			\hline
			\textbf{Time Gap (day)} & 1 & 7 & 15 & 30 & 60 & 90 & 180 & 270 & 365\\
			
			\hline
			
		\end{tabular}
%	\end{threeparttable}
\vspace{2mm}
\end{table}

%\secspace
\vspace{-3mm}
%\vspace{-1mm}
\subsection{Algorithm Attack: Attacking Prominent Fingerprinting Based Techniques} 
%\vspace{-2mm}
%Since the fingerprint information can be easily changed and spoofed in its entirety (as demonstrated through the visual attack), 
We emulate the attack against the  browser fingerprinting algorithms by simply
copying the entire fingerprint, referred to as the \textit{algorithm attack}. To evaluate the performance of our algorithm attack, we employ three prominent browser fingerprinting algorithms -- \textit{Panopticlick, \RLA, \HLA} and launch the algorithm attack against them. 
We utilize the dataset from \cite{spirals-team}, referred to as
the \textit{\originaldata}, to evaluate the performance of the algorithm attack. Details on the dataset are provided in Section \ref{sec:f_d}. Each fingerprint in the dataset has following three timestamps: $created\_date$, $updated\_date$ and $expired\_date$, which denote the timestamps when 
the fingerprint is created/recorded, updated, and expired, respectively.
%When a user visits the website, the website would record this visit, obtain the device fingerprinting features, and store it as one tuple into the database. Each tuple contains all the features that would be used in the three device fingerprinting algorithms. We will give more details about this database in Section \ref{sec:f_d}.
% ZL 4/8: Adding explanation about the attack instances.
Utilizing the \originaldata, various datasets are 
generated based on different collect frequency, referred to as the \textit{\benigndata}. 

In a real-world setting, an adversary can capture the victim's browser fingerprint at any point in time. Given this, we consider that the attacker can spoof any of the fingerprints in the \originaldata. 
Therefore, we copy one fingerprint instance  of the given user at a time, update the creation date and order, consider it as a spoofed fingerprint, and inject it back into the \originaldata,  
forming the \textit{\attackdata}. Such an injection of copied fingerprint simulates the scenario where an adversary
acquires the victim's fingerprint, and then tries to impersonate the victim by spoofing  the fingerprint.
The fingerprinting algorithms are executed on the \attackdata to link together the browser fingerprints from the same user. The attack succeeds if the fingerprinting algorithm incorrectly marks the spoofed fingerprint as from the victim.

Since the browser fingerprint changes over time, 
the impact of the algorithm attack may vary based on the 
gap between the time when the fingerprint is acquired and the time when the attack is launched, referred to as ``\textit{time gap}''. 
In terms of the dataset, the time gap refers to the difference in the $created\_date$ between two fingerprints from the same user. 
%., e.g.,  
%the time gap of two fingerprints with the $created\_date$ of  04/01/2020 and 04/26/2020 is  25 days. 
To measure the effectiveness of the time gap in our algorithm attack, we design and build nine different attacks based on nine different time gaps. 
\redtext{The time gaps (referred attack number) considered in our attacks are -- 1 day (attack 1), 7 days (attack 2), 15 days (attack 3), 30 days (attack 4), 60 days (attack 5), 90 days (attack 6), 180 days (attack 7), 270 days (attack 8), and 365 days (attack 9). 
}The attack number and corresponding time gaps are presented in Table \ref{tbl:nine_attack}.

In the \originaldata, each user has more than one fingerprint collected over a long period of time. 
To execute the aforementioned nine different attacks,
we assume that the adversary captures  
the oldest of the fingerprints (the first one) of the user
and spoofs after each of the `n' days considered in nine different attacks, referred to as spoofed/copied fingerprint. Thus, we consider \AOSF setting for our nine attacks.
The $created\_date$ of the spoofed fingerprint is set as `n' days after its original $created\_date$. 
Similarly, the $expired\_date$ is set to 5 days after its $created\_date$. 
Since none of the three algorithms uses the $updated\_date$, we set its value to ``NULL''.
Although we employ \AOSF approach for all our attacks, 
the results are also applicable to \AOSO, where the fingerprint is spoofed only once. If the fingerprint is acquired frequently over a period of time, our attack would have a higher chance to succeed. 
%Given that our algorithm attack with \AOSF is highly successful, the attack with \AFSF would also be highly successful. 

%\begin{table}
%\begin{threeparttable}
%	\centering
%	\scriptsize
%	\captionsetup{width=\linewidth}
%	\caption{The attacks executed in our evaluation for each user.}
%	\label{tbl:nine_attack}
%\begin{tabular}{ || l || l || }
%\hline
%\textbf{Attack Number} & \textbf{Time Gap} \\
%
%\hline
%Attack 1 & 1 day\\
%\hline
%Attack 2 & 7 days\\
%\hline
%Attack 3 & 15 days\\
%\hline
%Attack 4 & 30 days\\
%\hline
%Attack 5 & 60 days\\
%\hline
%Attack 6 & 90 days\\
%\hline
%Attack 7 & 180 days\\
%\hline
%Attack 8 & 270 days\\
%\hline
%Attack 9 & 365 days\\
%\hline
%
%\end{tabular}
%\end{threeparttable}
%\end{table}

%\bluetext{Those 9 attacks are under attack model \textit{\AOSF}. 
%For each attack instance, we set the $expired date$ as 5 days later than its $created date$. As the three algorithms would not use the tuple $updated date$, we set this value as ``NULL'' in attack records.}
%Using different time gap will allow us to see the attack effectiveness. 
% PS 4/7: What does it mean by "all other" features? It is all features, no?
% ZL 4/8: No. Three attributes are different, I explained in the previous paragraph. 

%\bluetext{
%All other features in 9 attack records remain the same as the benign case that is the acquired browser fingerprinting information.} 
% 4/7: What is ``auto-attack running code'', never defined. Define it.
 To evaluate our algorithm attack,
we utilize the exact same code as that of FP-Stalker, 
which is made publicly available in the GitHub repository by its authors~\cite{spirals-team}. 
\redtext{We note that the authors of FP-Stalker }They have implemented all three 
algorithms, namely Panopticlick, \RLA, and \HLA, considered in our study, and can be found in their code repository.
%The code  give results of different 3 tracking algorithms under same dataset. 
For each user in the dataset, we run these algorithms in two different settings --  i) the benign setting without any spoofed fingerprints, and ii)  the attack setting with nine different spoofed (or attack) fingerprints. 
% PS 4/7: Why do we need the following description?
%The results contain Panopticlick - benigncase, Panopticlick - attackcase, Rulebased - benigncase, Rulebased - 9attackcase, Hybrid - benigncase and Hybrid - 9attackcase, totally 6. 
% PS 4/7: Why is the following recommendation important and how you follow it?
%We follow the recommendation of the authors of FPStalker that there should be at least six fingerprints in the dataset. 
%The authors of FP-stalker recommended that 
%the minimum number of fingerprints in the dataset should be 6, so we follow their advice. 

% ZL 4/8: Explained here.
\redtext{In our experiment, our algorithm attack was automated 
through a python script that injects the nine spoofed fingerprints corresponding to nine different attacks against a user into the \originaldata, executes the three algorithms, and deletes the injected nine fingerprints. The script repeats the process for each user.}
%When we add the attack instances into the database, we use our ``auto-attack running code''. 

% PS 4/7: What does it mean by other existing data? Other user instances?

% ZL 4/8: I commented this sentence. No use now.
%It will make sure the attack insertions does not impact any other existing data in the database. 
% PS 4/7: BTW, you never mention which benign instance of the user you used to spoof? Are all benign instances used for the attack, or 

% ZL 4/8: I added the explaination.

%\vspace{-5mm}
\section{Dataset \& Evaluation Methodology}
\label{sec:eval_methodology}
%\secspace

%In this section, we first provide a brief description of the dataset from FP-Stalker that we utilize in our study. Then, we present the details of the methodology we followed to evaluate our spoofing methods and the performance of our attack against the browser fingerprinting techniques. We also present the evaluation metrics that we used to evaluate the performance of our attacks.
% introduce the details of our dataset, how we add attacks into our dataset, the evaluation scenario and the evaluation metrics.

%PS 4/8: You should "briefly" discuss how FPStalker collected data samples.

%\begin{figure}[]
%	\centering
%	\includegraphics[width=.9\linewidth]{macos_chrome.pdf}
%	\vspace{3mm}
%	\caption{Attacker (before the attack): The original features of combination ``Mac+Chrome''. Testing website: \textbf{Panopticlick}}
%	\label{fig:mac_chrome}
%	\vspace{1mm}
%\end{figure}

%\vspace{-1mm}
%\vspace{-2mm}
\subsection{FP-Stalker Dataset} 
\label{sec:f_d}
%\vspace{-2mm}
%PS 4/7: What does it mean by record here, should define that it is fingerprint of the users' device that contains several device information.
%\vspace{-2mm}
We use the FP-Stalker dataset \cite{spirals-team} to evaluate the performance of \GB against browser fingerprinting techniques.
\redtext{The dataset and its implementation of fingerprinting algorithms are publicly available in the GitHub repository. 
}The authors of FP-Stalker designed and built two extensions, one for the Firefox browser and the other for the Chrome browser, and used the AmIUnique website to collect the browser fingerprints.
Although they noted that their dataset consists of 98598 browser fingerprints from 1905 users collected over a period of two years in their paper, their public dataset contains only 15000 fingerprints collected from 1819 users.
Each fingerprint in the dataset contains 40 variables. 38 of them correspond to browser fingerprinting attributes. 
The remaining two variables are ``Counter'' and ``ID''.  The counter denotes the order of the fingerprint based on the 
created date of the fingerprint. 
ID uniquely represents an individual user, referred to as ``\textit{original ID}'' in our analysis. 

%  On their GitHub repository , the dataset they provided contains 15000 records collected from 1819 different users, which is what we use to evaluate our attack. 
% PS 4/8: What are these 38 features? did you defined those features in earlier section, if so, should refer back where you have defined.

% PS 4/8: What exactly is the inconsistency? Can  you please clarify?
% ZL 4/10: I added explanation in the following paragraph.

We observed that the fingerprints in the dataset have inconsistency, i.e., the  fingerprints from the given user do not have consistent browser attributes, e.g., different operating systems, the newer fingerprint having older browser version/vendor than the older fingerprint.  
As such inconsistency in the dataset may impact 
the performance of the browser fingerprint algorithms as well as that of our attack, we removed all 
inconsistent fingerprints resulting in the
dataset with the fingerprints from 275 users. Further, we remove the user having less than seven fingerprints, which is considered insufficient for the three fingerprint algorithms, dropping the user counts in the dataset from 275 to 239. This dataset is what we use to evaluate our attack. 
%After deleting the user who has inconsistency records in the dataset, there are totally 275 users left. 

%\vspace{-1mm}
%\smallskip
\noindent{\textbf{\textit{Collect frequency:}}}
{We sample the dataset using a configurable collect frequency similar to FP-Stalker.  
	Collect frequency indicates how often a browser is fingerprinted. 
	The lesser the fingerprinting frequency (or the higher collect frequency), the harder it would be to track the user. 
	We use 11 different collect frequencies -- 1, 2, 3, 4, 5, 6, 7, 8, 10, 15, and 20, in terms of days.
	To generate a dataset for a given collect frequency, we employ the approach as suggested in FP-Stalker. 
	When a dataset is sampled using a collect frequency, the approach usually extends the dataset by copying (or replicating) the fingerprints 
	at missing dates, therefore, we refer to it as the \textit{expansion algorithm}.
	The expansion algorithm iterates in time with a step of collect frequency days and creates (or recovers) the browser fingerprint at each time step $(t \pm f_c * i)$, where $t$ is the fingerprint creation date, $f_c$ is collect frequency, and $i$ is a natural number. The iteration continues until the expired date of the previous and the current fingerprint is reached.
The process is repeated for each of the fingerprints collected from the given user. 
Thus, the expansion algorithm generates a new dataset with the fingerprints sampled at a consistent frequency of collect frequency days.} 

% PS 4/8: What does it mean by this? raw or processed features?
% ZL 4/10: I commented them. No use now.

%no matter the raw features or processed feature. 
%The remaining two variables are ``Counter'' and ``ID''. Counter denotes the order of the record based on the date of creation 
%of the record. 
%So the higher the value of the counter, the later the create date in the record. 

\vspace{-3mm}
%\vspace{-2mm}
%\vspace{-1mm}
\subsection{Evaluation Methodology}
%\vspace{-2mm}

\subsubsection{Visual Attack}
%\vspace{-2mm}
%As mentioned earlier, we leverage the Panopticlick website \bluetext{and the FingerprintJS demo website} to visually evaluate the effectiveness of various spoofing methods employed in \GB. Specifically, 
We leverage the Panopticlick website and the FingerprintJS demo website and use various combinations of the terminal and the browser that the victim user may use to visually assess the spoofing methods.  
We employ a Mac laptop running macOS 10.14
Mojave, an Android phone running Android OS Pie 9.0, a Windows
desktop running Windows 10 OS as the terminal, while we use Google Chrome, Mozilla Firefox, Microsoft Edge, and Tor as the browser.  
%We consider that the adversary uses Google Chrome browser on a Mac laptop to spoof the victim's browser fingerprint. In other word, the adversary aims to 
%expose its Google Chrome browser as the victim's browser.    
% PS 4/8: Please check the OS version?
% ZL 4/10: checked. Changed macOS version.
%
%In the website attack to the website "https://panopticlick.eff.org/", we used the device-browser combination: ``MacOS+chrome". And the users' combinations are: ``MacOS+Firefox", ``MacOS+Tor", ``Android+Chrome", ``Win10+Chrome", ``Win10+Firefox" and ``Win10+Edge". 
Using the Panopticlick website, we note all the fingerprinting features when using different terminal-browser combinations.

For the purpose of our evaluation, we consider that the attacker uses the Google Chrome browser on the Mac laptop, 
i.e., ``Mac+Chrome'' to launch the attack.
% against the fingerprinting algorithms. 
We believe that this is a very standard setup, and the attacker can just use this setup to launch the spoofing attack.
%At first, we used ``MacOS+chrome" to visit ``https://panopticlick.eff.org/", checking this combination browser fingerprinting features. 
Since the user may use different combinations of the terminal and the browser to access the target website, we consider the browser fingerprint obtained from all the remaining combinations of the terminal and the browser as the victim's browser fingerprint.
 %Then, we used different combinations to visit ``https://panopticlick.eff.org/", to obtain users' browser fingerprinting features. 
We spoof each of the victim's fingerprints on the attacker's 
%choice of the terminal and the browser, specifically 
%Mac laptop with Google Chrome browser in our case, 
Mac+Chrome setup
% in our case
using various spoofing methods detailed in Section \ref{sec:spoofing_methods}. 
\redtext{The goal of spoofing is to orchestrate the attacker's 
fingerprint such that it appears as the victim's fingerprint. }To validate if the spoofing methods have indeed succeeded or not, we compare the fingerprint shown on the attacker's browser after spoofing with the previously noted victim's fingerprint.
 %We used those combination browser fingerprinting features, and put them into the `MacOS+chrome" combination. 
%At last, we make comparison with the users' original combination browser fingerprinting features on the ``https://panopticlick.eff.org/" with the spoofed combination browser fingerprinting features on the ``https://panopticlick.eff.org/". 
%\textcolor{blue}{We consider that the attacker uses the same device and browser as in attacking \textit{Panopticlick} website to attack against \textit{FingerprintJS} website. However, not all of the fingerprint features are displayed on the \textit{FingerprintJS} webpage like \textit{Panopticlick}. But \textit{FingerprintJS} website provides the victim browse history which can prove that our spoofing is successfully.}

%\begin{figure}[]
%\centering
%\includegraphics[width=.9\linewidth]{appendix_figure/fjsattacker.png}
%\vspace{3mm}
%	\caption{Attacker (before the attack): The original features of combination ``Mac+Chrome''. Testing website: \textbf{FingerprintJS}}
%	\label{fig:fjsattacker}
%\end{figure}
%\secspace

%\subsubsection{Benign Setting}

\vspace{-3mm}
%\vspace{-1mm}
\subsubsection{Algorithm Attack}
\label{sec:e_c}	
%\vspace{-2mm}
%\vspace{-1mm}

\noindent\textbf{Evaluation Scenarios:}
As mentioned earlier, to emulate our attack against 
the three fingerprinting algorithms, we insert
nine spoofed fingerprints, each corresponding to our nine different attacks, to the \originaldata. 
We inject the spoofed fingerprint after the latest fingerprint in the dataset that has the smaller (or same) created date as that of the spoofed fingerprint. 
As the counter in the dataset represents the order of the fingerprint based on its created date, when injecting 
the spoofed fingerprint, the dataset is re-organized for the counter. 
% PS 4/8: Have you defined the creation time earlier? If not need to define it.
%20210619For instance, suppose the spoofed fingerprint has the created date of {2015-10-15} in the \textit{year-month-day} format, the fingerprint with counter 6221 has the created date of 2015-10-15, and the fingerprint record with the counter 6222 has the created date of 2015-10-15. 
%20210619With these records, the counter 6222 is assigned to the spoofed fingerprint, and the counter of all the fingerprints having counter $>$6221 is increased by one. 
%The fingerprint with the counter 6222 becomes 6223, 6223 becomes 6224, and so on. 
%20210619Following the same approach, the position of our next spoofed fingerprint is determined and the counter of the fingerprint after the spoofed fingerprint is incremented by one. 
%20210619If  existing fingerprints have the same created date as that of the spoofed fingerprint, it is inserted at the end of such benign fingerprints 
%We will put the attack records after the last same time benign record. 
%For instance, suppose two benign fingerprints with the counters 7000 and 7001 have the same created date  as that of the spoofed fingerprint. In such a scenario, the counter 7002 is assigned to the spoofed fingerprint 
%20210619and all the counter of subsequent fingerprints are incremented by one.  
Thus, after injecting all our nine spoofed fingerprints, the new dataset would contain 15009 fingerprints (the \originaldata had 15000 fingerprints), with a different and corrected order in terms of the counter. 

In our evaluation, we choose one user as a victim at a time and evaluate our attack against the three fingerprinting algorithms, i.e., nine spoofed fingerprints corresponding to the chosen user are injected into the \originaldata generating the \attackdata. 
The \attackdata is then reverted back to the \originaldata. 
 We repeat the process for each user in the dataset,
 resulting in a total of 239 attacks (for 239 users). %against each of the 239 users. 
 
% Each attack contain 9 sub-attacks (attack 1 to attack 9), which is defined in the section \ref{sec:attack_design}. After finishing one attack for one user, we will recover the database to the benign setting. Then do the attack on the next user. 

% PS 4/8: 60% as testing or training? Please confirm.
% ZL 4/10: 60% testing

% PS 4/8: Why larger number of records is a issue to split with 60-40? Not clearr. 
% ZL 4/10: Because the I want to remain the number of records in the training set in the benign dataset is the same as the number of records in the training set in the dataset that has attack instances.
FP-Stalker uses 40\% of the total fingerprints as a training dataset and the remaining fingerprints as the testing dataset. 
% PS 5/3: I think the para below is not needed.
%%%%%%%%%%%
%Since the nine spoofed fingerprints result in a larger number of records, so (60-40)\% split cannot be used directly, because we want to make sure the number of records in the training set in the benign dataset is the same as the number of records in the training set in the dataset that has attack records. Therefore, we set it to 60.0748\% so that the first record in the nine attack cases in the testing dataset would remain the same as the benign dataset.
%%%%%%%%%%%%%%
% PS 4/8: Not clear what you are trying to convey here? CLarify it.
% ZL 4/10: Added the explanation in the previous paragraph.
% PS 4/8: Why we cannot used testing dataset directly, not clear at all. Why it is not sufficient and low in numbers?
% ZL 4/10: Explained now.
\redtext{The dataset cannot be directly used to evaluate the performance of browser fingerprinting algorithms due to the lack of sufficient fingerprints needed to execute the said algorithms, especially after the removal of inconsistent fingerprints. 
Therefore, }The fingerprint dataset is extended
% PS 4/8 How does the expansion algorithm work?
%Since the FP-Stalker code offers
leveraging the expansion algorithm (which is based on the collect frequency provided in FP-Stalker) resulting in a sufficiently large fingerprint dataset. 

\noindent\textbf{Evaluation Metrics:}
%We use Evaluation Metrics to measure the effective of our attacks in the user tracking. 
%The benign setting means the website is not subject to the spoofing attack, which means the website database only contains the regular users' records, and does not have any attack records.
% PS 4/8 Did you define the benign and attack case earlier? If not, then should define. 
% ZL 4/10 Yes, Defined in the last Paragraph of section 4.3
To evaluate the performance of 
fingerprinting algorithms in the benign setting (using \benigndata), we use true positive rate (TPR), whereas, to evaluate the performance of our attack against  fingerprinting algorithms, we use false positive rate (FPR). 
TPR measures how often the legitimate fingerprints have been correctly identified as belonging to the correct user's device. 
FPR measures how often the spoofed fingerprints are incorrectly identified as belonging to the victim.
%In the benign setting, we have the TPR ( true positive rate)=correctly identified instances of fingerprints from all users / total fingerprint instances from all users. 

In our evaluation, since we consider the tracking of the 
user over a period of time, we compute TPR and FPR for each day separately. 
When computing TPR and FPR for a given day, we consider only the fingerprints from that particular day. 
\redtext{TPR is computed as follows. 
\begin{equation}
\small
\setlength{\abovedisplayskip}{2pt}
\setlength{\belowdisplayskip}{2pt}
TPR  = \frac{TP}{N}
\end{equation}
where, $TP$ represents the number of fingerprints that have been correctly identified as from the legitimate users and $N$ represents the total number of legitimate fingerprints from all the users. Similarly, FPR is computed as follows.
\begin{equation}
\small
\setlength{\abovedisplayskip}{2pt}
\setlength{\belowdisplayskip}{2pt}
FPR  = \frac{FP}{\overline{N}}
\end{equation}
where, $FP$ indicates the number of spoofed fingerprints 
that have been incorrectly identified as from the victim user, and $\overline{N}$ indicates the total number of spoofed fingerprints.
}We expect the TPR to be high, close to 1, which indicates  the benign user is
being tracked well even in the presence of the \GB attack. We also expect FPR
to be close to 1, which denotes the attack is highly successful.

\section{Results}
\label{sec:result}
%\secspace

%In this section, we present the results of our evaluation with \GB, specifically the visual attack with Panopticlick, and the algorithm attacks against the three fingerprinting algorithms.

%\vspace{-2mm}
%\vspace{-1mm}
\subsection{Visual Attack Results}
%\vspace{-2mm}
%The results are: we spoofed ``https://panopticlick.eff.org/" with all of the users' combination browser fingerprinting features.
%
%I listed
%several screenshots of my ``MacOS+chrome" original features, and the comparison of different combinations with spoofed combination features: Figure 
%\vspace{-2mm}
We have successfully spoofed all the fingerprinting information on Panopticlick and FingerprintJS website. The full details of spoofing results are listed in Appendix \ref{sec:vpfp}.

%20210619\subsection{Attacking Real-Life Fingeprinting \\Services results}

%20210619In attacking \textit{Testing site} that deployed the FingerprintJS open source code, our two attack methods \textit{Script Injection} and \textit{Script Modification} all successfully spoofed all the 29 fingerprinting features listed in the script, and passed four \textit{haslied} detection functions which are used to detect if the current visit used spoofed fingerprinting features or not. We picked the \textit{Windows-Firefox} as the victim device operating system and browser setting, and used \textit{Mac-Chrome} as the attacker device.

%\textcolor{blue}{Appendix Figure \ref{fig:fjsattacker} display the snapshot of \textit{actual} device information from the https://fingerprintjs.com/demo when we use the \textit{Mac-Chrome} setting. Appendix Figure \ref{fig:fjsspoof} shows the snapshot of spoofing \textit{Windows-Chrome} using \textit{Mac-Chrome} setting. After spoofing, we could even check the victim browser visiting history.}

%\begin{figure}[]
%\centering
%\includegraphics[width=.8\linewidth]{appendix_figure/fjsspoof.png}
%\vspace{2mm}
%	\caption{Attacker (after the attack): The spoofed features of combination ``Win+Chrome'', showed in Google Chrome of combination ``Mac+Chrome”. Testing website: \textbf{FingerprintJS}}
%	\label{fig:fjsspoof}
%\end{figure}

%\secspace
\vspace{-3mm}
%\vspace{-1mm}
\subsection{Algorithm Attack Results}
%\vspace{-2mm}
%PS 4/9: I do not understand what you are trying to say here? Please clarify, and try to be formal when writing. This is a scientific paper, not a lab report. 

%\bluetext{In the attack of 3 tracking algorithms: with 11 different collect frequency days, 210 9in1 attacks, 1 benign case, 3 different browser fingerprinting tracking algorithms, we first calculated the average of average TPR and FPR score for attacks. So we finally got 11*(attackTPR+attackFPR+benignTPR)*3=99 graphs. We only show 9 of them here. Thoes graphs are from Figure \ref{fig:h_benign} to Figure \ref{fig:p_fpr}.}

\vspace{-2mm}
\subsubsection{Benign Setting}
%\vspace{-2mm}

\begin{figure}[t]
	\vspace{-6mm}
	\centering
	\subfloat[FP-Stalker \cite{spirals-team}]{
%		\vspace{-7mm}
  \includegraphics[width=.35\linewidth]{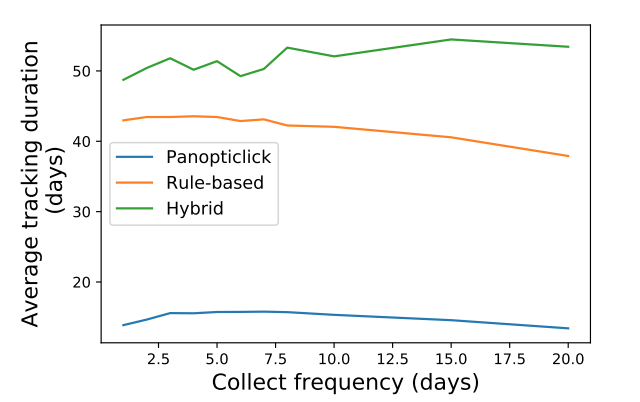}
    \label{fig:av1}}
  %\hfill
  \subfloat[Our Result]{
%		\vspace{-2mm}
  \includegraphics[width=.32\linewidth]{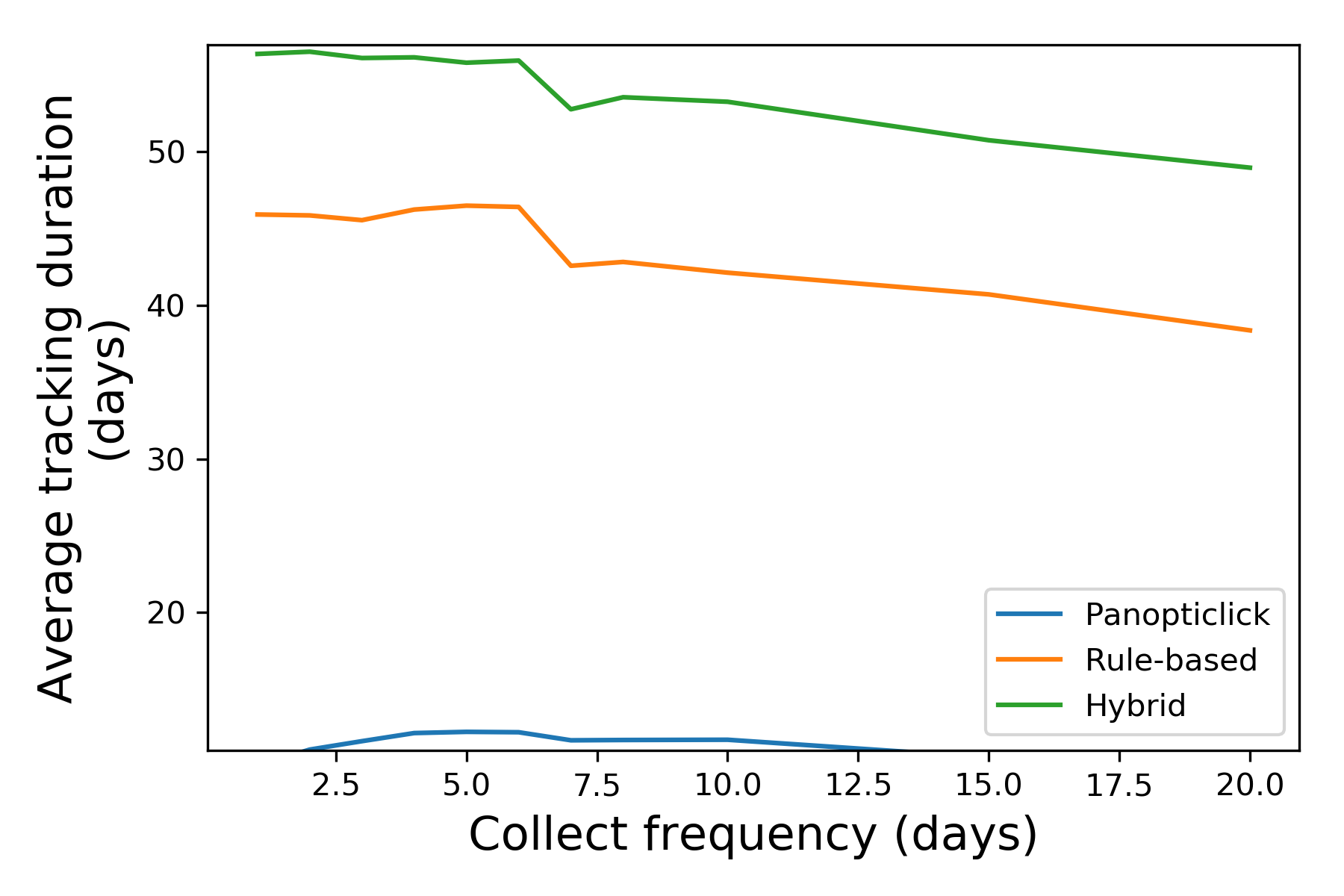}

  \label{fig:av2}
}
%\vspace{-1mm}
	\caption{Average tracking duration as a function of collect frequency for three different algorithms.}
%		\vspace{-3mm}
	\label{fig:av}
%\end{figure}

%\begin{figure}[]
	\vspace{-5mm}
	\centering
	\subfloat[FP-Stalker \cite{spirals-team}]{
  \includegraphics[width=.37\linewidth]{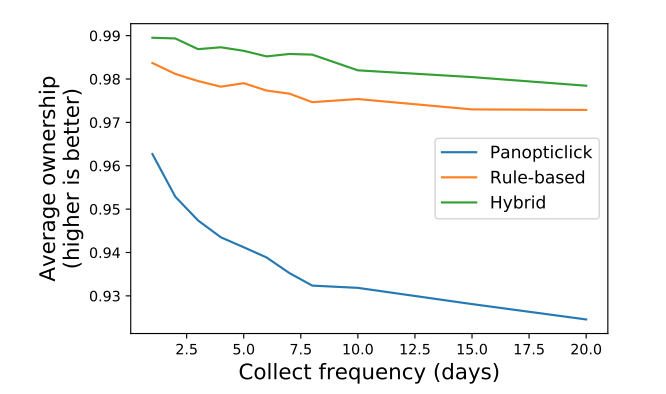}    \label{fig:av_ow1}}
  %\hfill
  \subfloat[Our Result]{
  \includegraphics[width=.32\linewidth]{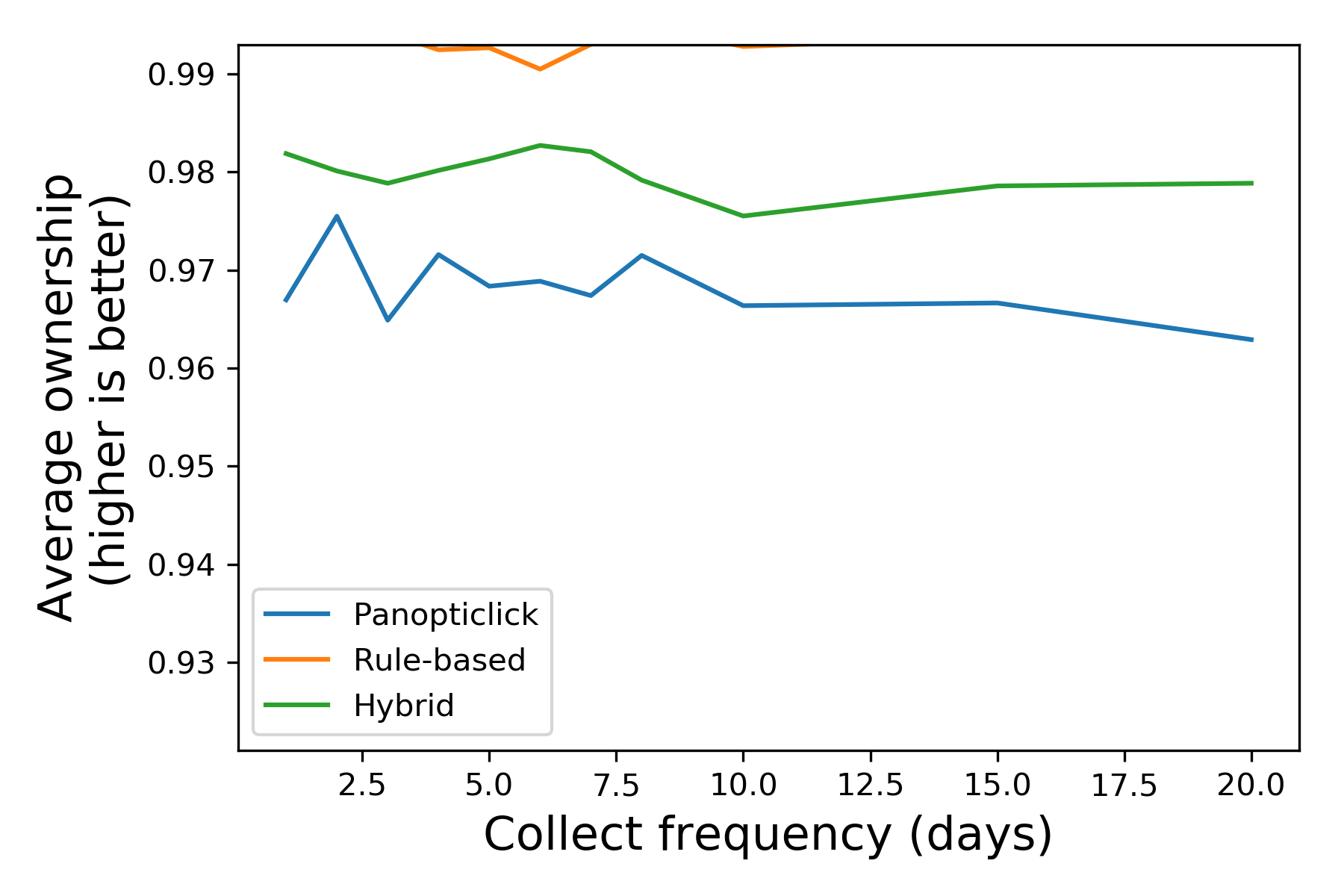}

  \label{fig:av_ow2}
}
%\vspace{-1mm}
	\caption{Average ownership as a function of collect frequency for three fingerprinting algorithms.}
%		\vspace{2mm}
	\label{fig:av_ow}
\end{figure}

%\vspace{-2mm}
To validate the implementation of the three algorithms (obtained from FP-Stalker repository), 
%that we use to evaluate our algorithm attack, 
%follow the algorithms presented in FP-Stalker~\cite{vastel2018fp}, 
similar to FP-Stalker, we plot various graphs on the performance of these algorithms for tracking the users. 
%The following graphs Figure \ref{fig:avg_maximum} to Figure \ref{fig:no_of_ids} are our recovery of the 3 algorithms in FP-stalker project. The x-axis is the collect frequency.
Figure \ref{fig:av} shows the average tracking duration (and Appendix Figure \ref{fig:av_ma} shows the average of \textit{maximum}  tracking duration) as a function of collect frequency for the three different fingerprinting algorithms. 
The tracking duration indicates the time duration (in terms of days) that the fingerprinting algorithm can track the user. 
The higher value of average tracking duration is considered good for user tracking. 
Figure \ref{fig:av} (and Appendix Figure \ref{fig:av_ma})
% in Appendix \ref{sec:bsta} 
shows that the 
\HLA outperforms Panopticlick and \RLA at tracking the user, which is inline with the one reported in FP-Stalker. 
Further, we achieved similar results as those reported in FP-Stalker for each of the three fingerprinting algorithms.

Figure \ref{fig:av_ow} shows the average ownership as a function of collect frequency.  
Ownership indicates how often the fingerprints were correctly associated with their actual users by the fingerprinting algorithms.
The higher the ownership score, the better would be the performance of the fingerprinting algorithms. 
We achieved average ownership of above 0.95 for all the three fingerprinting algorithms, which is inline with that reported in FP-Stalker \cite{spirals-team}. 
Appendix Figure \ref{fig:nu} 
shows the number of new IDs assigned to each user as a function of collect frequency for three different fingerprinting algorithms. If the number of new IDs assigned to a user is `1',  this means all his fingerprints have been identified as from the original user (the best result).  
If the number of new assigned IDs is larger than `1', say `n', 
this means the user's fingerprints are still tracked correctly, but as `n' separate tracking durations, which 
%These three tracking durations cannot be connected, and
 can be seen as from three different users. Although we used the exact same implementation of the three algorithms from FP-Stalker, we achieved slightly different 
results compared to those in FP-Stalker. We attribute this difference \redtext{in the results }to the difference in the volume of our dataset (239 users) compared to that used in FP-Stalker (1905 users).

\begin{figure}[t]
%\begin{figure*}[]
	\centering
\vspace{-6mm}
	\subfloat[Panopticlick]{
  \includegraphics[width=.25\linewidth]{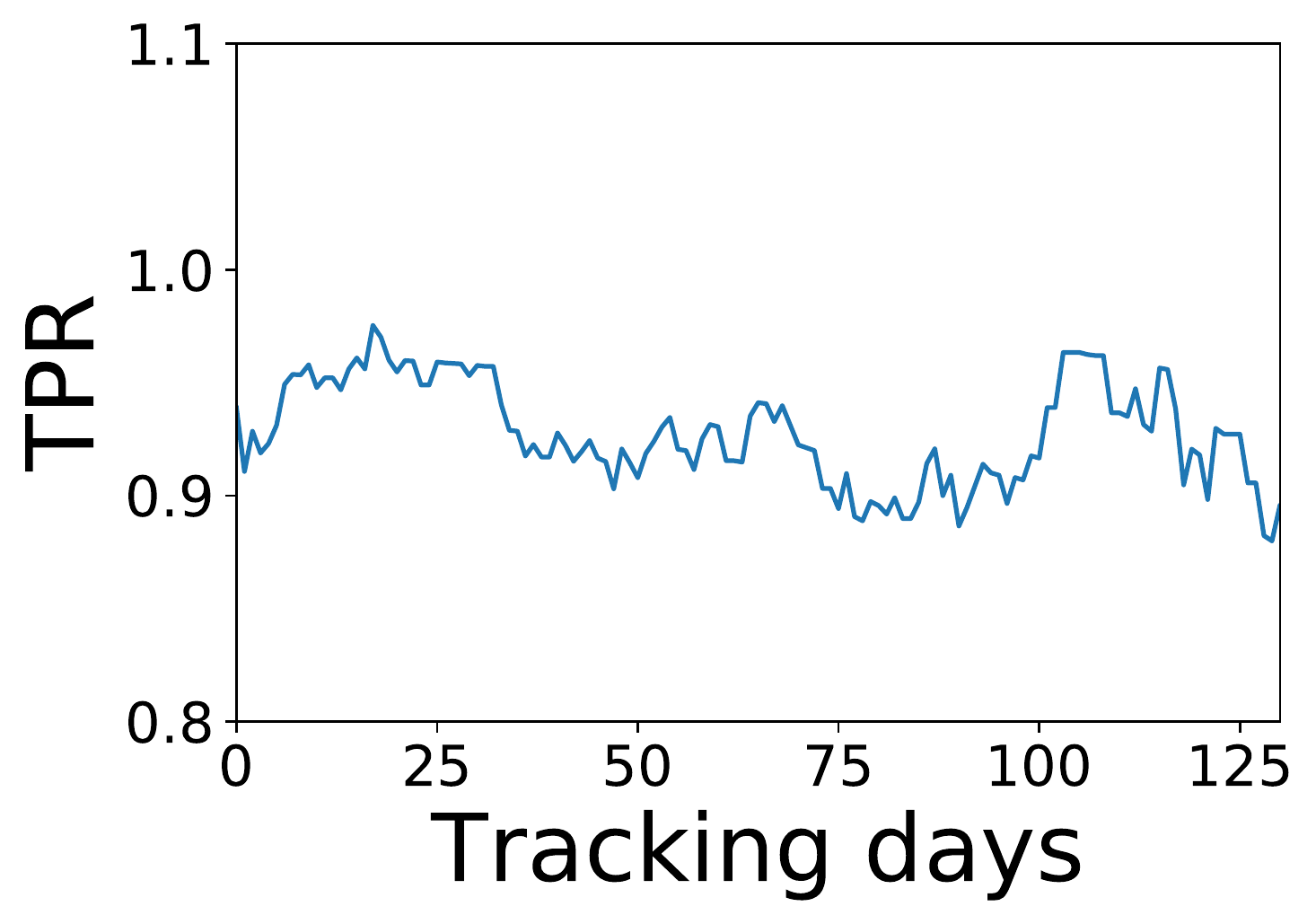}
    \label{fig:b1}}
  %\hfill
  \subfloat[\RLA]{
  \includegraphics[width=.25\linewidth]{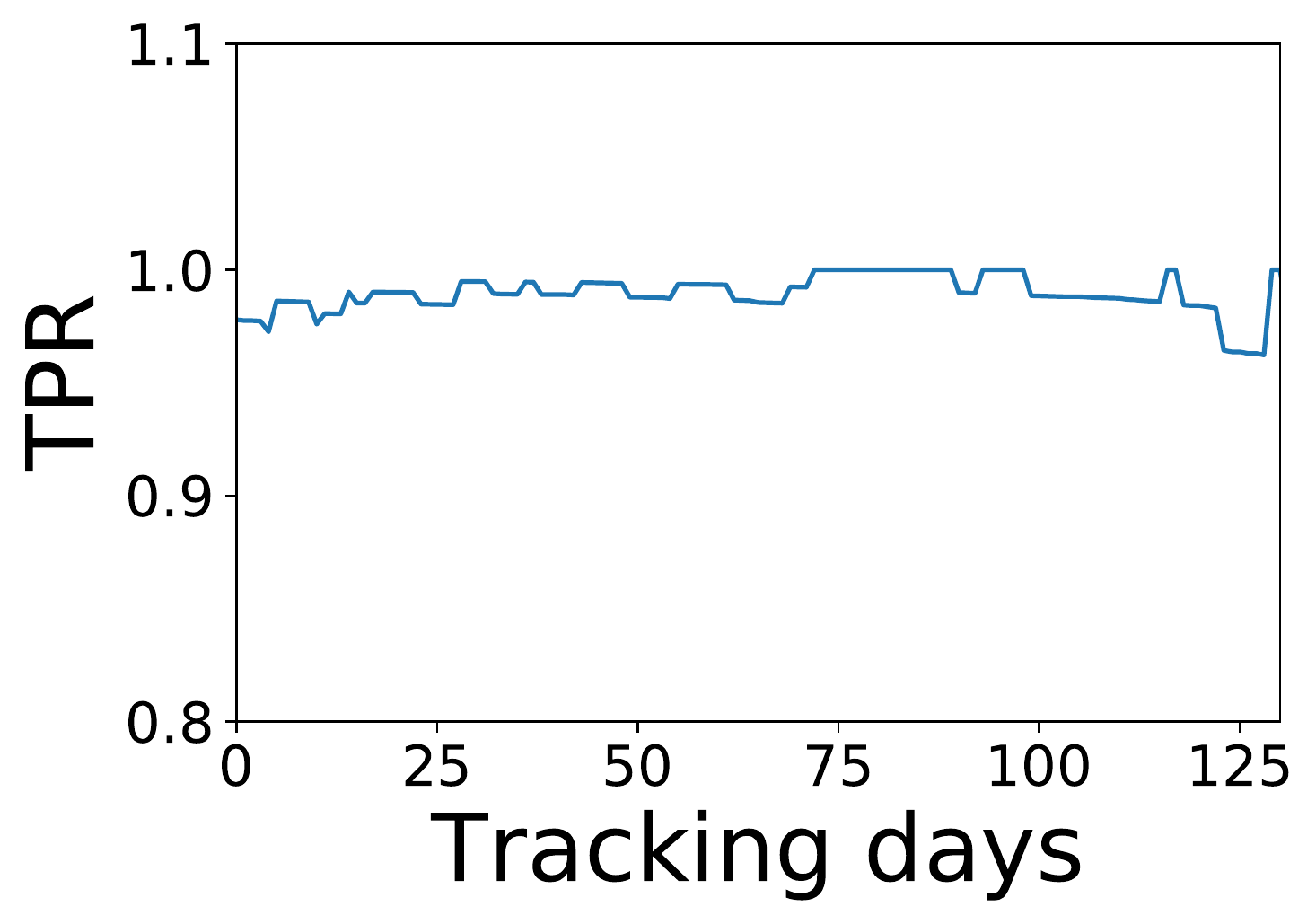}
    \label{fig:b2}}
  %\hfill
  \subfloat[\HLA]{
  \includegraphics[width=.25\linewidth]{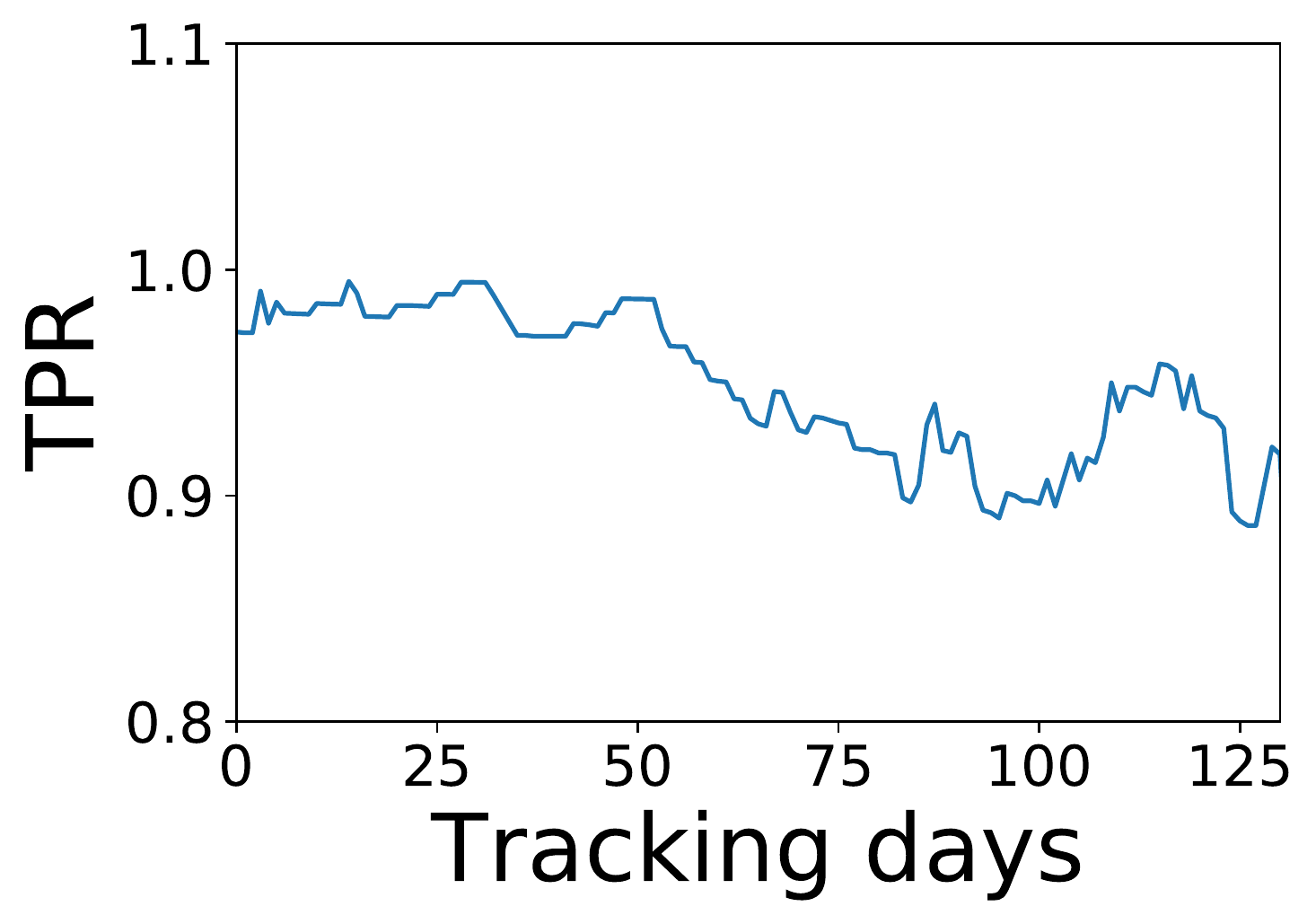}

  \label{fig:b3}
}
%		\vspace{-1mm}
	\caption{True positive rate (TPR) as a function of tracking days in the benign setting when the collect frequency is set as 1. 
		\vspace{-3mm}
%		  of Algorithms. (a) Panopticlick, (b) \RLA, (c) \HLA Collect frequency: 1
	  }
	\label{fig:b}
	
%\vspace{-3mm}
	\subfloat[Panopticlick]{
  \includegraphics[width=.25\linewidth]{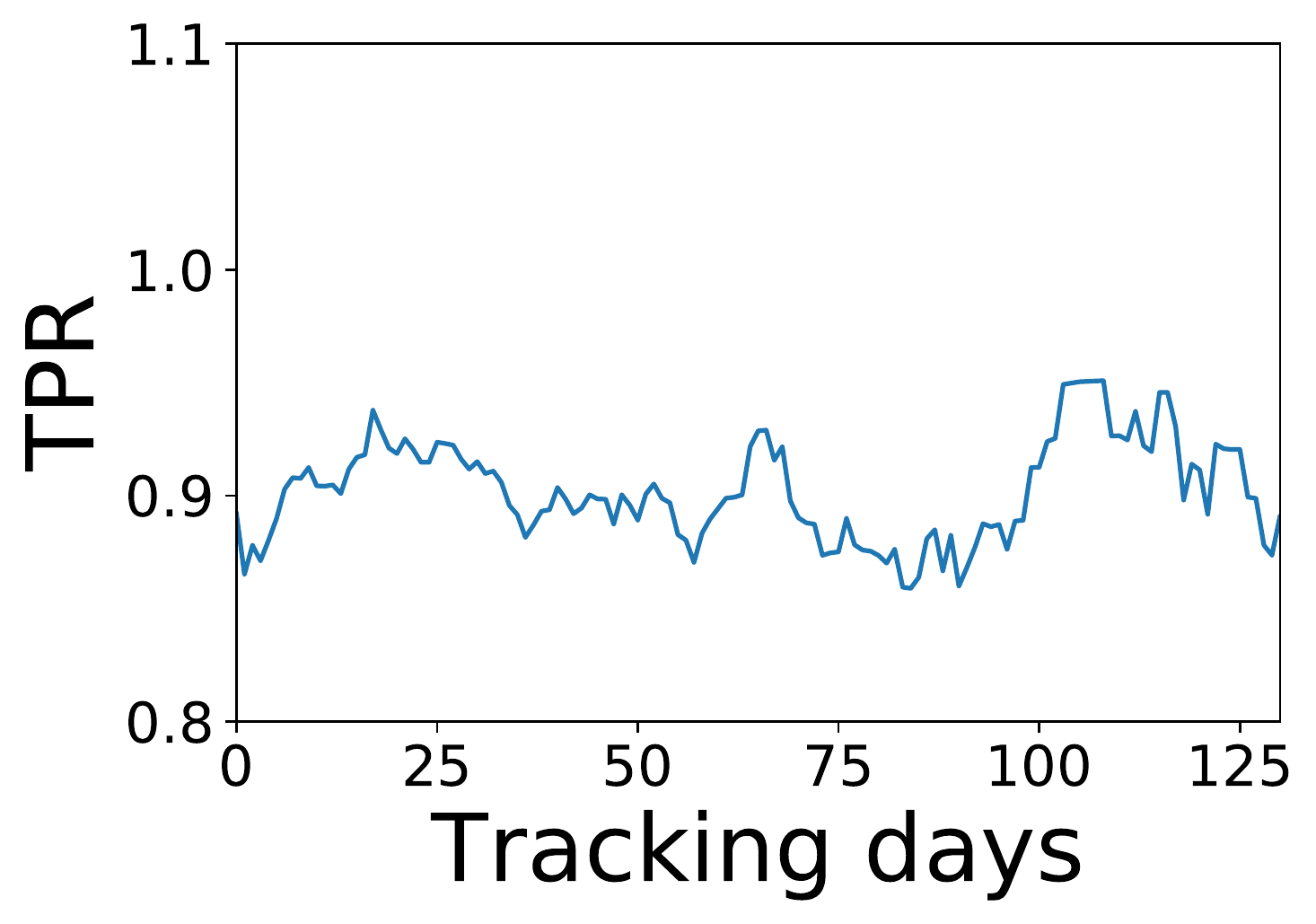}
    \label{fig:at1}}
  %\hfill
  \subfloat[\RLA]{
  \includegraphics[width=.25\linewidth]{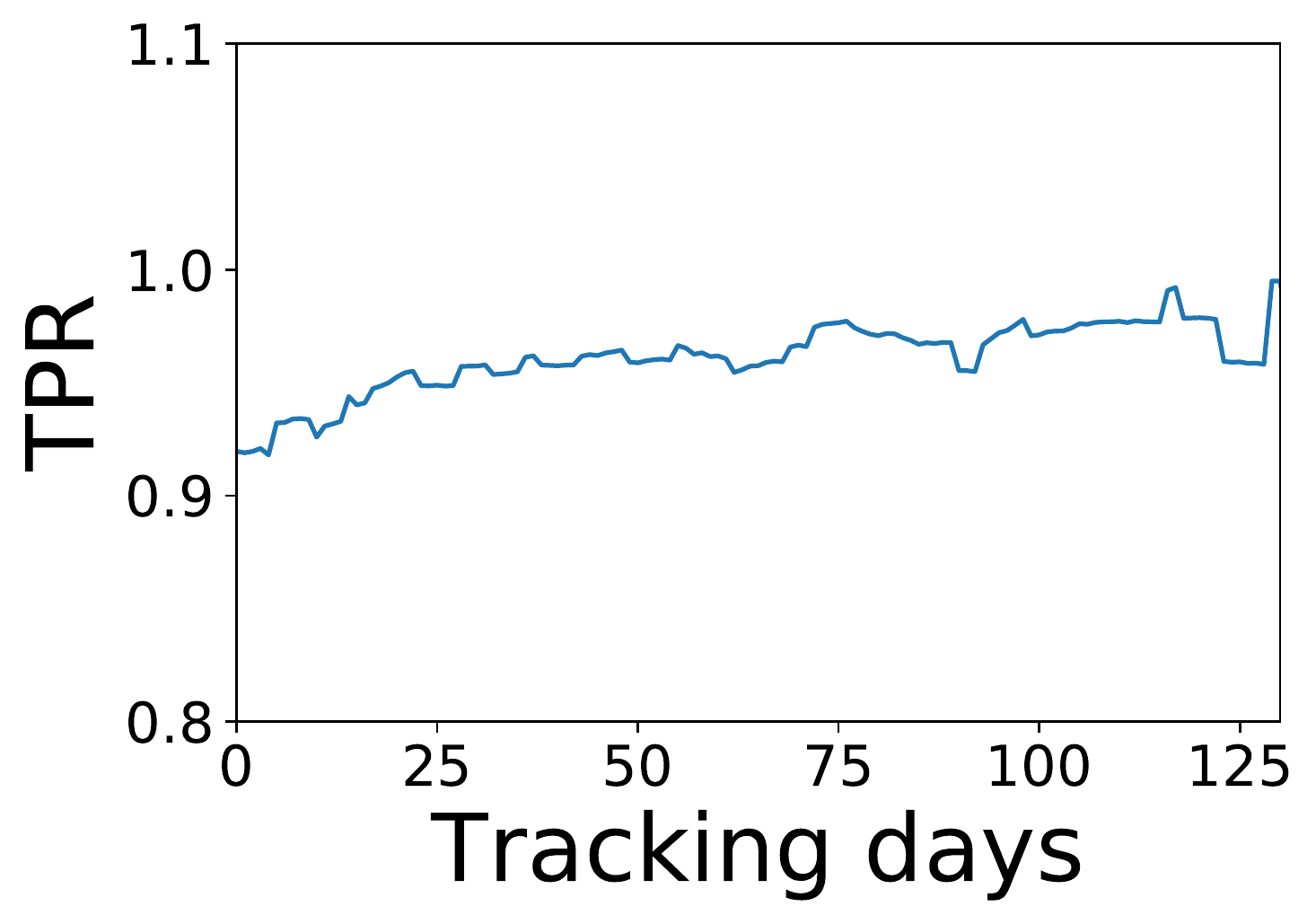}
    \label{fig:at2}}
  %\hfill
  \subfloat[\HLA]{
  \includegraphics[width=.25\linewidth]{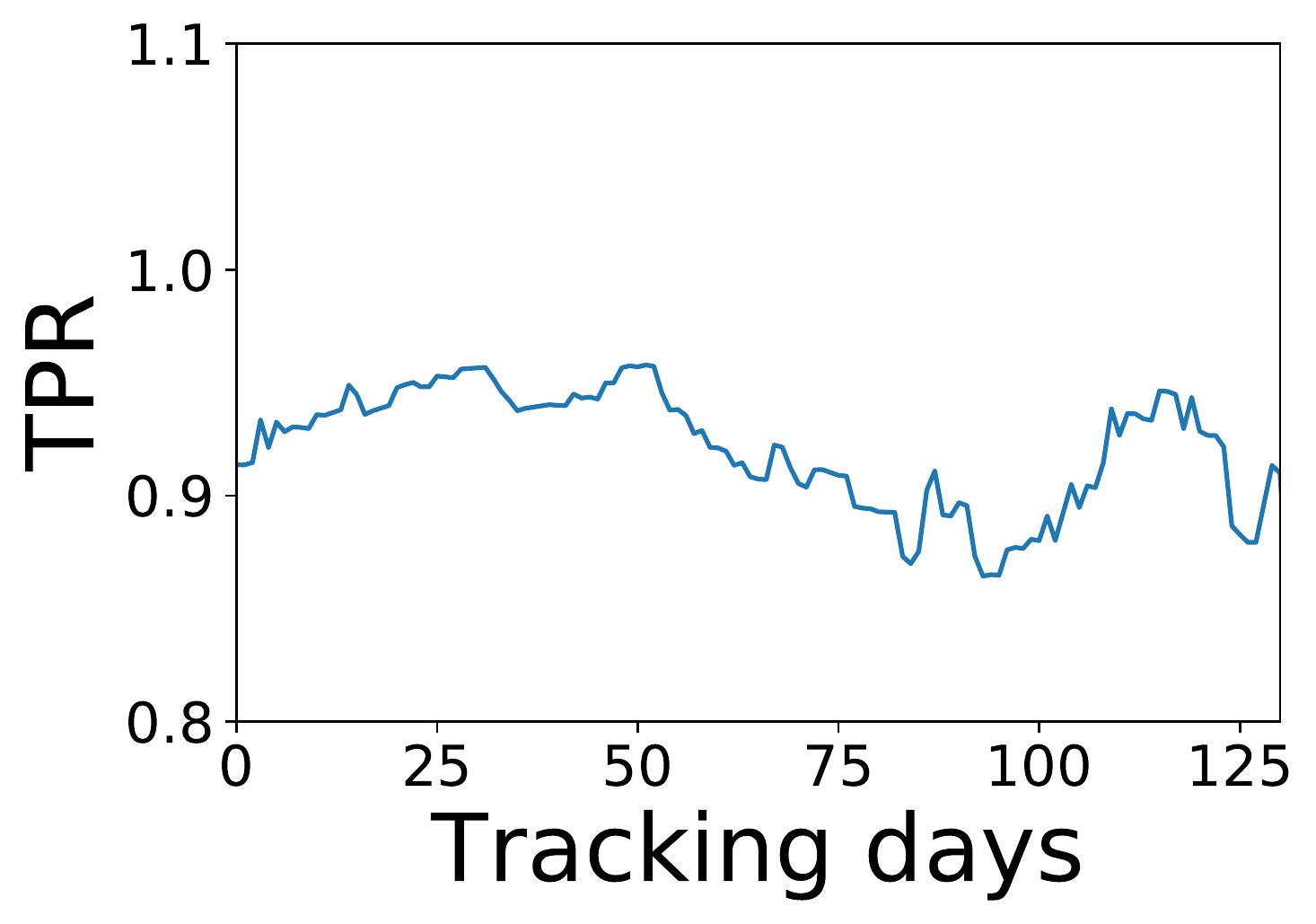}

  \label{fig:at3}
}
	\caption{True positive rate (TPR) as a function of tracking days in the attack setting when the collect frequency is set as 1.}
		\vspace{-3mm}
	\label{fig:at}

		\subfloat[Panopticlick]{
  \includegraphics[width=.25\linewidth]{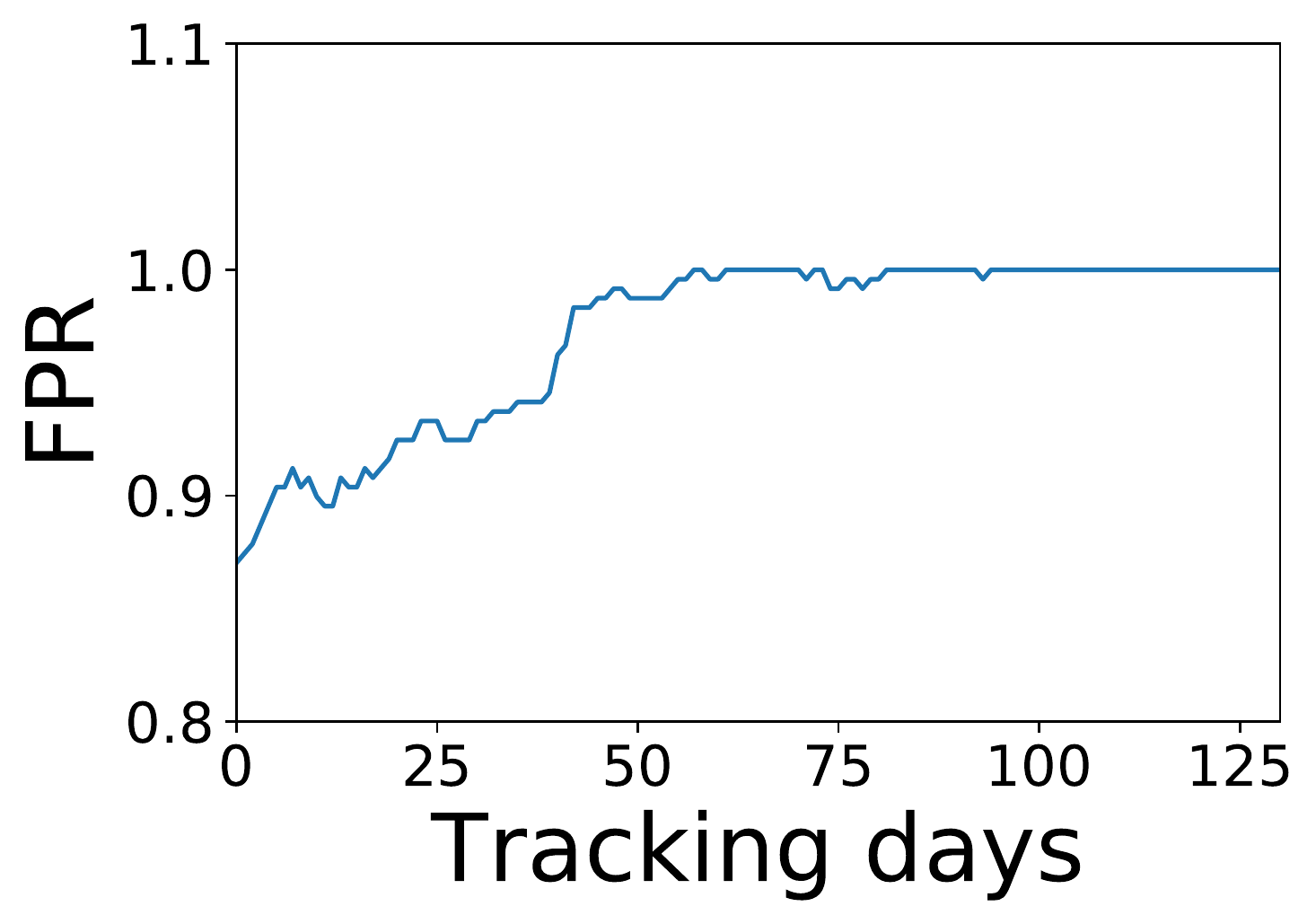}
    \label{fig:af1}}
  %\hfill
  \subfloat[\RLA]{
  \includegraphics[width=.25\linewidth]{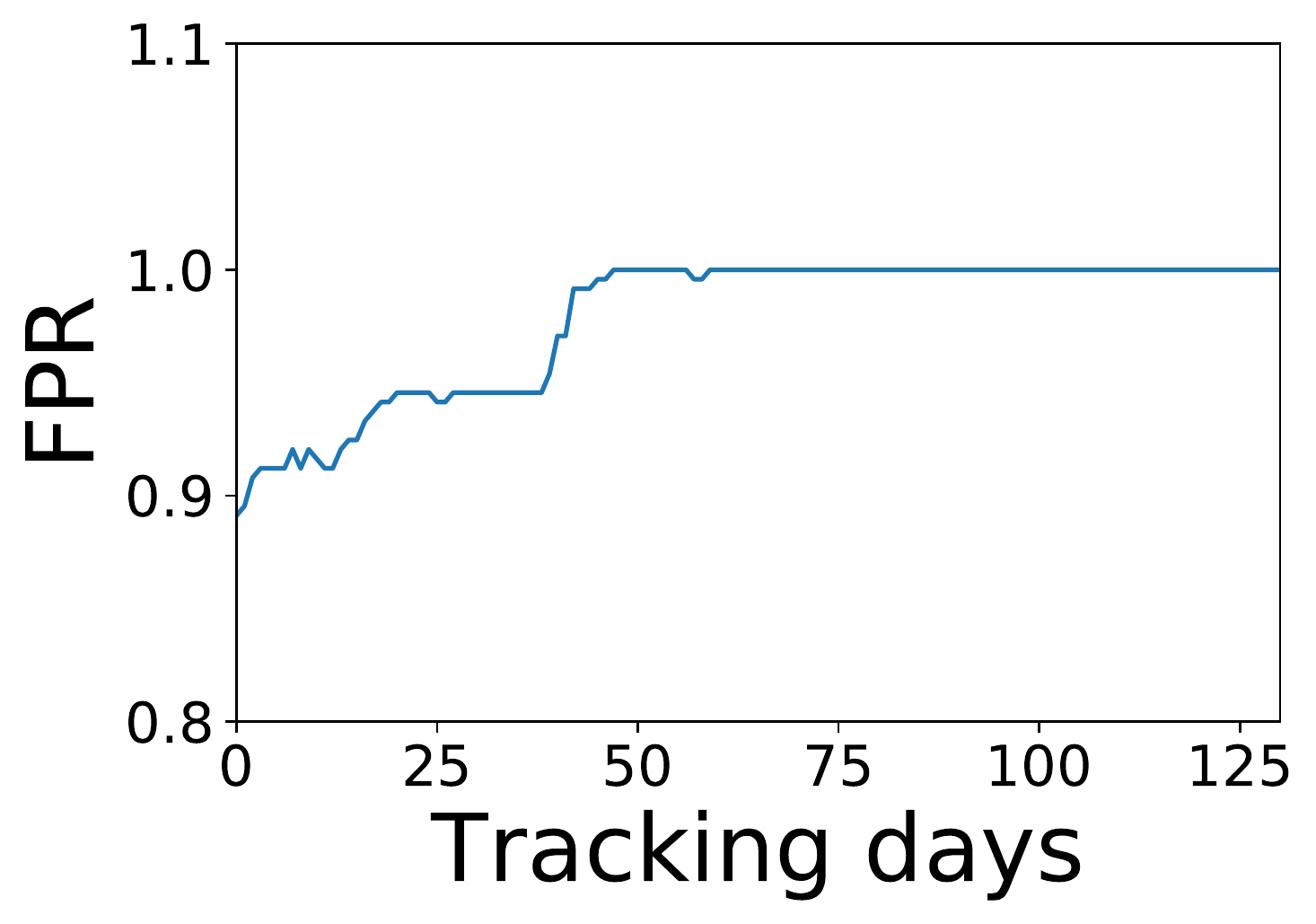}
    \label{fig:af2}}
  %\hfill
  \subfloat[\HLA]{
  \includegraphics[width=.25\linewidth]{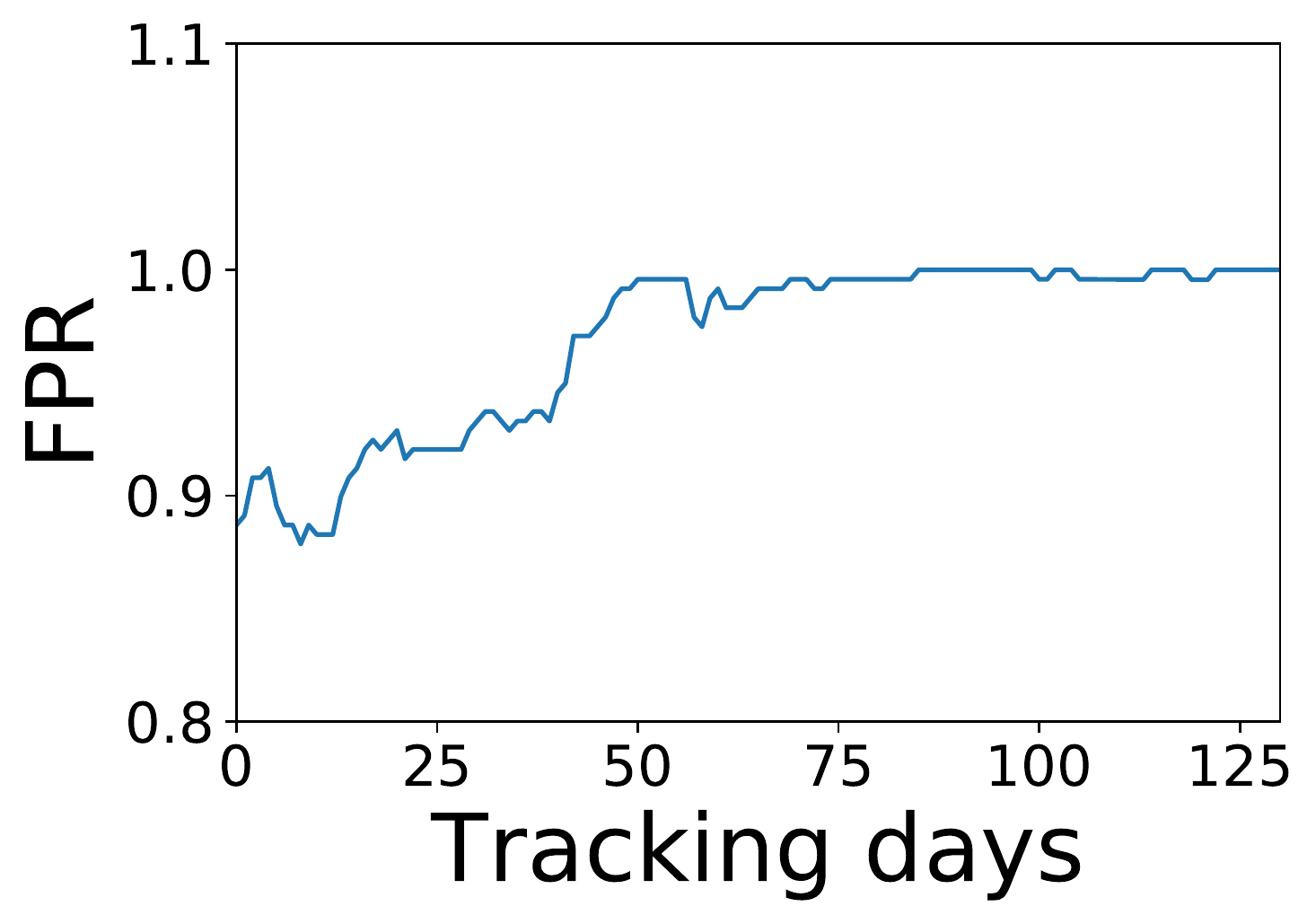}

  \label{fig:af3}
}
	\caption{False positive rate (FPR) as a function of tracking days in the attack setting when the collect frequency is set to 1.}
%	\vspace{-5mm}
	\label{fig:af}

%\end{figure*}
\end{figure}

%\begin{figure}
%\includegraphics[width=\linewidth]{pano-benign.png}
%\caption{Panopticlick, Collect Frequency = 1. True Positive Rate (TPR) over a time duration.}
%\label{fig:p_benign}	
%\end{figure}

%\begin{figure}
%\includegraphics[width=\linewidth]{rule-benign.png}
%\caption{\RLA, Collect Frequency = 1. True Positive Rate (TPR) over a time duration.}
%\label{fig:r_benign}	
%\end{figure}

%\begin{figure}
%\includegraphics[width=\linewidth]{hybrid-benign.png}
%\caption{\HLA, Collect Frequency = 1. True Positive Rate (TPR) over a time duration.}
%\label{fig:h_benign}	
%\end{figure}

%We also listed the figure \ref{fig:av_ma1} which are from FP-Stalker paper \cite{vastel2018fp}. In the paper, they used 59159 fingerprints in the fingerprinting algorithms, and we used 9000 fingerprints in the fingerprinting algorithm, so the graphs are a little different. 

\redtext{As mentioned earlier, to measure the performance of the tracking algorithms in the benign setting, we compute TPR for each tracking day. 
}Figure \ref{fig:b1}, \ref{fig:b2}, and \ref{fig:b3} show the performance of the tracking algorithms in the benign setting. Specifically, they show the TPRs as a function of tracking days (when the collect frequency was set as 1) for Panopticlick, \RLA, and \HLA, respectively. 
Like earlier, \RLA and \HLA perform better than Panopticlick.  
\vspace{-5mm}
\subsubsection{Attack Setting}
\label{sec:as}
%\vspace{-2mm}
% PS 4/9: The first two sentence should be explained in detail under Evaluation Metric subsection. 
% ZL 4/13: Commented the two sentences, moving it to the Evaluation Metric subsection with more details now. Also adding the explanation about the benign setting.

%In our analysis, we launch our attack against each user independently. In other word, we launch our attack against one user at a time, and hence total of 239 attacks were executed, each against one of the 239 users.
%\vspace{-2mm}
 To evaluate the performance of our attack and its impact on the 
 tracking of legitimate users, we compute the average of TPRs and the average of FPRs over 239 attacks. 
% TPR measures how often the fingerprint records have been correctly identified as belonging to the correct user's device. `1' is the value of TPR we want, meaning all the fingerprint records are correctly classified to the original user. FPR measures how often the attack records are incorrectly identified as belonging to the victim user's device. `1' is the value we hope to get for the FPR, meaning all attack records are classified to the victim.
%We also show the average attacks TPR and FPR. 
Figure \ref{fig:at1}, \ref{fig:at2}, and \ref{fig:at3}  show the average TPRs as a function of tracking days in the attack setting for Panopticlick, \RLA, and \HLA, respectively, when collect frequency is set to 1. 
When comparing these TPRs with those in the benign setting, 
we see only a very minor difference in the TPR scores, potentially
because of the addition of the spoofed fingerprints in the attack setting. This indicates that our attack does not have any significant impact on the performance of fingerprinting algorithms. 
%The scores in three algorithms are no lower than 0.875 with collect frequency 1, and the score is generally approaching 1, which means that the algorithms are doing a good job at tracking. 

Similarly, Figure \ref{fig:af1}, \ref{fig:af2} and \ref{fig:af3} show the average FPRs as a function of tracking days
in the attack setting for Panopticlick, \RLA, and \HLA, respectively, when the collect frequency is set to 1. 
% for Panopticlick, \RLA and \HLA, respectively. 
 We achieved average FPRs of greater than 0.95, mostly close to 1.00, which indicates that most of the spoofed fingerprints were
 misrecognized as the legitimate ones. 
 In other words, these results show that our attacks were highly successful in fooling the fingerprinting algorithms into believing 
 the spoofed fingerprints as the legitimate fingerprints. We note that similar results were achieved in both the benign and attack settings when the collect frequency was set to values other than 1.
\section{Implications of Our Attack}
\label{sec:implications}
%\secspace

%\vspace{-2mm}

As the browser fingerprinting is processed at the backend (i.e., the remote server) of the website and no web services are claiming that they are using any browser fingerprinting approaches, we could not verify the actual impact of our attacks without inspecting the backend codes of the website. However, our results show that if they were to implement only fingerprinting techniques (without integration with any of the cookies, caches, or authentication mechanisms), our attack can have a significant impact on the user's privacy and security applications as described below.  

%\textcolor{blue}{As the browser fingerprinting would be processed at the backend of each website, and there are no websites claiming they are using any browser fingerprinting features on specific web page, we could not verify the actual results of our attacks by inspecting the website backend codes. However, we can understand if our attacks are successfully, based some feedbacks when those websites are only using the browser fingerprinting. (No cookies, no caches, no other authentication method like 2 Factors authentications.)}

%\vspace{-0.5mm}
\noindent{\textbf{Compromising Ad Privacy}}: A prior study~\cite{das2016tracking} has shown that by simply monitoring the user's personalized ads, one can build the user's personal profile.
% people including attackers can know someone personal profiles (the paper said the Google personal profiles) by personalized ads. 
In our attack, the attacker is successful at presenting his device to a target website as if it is the victim user's device through various spoofing methods. 
If the target website only uses browser fingerprinting to track the user and for personalized ads, the same or similar ads, or the ads from the same category would show up on the attacker's device.
 %In our attack, we own the users' browser fingerprinting features. So when we use those features to spoof some websites which are using the browser fingerprinting algorithms in the personalized ads service, we can say the same ads, at least the same category ads as the users. 
 Given this, the attacker may learn various sensitive information about the user, including his gender, age group, the potential location of the user, his habits, and many more. 
%20210617If most ads are related to makeup products, pregnancy, 
%20210617dressing, or skirts, the user's gender might be `female'. If the ads are kids related products, like toys or games, the user may be a minor child. If the ads show service or products based on location information, the potential location of the user may be inferred. 
% If the most ads are related to makeup, pregnant dressing, or skirts, this user gender might be `female'. If the ads are all kids related products, like kids toys or kids games, the user may be a kid. If the ads show service or product with address, these addresses may in the same area, and the user may live in this area. 
%20210617 The ad categories can leak what kinds of websites or brands of products the user
%20210617 most viewed in the recent past, which will leak the user's personal habits or even income level \cite{income-brands}.
% Ads categories can leak what kinds of website or product the user most viewed before, which leak the user personal habits. 
 Further, the attacker can sell such user's information for the purpose of personal and financial gain.
 
% We may have possibilities to know who are these user, where they are living, or their personal habits. We may locate their home addresses or working locations by the location related ads, or the addresses on the ads. Let's talk this more serious: the attackers may sell these users' information as long as they obtained the victims' enough personal details. 
% PS 4/9: Not sure what you are trying to convey with following sentence.

% ZL 4/12: This sentence means with the leak of personal information, theif may steal property from this user's houser, or user's personal information are used by some people to lease phone or apply credit card, which may affect user's credit score.

 %This would give these users' property or credit loss.

%\vspace{-1mm}
%\smallskip
\noindent{\textbf{Defeating User Authentication}}: 
\redtext{The authentication schemes may not work well all the 
time if they are solely based on browser fingerprinting.
The browser fingerprinting based authentication may not 
fit well for login from a new device. Thus, }The purpose of browser fingerprinting in authentication is to remember the old device\redtext{, not the new one,} and enhance the security of traditional authentication methods such as passwords. For account login, since the attacker exposes
his device as the victim's device in our attack, the target website will misrecognize
the attacker as the victim who is using an old device, assuming that
the attacker has obtained the victim's login credentials (i.e., the
user's username and password). The authentication mechanisms
only based on browser fingerprinting cannot block such an attack.

%\vspace{-1mm}
%\smallskip
\noindent{\textbf{Bypassing Fraud Detection}}: Given the fact that many of the fraud detection techniques use browser fingerprinting information, the attacker can circumvent the detection by exposing his device as the victim's device leveraging various spoofing methods. 
%As we can always use the users' browser fingerprinting information to visit the websites which are using the browser fingerprinting based fraud detection, we can pass this detection in most of time as long as the fraud detection is only browser fingerprinting based. 
Unless the victim user does not make any major big changes on his device
% PS 4/9: Need more exaomples of such big changes
% ZL 4/12: Added two more examples here.
%20210619(e.g., changing the device operating system from Windows to Linux, reinstall a lower version of a same operating system, from Win 8.1 to Win 8.0, or replace hardware, like CPU from Intel core 4 to Intel core 2, or downgrade the graphic disk),  
(e.g., changing to a different operating system, downgrading system version, or replacing hardware)
the attacker can impersonate the
victim and bypass the detection. 
Generally, the attacker would be unaware of such big changes.
However, the attacker can always pull the most recent browser fingerprint by simply fooling the user into visiting an attacker designed website.
%In the detection algorithm, we are always the same personal as the users, 
%unless the users make some big changes in their devices, like changing the 
%device system from Windows to Linux (usually the attackers would not know 
%this). 
%So at this moment, the attacker may use another user's browser 
%fingerprinting to continue visit this website to pass the fraud detection 
%and plunder valuable information, or try to obtain the current users' 
%updated browser fingerprinting information. 
Given this, the fraud-detection algorithm cannot thwart our attack
solely based on browser fingerprinting. It needs some additional metrics 
to detect fraud.
%The most dangerous of this 
%implication is: the attackers may not be blocked by the browser 
%fingerprinting-based only fraud detection forever.

%\vspace{-1mm}
\section{Discussion}
\label{sec:discussions}
%\secspace

%\vspace{-1mm}
\noindent{\textbf{Potential Attack Detection}}: 
The web service may detect our attack 
if the adversary does not follow the correct data format, provides invalid data, or takes time longer than the set time limit. However,  the attack can remain undetected if the adversary carefully 
provides the correct and valid spoofed data within the set time limit. 
To use the script injection approach, the attacker should use a
valid value to replace the Javascript API values, e.g., in the $\mathtt{Date()}$ object, `year' should be replaced with `year' (not  `month'). 
When employing the script modification approach, the attacker 
has to use the  correct data format in the return value, e.g., `2020-04-12' can be replaced with `2020-03-29', but not with `2020.04.12'. 
We note that the spoofed date should not be older than the current date.
To detect our attack, the web service may periodically request
a response from the website running in the client machine, e.g., request the current time for every 5 seconds. 
%
%The attacker should also be careful for the function which is set by time to send data to the web service, like sending the current date every 5 seconds. 
When modifying the script, the adversary 
 needs to stop all the scripts on the website, 
and thus prevent the website from sending the response to  
the web service. When the web service does not receive 
the expected response from the client machine in a timely manner,  it can detect the potential attack. 
%all the functions in the script would stop working, 
However, the attacker can use a pre-designed script to overwrite the  \redtext{(part of the)
}existing scripts in the targeted website. The use of such a pre-designed script automates the script modification process, thereby defeating the above detection approach.
% When the attacker spoof some features, he should be careful to other features which have relations to the spoofed ones. For example, the `Navigator.platform' would give the platform information of the device. The possible value could be `MacIntel', `Win32', `Linux i686', etc. The `User Agent' in HTTP header contains the Operating System information, like `Win 10' or `MacOS'. So when the attacker spoof `Platform', he also should spoof `User Agent' to prevent the web service possible detection. 
 
% If the data format is correct, the data value is reasonable, and script response time is the same as the one set by the website, the attack can not be noticed.
\label{sec: PAD}

\noindent{\textbf{Limitations and Future Work}}:
 \redtext{As mentioned earlier, there exist several browser fingerprinting techniques that leverage multitude of browser and system attributes for fingerprinting. 
}Although the fingerprinting techniques, considered in our study, utilize many of the attributes, they exclude several attributes used in other fingerprinting algorithms, such as the ones related to network and protocols (e.g., TCP/IP stack fingerprinting \cite{glaser2000tcp}, DNS resolver \cite{kim2011effective}), and the hardware sensors \cite{das2016tracking} on the performance of our attack. 
%\bluetext{Hardware sensors can be fingerprinted based on the variation in the calibration error of motion-position sensors (e.g., accelerometer) or frequency-response of audio sensors (e.g., speaker, microphone).}
% PS 4/10: Need more examples of features that are not already included in our study.
% ZL 4/12: Added.
The impact of these attributes on the performance of our attack has not been assessed in our study. 
Further investigation is needed to explore this direction.
%In the future work of website attacks, we will test more features, like another feature set: “Network and Protocol level Techniques” (Application-level APIs on client devices. Network- and protocol-level techniques can also be used as the features.). Those features are not included in our current website attacks. Our current website attacks  are based on the browser level.
As noted earlier, 
the current dataset is insufficient 
to evaluate the performance of fingerprinting algorithms and that of our attacks after 130 tracking days. 
Further study with a larger dataset would be needed to 
assess the performance of our attack for a longer tracking period.
%In the future work of algorithm attacks, we first solve the weaknesses of our experiment -- using a way more large dataset to do our experiments. We will use the appropriate dataset to calculate the correct TPR curve at least after the 130 days. 
Our study assumes that the attacker can fully spoof all  browser information obtained from the victim's device. 
In some scenarios, the spoofed information may be outdated. 
% PS 4/10: More examples needed.
% ZL 4/12: Added.
%20210619For instance, the victim user may update his operating system, install new fonts, move to another city, or replace the hardware component of the device (e.g., CPU or graphic disk) after the attacker has obtained the victim's browser information. 
In such a scenario, only partial browser information is correctly spoofed that may impact our attack.
Future work would be needed to evaluate the impact of partial spoofing on the performance of our attack.
%The second work would be using parts of the user instances value in the attack instances. Our current experiments are still based on the fully copy of attacked users' instances value. Like the second paragraph in 5.2, we would set different combination of the values in the attack instances and run the tracking algorithm to evaluate the results, then do our attacks analysis. The analysis results may give us a purely different graph compared to the fully copy.
Furthermore, an ethically-sound study of attacking personalized ads, authentication
and fraud detection schemes that use fingerprinting in the real world
via \GB should be conducted in future work.
Our spoofing methods (detailed in Section  \ref{sec:VAAPS}) can also be extended as an evasion technique 
%20210619(e.g., to defeat targeted advertising based on fingerprinting) 
that can obfuscate the true user’s identity by creating and supplying a fake browser fingerprint to the visiting website. Similar to \GB, the evasion can be oblivious to the target website. The impact of such evasion and subtle difference between \GB and the evasion technique should be evaluated and discussed further in future work.
%\textcolor{blue}{Meanwhile, we are considering to use the spoofing methods in Section \ref{sec:VAAPS} as a potential evasion technique. We can create fake browser fingerprinting features in our browser if we don't want the website know who we are. Similar to the spoofing steps, we can evade the website detection. But the differences between spoofing and evasion should be discussed further in our future work.}
%\vspace{-3mm}

%\vspace{-2mm}
\section{Conclusion}
\label{sec:conclusion}
%\secspace

%\vspace{-1mm}
%\vspace{-2mm}
In this paper, we identified a novel and serious threat akin to the
well-studied and popular notion of browser fingerprinting.  Specifically, we
showed that an attacker can make its own browser appear as the victim's browser
by simply capturing (through an attacker-controlled or a malicious website) and
mimicking the browser fingerprint (through script injection/modification or the leveraging browser's built-in settings and debugging tools).  By exploiting
this threat, we introduced and designed \textit{\GB}, an attack system that
would enable a malicious entity to subvert any web application that uses
browser fingerprinting, for example, to glean various sensitive information
about the user in a targeted advertising application and to compromise the
security of online defensive schemes, such as user authentication and fraud
detection.  We employed state-of-the-art browser fingerprinting techniques,
Panopticlick and FP-Stalker, and evaluated the performance of \GB against these
algorithms.  Our results showed that \GB can successfully impersonate the
victim's browser transparently almost all the time without affecting the
tracking of legitimate users.  Since acquiring and spoofing the browser
characteristics is oblivious to both the user and the remote web-server, \GB
can be launched easily while remaining hard to detect.  The impact of \GB can
be devastating and lasting on the online security and privacy of the users,
especially given that browser-fingerprinting is starting to get widely adopted
in the real world. In light of this attack, our work raises the question of whether browser fingerprinting is safe to deploy on a large scale.

\bibliographystyle{splncs04}

\bibliography{references}

%\newpage
%\vspace{5mm}
%\section*{Appendix}
%\input{appendix}
%=======
%\newpage
%\vspace{5mm}

%\vspace{5mm}
\newpage
\section*{Appendix}
%\clearpage

\appendix

\section{Examples of Applying Spoofing Methods}
\label{sec:example}

Below we describe how we spoof different attributes listed in Table  \ref{tbl:features_categories}.

\noindent{\textbf{{C1: Browser-provided Information}:}}
To spoof  user-agent (C1-1), cookie enabled (C1-5), and do not track (C1-12) attributes, we used all three spoofing approaches mentioned earlier.
A separate script can also be designed to overwrite respective JavaScript APIs -- $\mathtt{navigator.userAgent}$, $\mathtt{navigator.cookieEnabled}$ and $\mathtt{navigator.doNotTrack}$ through script injection such that they return the spoofed value when they are invoked. 
We employ the following approach when using browser setting and debugging tool.
In the Google Chrome browser, we set the user-agent through its debugging tool. In the Firefox browser, we set the user-agent in its $\mathtt{about:config}$ setting page. 
The cookie enable and do not track attribute can be easily spoofed by simply going through the browser setting.
These three attributes can also be spoofed by changing their associated variables in the object being sent to the remote server, i.e. through script modification.

We utilize the script injection and script modification methods to spoof  system time (C1-3), battery information (C1-4), Platform (C1-8), and resolution (C1-11). \redtext{Specifically, we built a script to overwrite their respective JavaScript APIs -- $\mathtt{Date()}$, $\mathtt{navigator.battery}$, $\mathtt{navigator.platform}$, and $\mathtt{screen}$. 
They can also be spoofed by replacing related variables in the return object with the spoofed values.
}To spoof WebGL (C1-2) information, WebRTC characteristics (C1-6), 
password autofill feature (C1-7), and local storage (C1-10) attributes, we employ the script modification method. 
\redtext{Specifically, we looked for related variables in the response object and 
replaced them with corresponding spoofed values before sending it to the remote server. 
}The remote server may verify the validity of WebRTC information based on the browser version and the operating system. Fortunately, these attributes can also be spoofed.
For spoofing the language, we used the browser setting option.

\noindent{\textbf{{C2: Inference based on Device Behavior}:}}
We used script modification approach to spoof all the attributes in this category, such as canvas information (C2-1), system performance (C2-2), font list (C2-3), scroll wheel related attributes (C2-4), and CSS feature (C2-5). Specifically, we changed the variables related to these attributes in the return object before sending it to the remote web server.

\noindent{\textbf{{C3: Extensions and Plugins}:}}
%\noindent\textit{\textbf{Browser plugins fingerprinting (C3-1)}}: 
For spoofing the plugins information (C3-1), we used the script injection and script modification methods. 
We  overwrite the JavaScript $\mathtt{navigator.plugins}$ API to modify the plugins attributes with the spoofed values. 
We also changed the related variable in the server's return object for spoofing.
We used script modification approach to spoof the browser extension related attributes (C3-2). 
%\noindent\textit{\textbf{Browser extension fingerprinting (C3-2)}}: 
The website script can inspect if an extension has been installed in the browser. If the ads field in the website is replaced or deleted, it shows that the browser may have the extension to block the ads. We used script modification to change these results.
\redtext{in the server's return object.}

%\begin{enumerate}[leftmargin = *]
%\item  \textit{\textbf{Snapshot of Visual Attack}}
\section{Visual Attack and Results}
%\section{Snapshot of Visual Attack}
\label{sec:vpfp}

%\vspace{-2mm}
%The results are: we spoofed ``https://panopticlick.eff.org/" with all of the users' combination browser fingerprinting features.
%
%I listed
%several screenshots of my ``MacOS+chrome" original features, and the comparison of different combinations with spoofed combination features: Figure 
%\vspace{-2mm}

\subsection{Visual Attack}

We have introduced the visual attack against Panopticlick site. In this section, we will describe how we attack FingerprintJS site and a real-life fingerprint service.

\noindent\textbf{Attacking FingerprintJS Site:}
Unlike the Panopticlick website, the FingerprintJS \textcolor{blue}{}pro service website does not show all the fingerprint features, but it provides the user's browsing history which can help to prove that our spoofing is successful. We use the browser setting and debugging tool to spoof the user-agent and languages, and use the script modification approach to spoof other features collected by the FingerprintJS website. Figure \ref{fig:fjsvictim} in Appendix \ref{sec:vpfp} presents a snapshot of the FingerprintJS dashboard showing fingerprint information when a user uses the Google Chrome browser on a Windows machine, i.e., “Win+Chrome”. Since we were not sure if all values in the return object
(e.g., `rid', `cv', `url', etc) are used to construct the unique ID at the FingerprintJS remote server, we spoofed all the variables extracted from the browser.

\noindent\textbf{Attacking Real-Life Fingerprinting Service:}
Attacking a real-world fingerprinting based service
can test the strength of our spoofing attack. Although FingerprintJS pro service did not provide all the fingerprint features they used in their service on the result page, FingerprintJS pro service website has the open source code \cite{fingerprintjsopensource} which can output all the fingerprint features in browser console. We deployed this open source script on our own server without making any changes and called this website \textit{Testing Site}, then used \textit{Script Injection} and \textit{Script Modification} to do the attack. We did not combine those two methods together. Each attack method can be seen as an independent attack. In \textit{Script Injection}, we used selenium to change all the fingerprint features that are used in the FingerprintJS open source code except fonts, and then visit the \textit{Testing Site}. We changed supported fonts in the operating system language setting. In \textit{Script Modification}, we first set break point at the beginning of the \textit{Testing Site}. This break point is set in the browser debugging tool, not the server. Then we spoofed all the values in the \textit{return value} in the script main function, as those values are all fingerprint feature values and \textit{haslied} detection values. \textit{haslied} detection functions include \textit{hasLiedLanguagesKey}, \textit{hasLiedResolutionKey}, \textit{hasLiedOsKey} and \textit{hasLiedBrowserKey}. \textit{hasLiedLanguagesKey} checks the consistency of values in two APIs ``navigator.languages'' and ``navigator.language''. \textit{hasLiedResolutionKey} compares if value of APIs ``window.screen.width'' is less than ``window.screen.availWidth'', or ``window.screen.height'' is less than ``window.screen.availHeight''. \textit{hasLiedOsKey} detects if the operating system value in APIs ``navigator.userAgent'', ``navigator.oscpu'' or ``navigator.platform'' are spoofed. \textit{hasLiedBrowserKey} extracts value in APIs ``navigator.userAgent'' and ``navigator.productSub'' to find the spoofed browser features.
This experiment demonstrates that we can effective execute the spoofing attack against any real-life service that deploys this FingerprintJS open source code. 

\subsection{Result}

\begin{figure}[H]
	\centering
	\includegraphics[width=.7\linewidth]{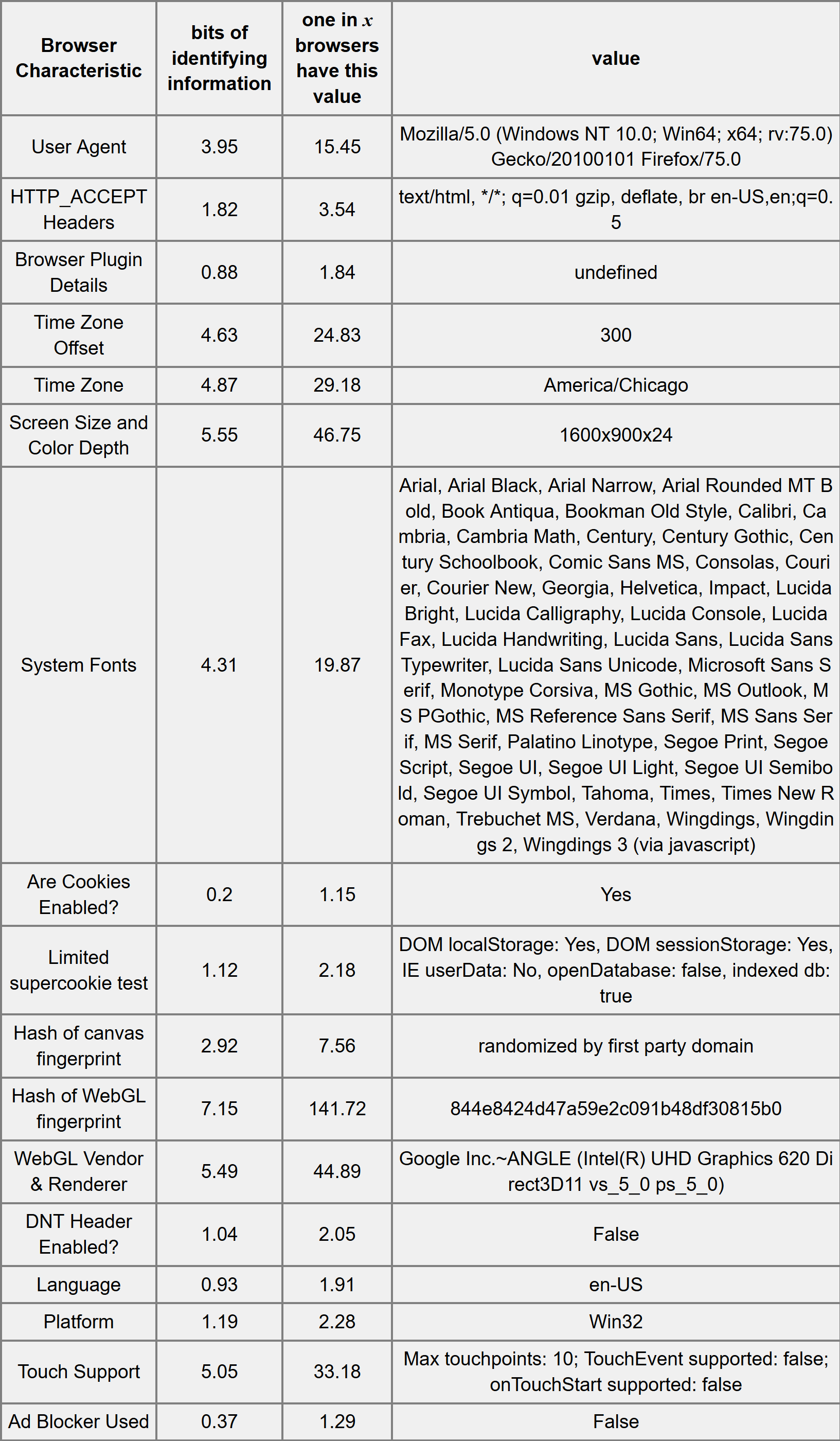}
%	\vspace{2mm}
	\caption{Victim User: The original features of combination ``Win+Firefox''. Testing website: \textbf{Panopticlick}}
	\label{fig:win_firefox}
	\vspace{1.5mm}
\end{figure}

\begin{figure}[H]
	\centering
	\includegraphics[width=.6\linewidth]{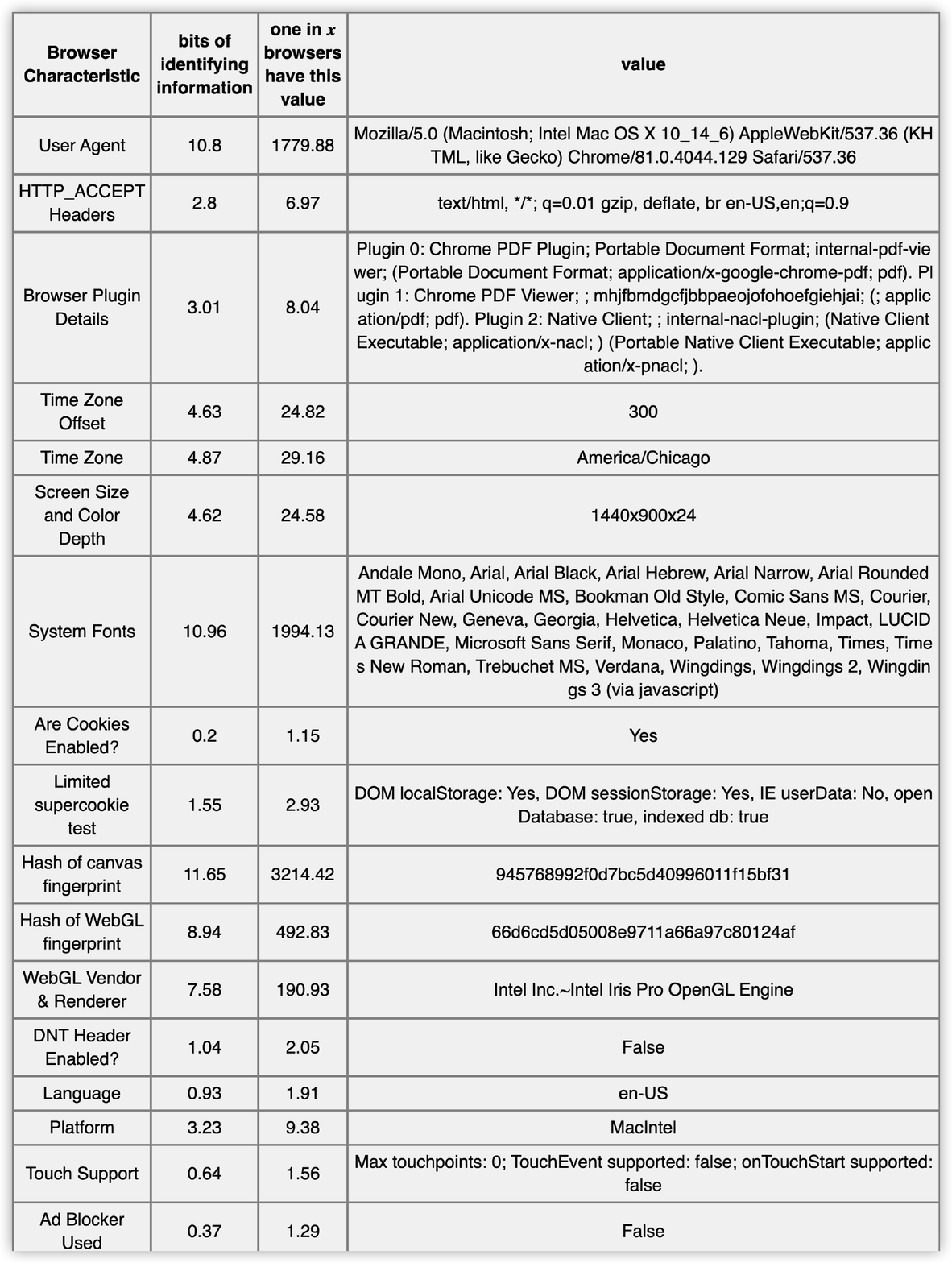}
%	\vspace{3mm}
	\caption{Attacker (before the attack): The original features of combination ``Mac+Chrome''. Testing website: \textbf{Panopticlick}}
	\label{fig:mac_chrome}
	\centering
\includegraphics[width=.7\linewidth]{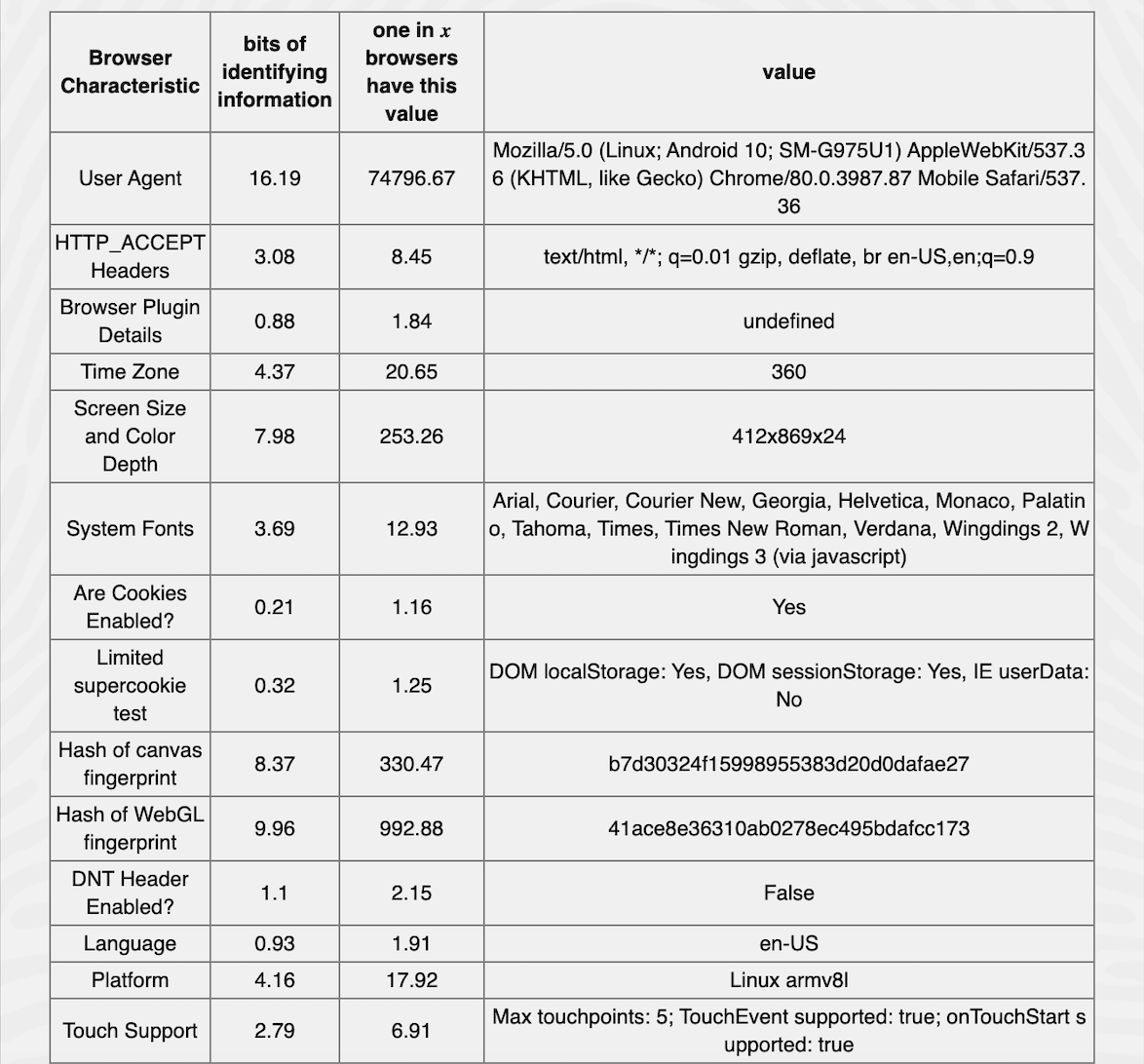}
\vspace{3mm}
	\caption{Attacker (after the attack): The spoofed features of combination ``Android+Chrome''. Testing website: \textbf{Panopticlick}}
	\label{fig:mobile}
%	\vspace{1mm}
\end{figure}

\begin{figure}[H]
	\centering
	\includegraphics[width=.6\linewidth]{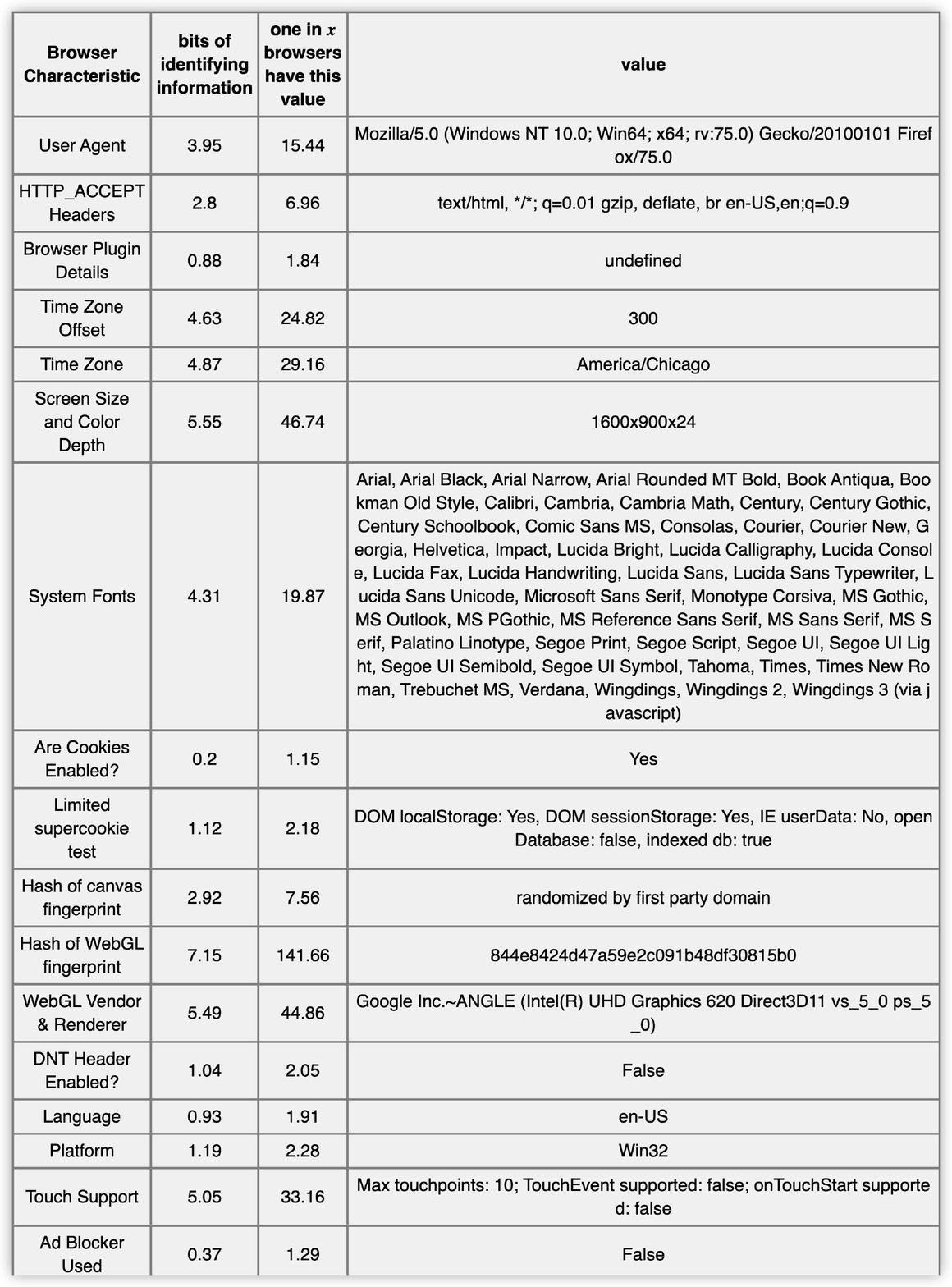}
%\vspace{3mm}
	\caption{Attacker (after the attack): The spoofed features of combination ``Win+Firefox''. Testing website: \textbf{Panopticlick}}
	\label{fig:m_c_to_w_f}
	\centering
\includegraphics[width=.7\linewidth]{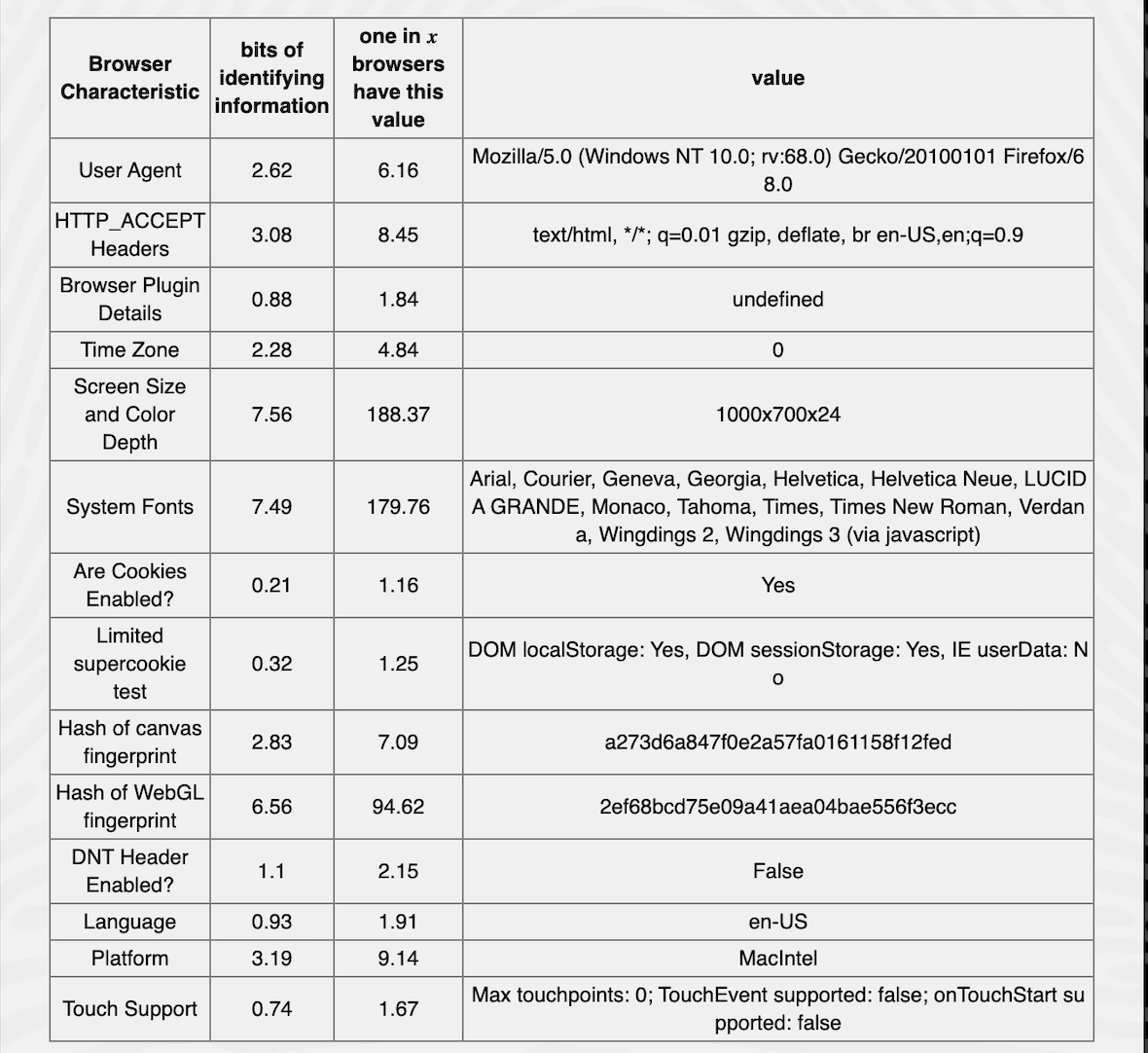}
\vspace{3mm}
	\caption{Attacker (after the attack): The spoofed features of combination ``Mac+Tor''. Testing website: \textbf{Panopticlick}}
	\label{fig:Tor}
%		\vspace{1mm}
\end{figure}

\noindent{\textbf{Result: Attacking Panopticlick Site:}} Figure \ref{fig:mac_chrome} shows a snapshot of 
the \textit{actual} device information shown by the Panopticlick website when using the \textit{Mac+Chrome} setting.
%\bluetext{Similarly, Figure \ref{fig:win_firefox} shows the \textit{actual} device information when Panopticlick website is visited through the Firefox browser in the Windows terminal, referred to as \textit{Win-Firefox} setting.
%}
To spoof the fingerprint of the victim's device, i.e., \textit{Win+Firefox}, (as shown in Figure \ref{fig:win_firefox}), 
we employed the three methods discussed earlier in Section \ref{sec:spoofing_methods} 
such that \textit{Mac+Chrome} (i.e., the attacker's device) appears as \textit{Win+Firefox} (i.e., the victim's device). 
After spoofing using \textit{Mac+Chrome}, the Panopticlick website shows the browser information as depicted in 
Figure \ref{fig:m_c_to_w_f}, which is exactly the same as that when using \textit{Win+Firefox} (i.e., the victim's device).
This indicates that our spoofing methods were successful in 
replicating the victim's browser fingerprint.
%In the process of spoofing Panopticlick website, we set the break point before the `post' function which contains the fingerprint object `whorls_v2' in the script `fetch_whorls.js'. In the `debugger', we change the fingerprint values in the object `whorls_v2'. Then we resume the breakpoint, and spoofed fingerprint object is sent to the web service, and the result page will display the feature value we just changed. 
We were even able to spoof the Google Chrome browser on the Android phone, the \textit{Android+Chrome} setting, and the Tor browser on the Mac laptop, the \textit{Mac+Tor} setting, using Mac+Chrome.  
Figure \ref{fig:mobile} and Figure \ref{fig:Tor} show the snapshots of spoofing \textit{Android+Chrome} and \textit{Mac+Tor} using \textit{Mac+Chrome}, respectively. 
We achieved similar results when spoofing the device information obtained from various other terminal-browser combinations using the \textit{Mac+Chrome} setting.
This indicates that our spoofing methods succeeded to spoof all the fingerprinting information, regardless of the victim's terminal-browser combination.

%Figure  \ref{fig:fjsattacker} provides the snapshot of actual device information from the FingerprintJS website in the \textit{Mac+Chrome} setting. Figure \ref{fig:fjsspoof} shows the snapshot of spoofing the \textit{Windows-Chrome} setting using the \textit{Mac+Chrome} setting. 

\begin{figure}[H]
\centering
\includegraphics[width=.65\linewidth]{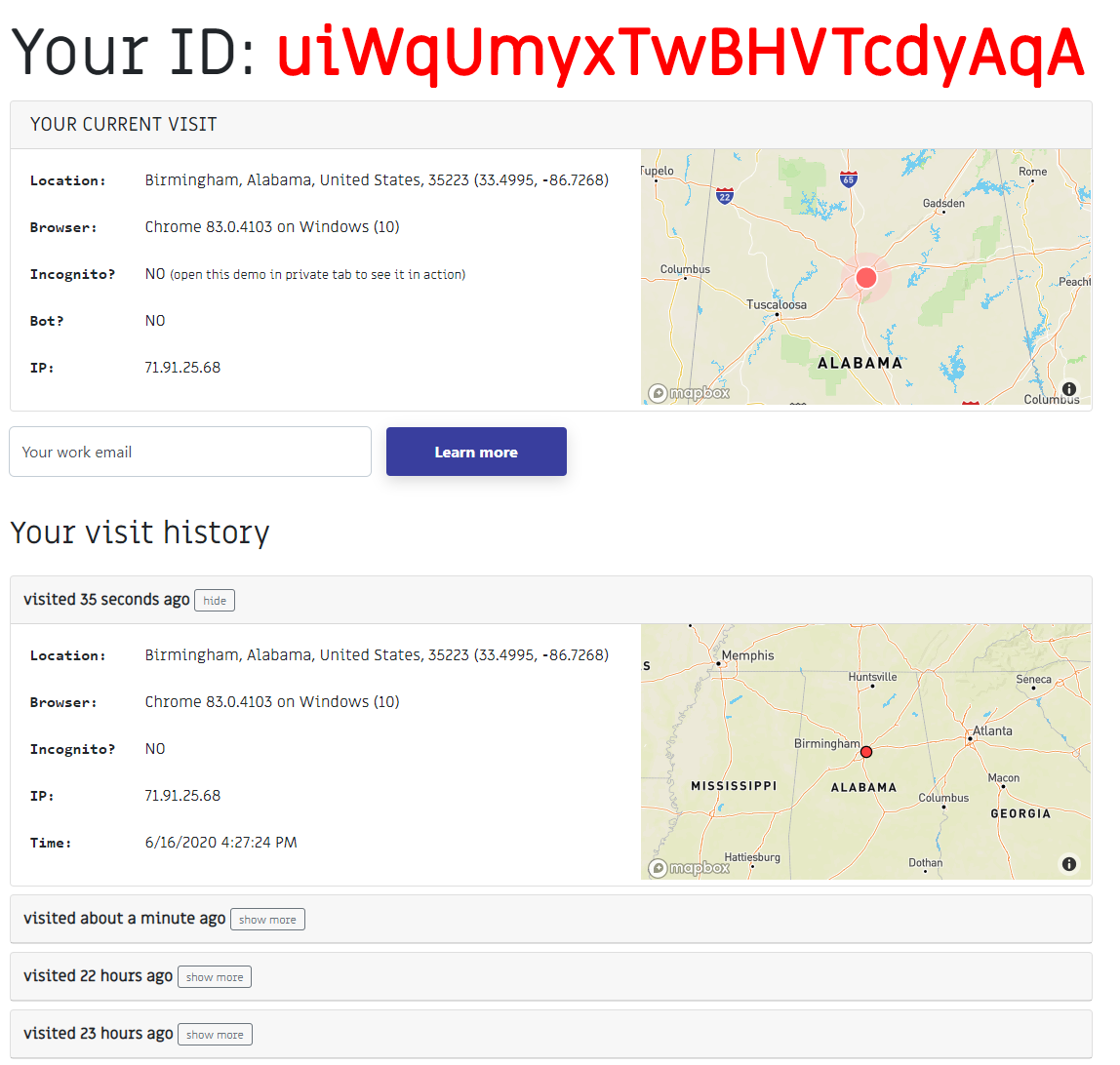}
%	\vspace{2mm}
	\caption{Victim User: The original features of combination ``Win+Chrome''. Testing website: \textbf{FingerprintJS}}
	\label{fig:fjsvictim}

\centering
\includegraphics[width=.65\linewidth]{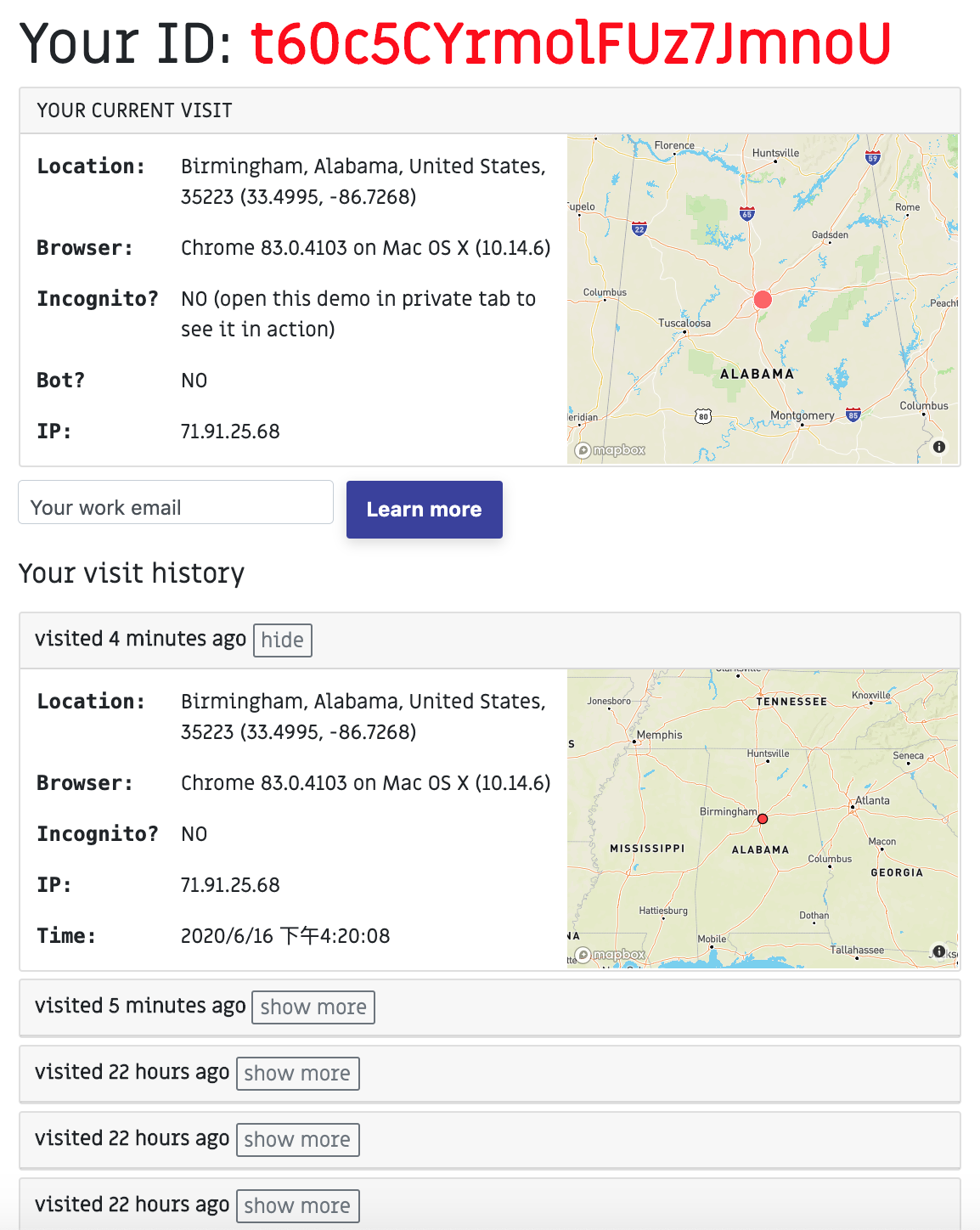}
%\vspace{3mm}
	\caption{Attacker (before the attack): The original features of combination ``Mac+Chrome''. Testing website: \textbf{FingerprintJS}}
	\label{fig:fjsattacker}
\end{figure}

\begin{figure}[H]
\centering
\includegraphics[width=.7\linewidth]{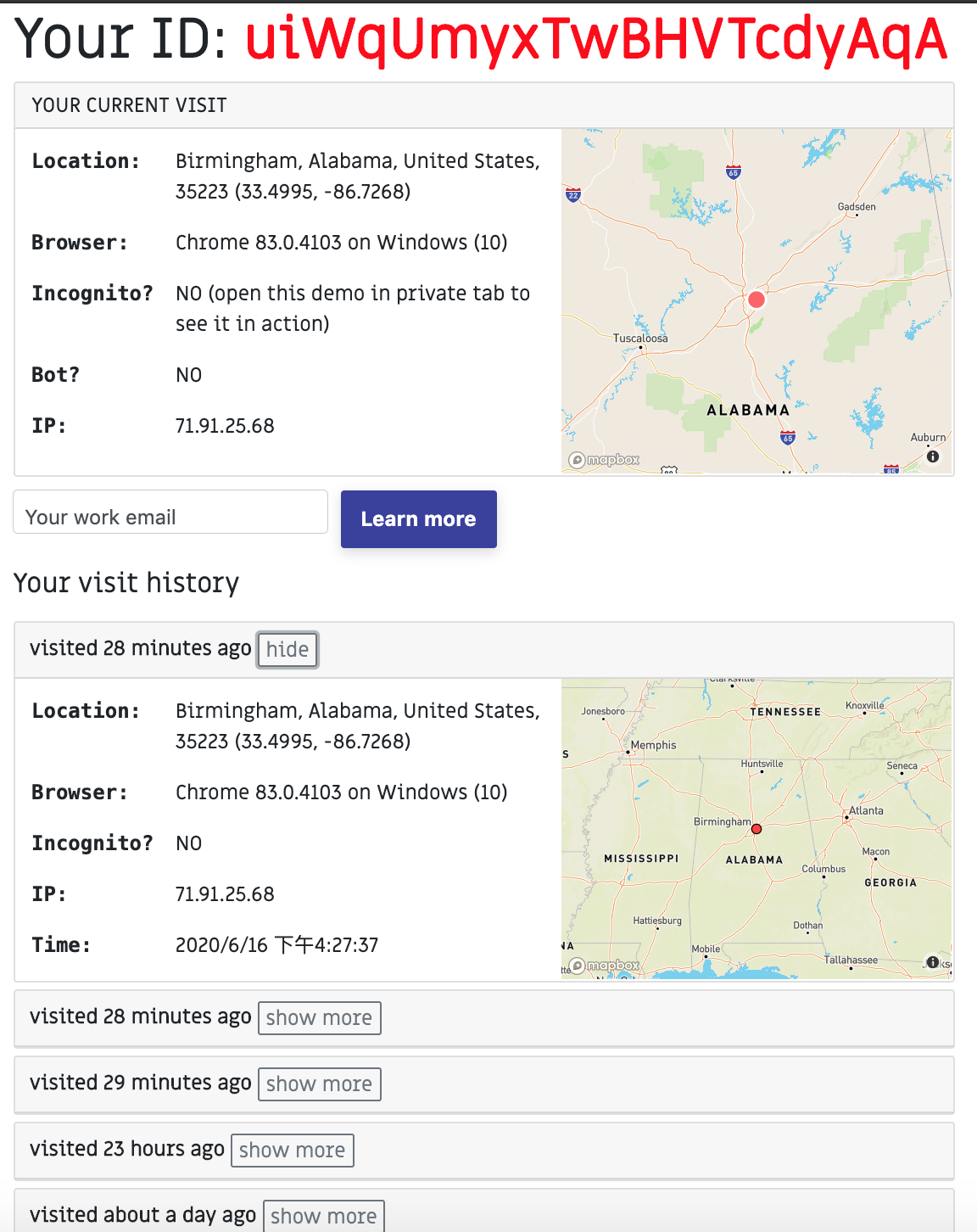}
%\vspace{2mm}
	\caption{Attacker (after the attack): The spoofed features of combination ``Win+Chrome'', showed in Google Chrome of combination ``Mac+Chrome”. Testing website: \textbf{FingerprintJS}}
	\label{fig:fjsspoof}
\end{figure}

\noindent{\textbf{Result: Attacking FingerprintJS Site:}} Figure \ref{fig:fjsattacker} and Figure \ref{fig:fjsvictim} present the snapshots of original device information corresponding to the attacker and the victim, respectively, when they visit the \textit{FingerprintJS} website. Figure \ref{fig:fjsspoof} shows the snapshot of the attacker's device information when he has spoofed the victim's device information. As can be seen from the figure, we could check the victim’s browsing history after spoofing.

\noindent{\textbf{Result: Attacking Real-Life Fingeprinting Services:}} In attacking \textit{Testing site} that deployed the FingerprintJS open source code, our two attack methods \textit{Script Injection} and \textit{Script Modification} all successfully spoofed all the 29 fingerprinting features listed in the script, and passed four \textit{haslied} detection functions which are used to detect if the current visit used spoofed fingerprinting features or not. We picked the \textit{Windows-Firefox} as the victim device operating system and browser setting, and used \textit{Mac-Chrome} as the attacker device.

\newpage

%Figure \ref{fig:win_firefox} and \ref{fig:mac_chrome} are the snapshots of victim and attacker device information from the website \textit{Panopticlick}. Figure \ref{fig:m_c_to_w_f} is the snapshot when the attacker spoofed the vicitm device information to visit the website \textit{Panopticlick}. 

%Figure \ref{fig:fjsattacker} and Figure \ref{fig:fjsvictim} present the snapshots of original device information corresponding to the attacker and the victim, respectively, when they visit the \textit{FingerprintJS} website. Figure \ref{fig:fjsspoof} shows the snapshot of the attacker's device information when he has spoofed the victim's device information.

%Figure \ref{fig:mobile} and Figure \ref{fig:Tor} show the snapshots
%of our visual attacks when spoofing Android+Chrome and Mac+Tor, respectively, using Mac+Chrome against the Panopticlick website.

%We executed visual attack using three instances of fingerprints from all the users in FP-Stalker dataset. Presenting all the snapshots of these attacks are not feasible. However, we provide snapshots of our visual attacks against three random users in the dataset in 
%Figures [\ref{fig:u1r1} -- \ref{fig:u3r3-2}] 

%present the snapshots of visual attacks when using three different fingerprints chosen from 
%the FP-Stalker dataset. 

%We used the Panopticlick website on the Mac-Chrome setting to launch all our visual attacks.

\section{Example of Script Injection}
%\section{Snapshot of Visual Attack}
\label{sec:efi}

	An example script to overwrite the platform information acquired through $\mathtt{navigator.platform}$ API~\cite{ref6} is presented below.
	
%	\vspace{-1mm}
{\scriptsize
\begin{lstlisting}[breaklines=true]
var Injectcode=``Object.defineProperty(navigator,`platform',{get:function(){return `my platform';}});";
var script = document.createElement(`script');
script.appendChild(document.createTextNode(Injectcode));
(document.head || document.documentElement).appendChild(script);
script.parentNode.removeChild(script);
\end{lstlisting}
}
%\vspace{-1mm}

In the example script above, a new object $\mathtt{navigator.platform}$ is defined with the value ``my platform''. 
The object is then injected at the start of all the scripts extracted from the target website.
%Then we inject this function at the start of all the scripts obtained from the website. 
Given this, when the script calls the $\mathtt{navigator.platform}$ API, 
it receives the spoofed value, i.e., ``my platform'', because 
the original API object has been overwritten by the injected object.

Thus, when using such codes (added as a browser extension or ran in Selenium), the remote web server would always receive the spoofed values added by the injected codes. In theory, any of the Javascript APIs can be self-designed and/or overwritten. However, the $\mathtt{getOwnPropertyDescriptor}$ API can be used to detect if any specific property exists in a web object or property. This API returns “undefined” if a property, say $\mathtt{navigator.platform}$, has not been defined or overwritten. It returns an object if the property exists and has been overwritten. Fortunately, we can also overwrite the $\mathtt{getOwnPropertyDescriptor}$ API itself such that it always returns “undefined”, indicating there has not been any manipulation on the object. Therefore, any detection mechanism solely based on JavaScript would not work on script injection and modification. Thus, our spoofing method is hard to detect as noted in Table \ref{tbl:features_categories}. As this method pre-changed the API values, attacks using this method can be finished automatically.

%\section{Performance of \GB at Different Collect Frequency}
%\label{sec:tprfpr}
%\input{tprfpr}
%\clearpage
%\item  \textit{\textbf{Performance Comparison of FP-Stalker Implementation}}

\section{Performance Comparison of FP-Stalker Implementation}
\label{sec:bsta}

\begin{figure}[t]
	\subfloat[FP-Stalker \cite{spirals-team}]{
		\includegraphics[width=.485\linewidth]{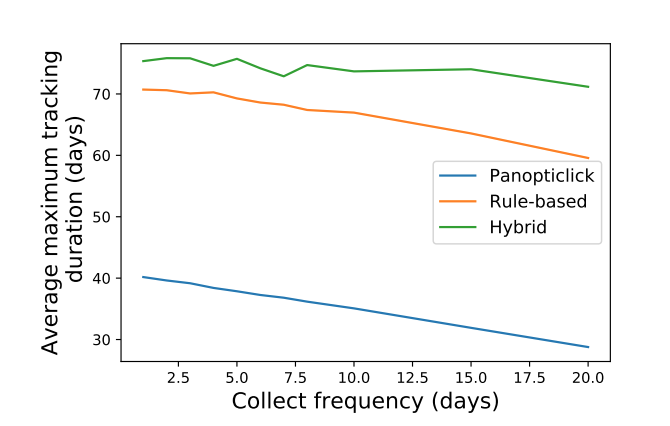}	
		\label{fig:av_ma1}}
	%\hfill
	\subfloat[Our Result]{
		\includegraphics[width=.44\linewidth]{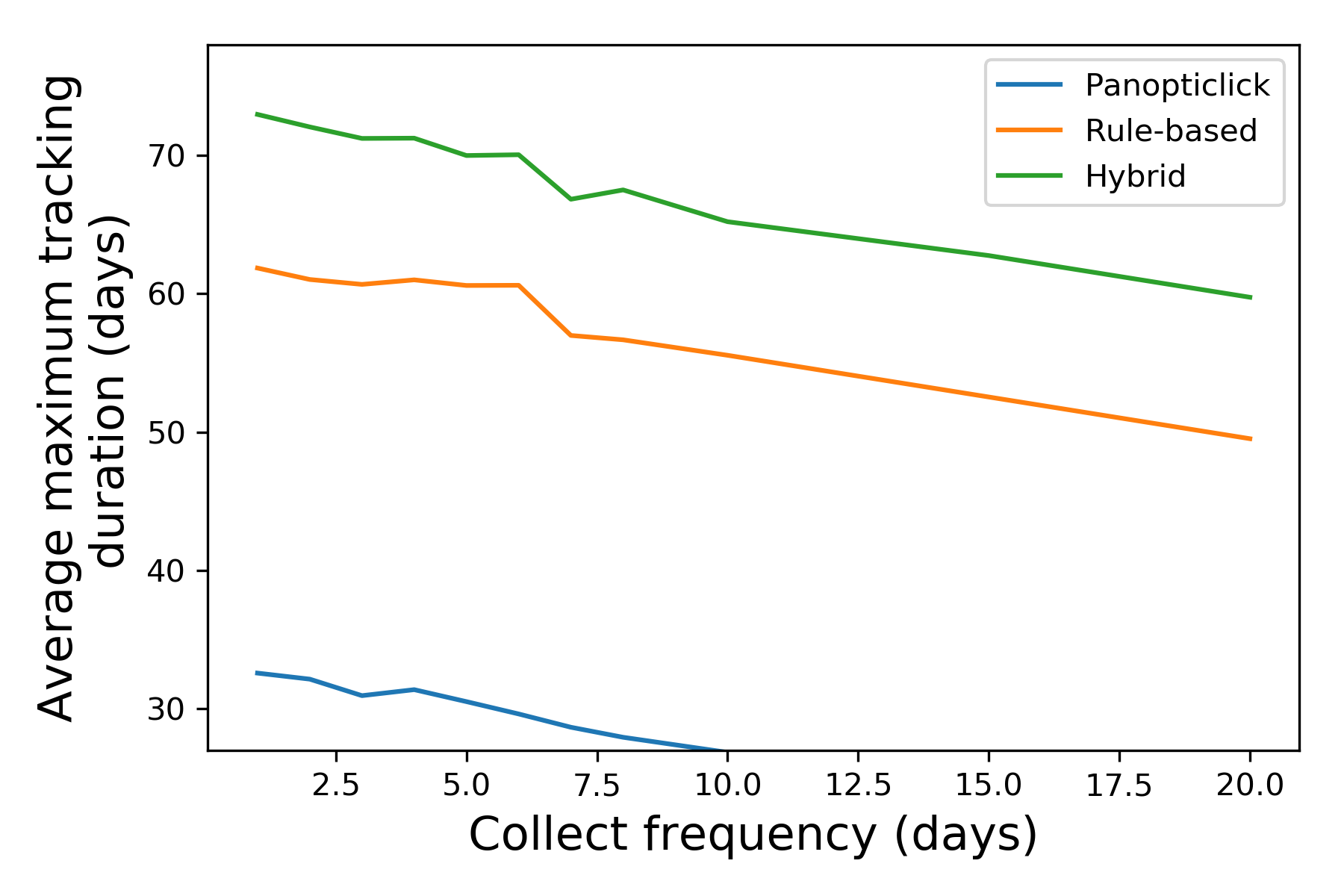}
		
		\label{fig:av_ma2}
	}
	
	\caption{Average of maximum tracking duration (in terms of days) as a function of collect frequency for three different browser fingerprinting techniques.}

	\label{fig:av_ma}

	\subfloat[FP-Stalker \cite{spirals-team}]{
  \includegraphics[width=.47\linewidth]{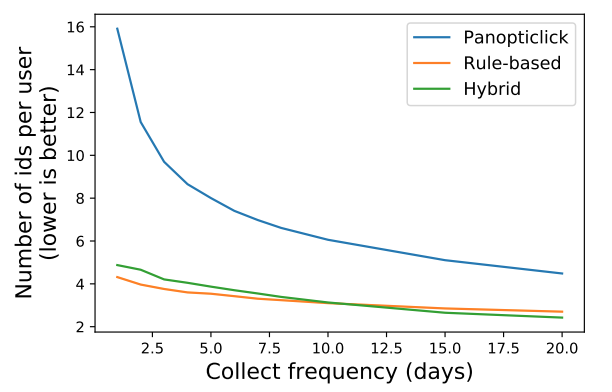}
  \label{fig:nu1}}
  %\hfill
  \subfloat[Our Result]{
  \includegraphics[width=.46\linewidth]{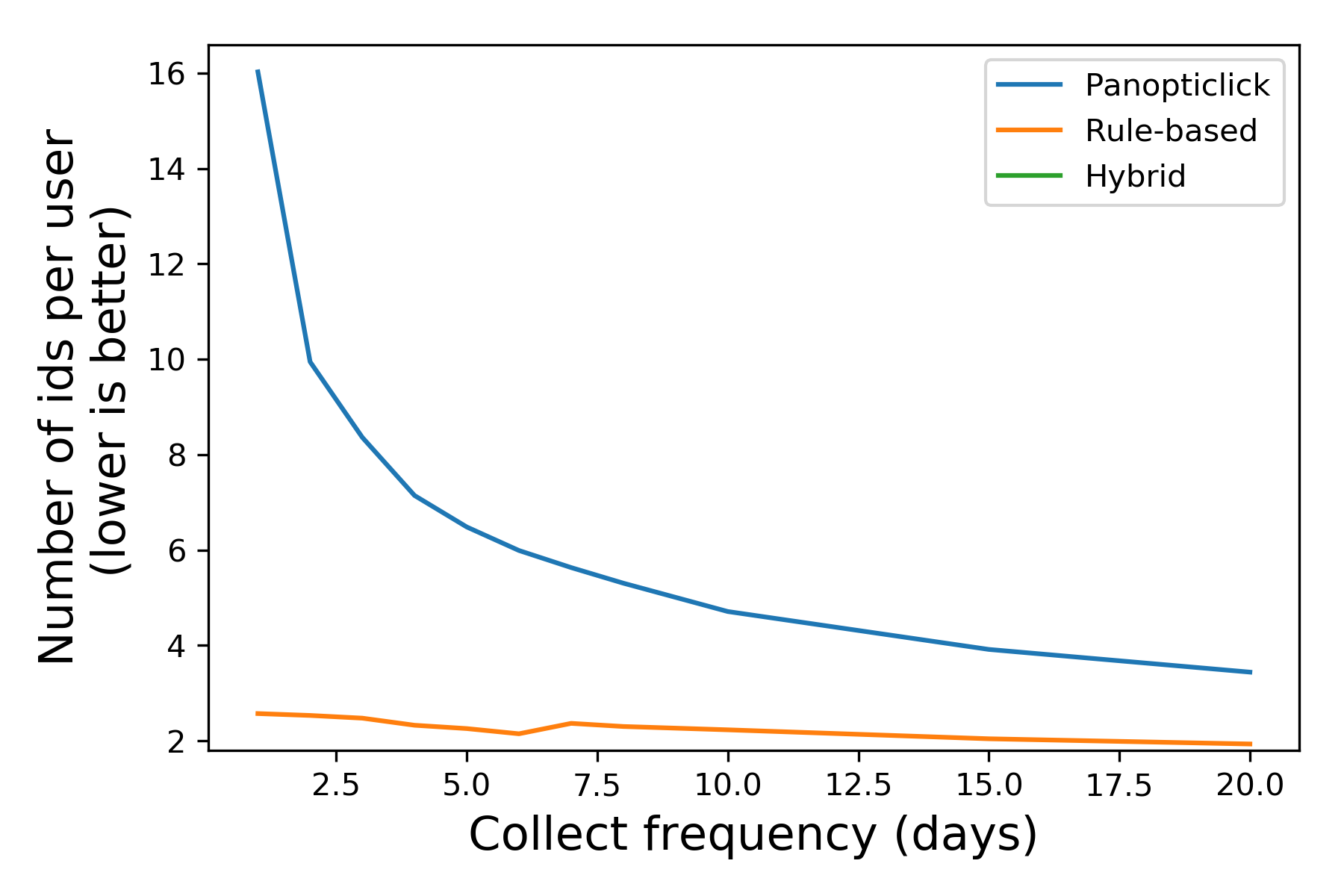}

  \label{fig:nu2}
}
	\caption{Average number of IDs assigned per user as a function of collect frequency for the three different fingerprinting algorithms.}
	\label{fig:nu}
\end{figure}

Figure \ref{fig:av_ma2} and Figure \ref{fig:nu2} show the average of maximum tracking duration and the average number of IDs assigned per user
 as a function of collect frequency for three different browser fingerprinting techniques, respectively.

%\end{enumerate}

\end{document}